\documentclass[aps,pre,twocolumn,superscriptaddress,preprintnumbers,nofootinbib]{revtex4-2}
\usepackage{graphicx,epstopdf,amssymb,amsmath,verbatim,mathdots}
\usepackage{xcolor}
\pdfoutput=1
\usepackage{ragged2e}
\usepackage[font={footnotesize}, skip=0pt,format=plain,singlelinecheck=false
]{caption}

\usepackage[hidelinks]{hyperref}

\newcommand{\kobe}[1]{{{\textbf{[KL:\ #1]}}}}
\newcommand{\sarah}[1]{{{\textbf{[ZF:\ #1]}}}}
\newcommand{\akshit}[1]{{{\textbf{[AG:\ #1]}}}}
\newcommand{\pankaj}[1]{{{\textbf{[PM:\ #1]}}}}

\newcommand{\clean}{
\renewcommand{\kobe}[1]{}
\renewcommand{\sarah}[1]{}
\renewcommand{\akshit}[1]{}
\renewcommand{\pankaj}[1]{}
}
\clean

\begin{document}

\preprint{MIT-CTP/5928}

\title{Emergent frequency-dependent selection predicts mutation outcomes\\ in complex ecological communities}

\author{Shing Yan Li}
\thanks{Equal contribution}

\affiliation{MIT Center for Theoretical Physics - a Leinweber Institute,  Cambridge, MA 02139, USA}

\affiliation{James Franck Institute, The University of Chicago, Chicago, IL 60637, USA}

\author{Zhijie Feng}
\thanks{Equal contribution}

\affiliation{Department of Physics, Boston University, Boston, MA 02215, USA}

\affiliation{NSF-Simons National Institute for Theory and Mathematics in Biology, Chicago, IL 60611, USA}

\author{Akshit Goyal}
\email{akshitg@icts.res.in}

\affiliation{International Centre for Theoretical Sciences, Tata Institute of Fundamental Research, Bengaluru 560089, India}

\author{Pankaj Mehta}
\email{pankajm@bu.edu}

\affiliation{Department of Physics, Boston University, Boston, MA 02215, USA}
\affiliation{Faculty of Computing and Data Sciences, Boston University, Boston, Massachusetts 02215, USA}

\date{\today}

\begin{abstract}
\akshit{rewrote parts of abstract; 214 words, within limit}
Ecological interactions can dramatically alter evolutionary outcomes in complex communities. Yet, most population genetics models neglect how interactions within complex communities shape evolutionary dynamics. Here, we bridge this gap by using dynamical mean-field theory to integrate community ecology into classical population genetics models. We show that near an ecological steady state, interactions generate emergent negative frequency-dependent selection between parents and phenotypically similar mutants. Strikingly, even in species-rich communities, this frequency-dependent selection can be characterized by a single parameter measuring the strength of ecological feedback. We find that these interactions strongly suppress the fixation of moderately beneficial mutations. This suppression arises because frequency-dependent selection promotes prolonged coexistence between parent and mutant lineages, creating a barrier to fixation. These effects strengthen with larger effective population sizes and more open niches in the community. Applying our theory to human gut microbiome data, we estimate the average strength of ecological feedback and find it consistent with the negative frequency dependence our theory predicts. Our study establishes a framework for integrating ecological feedback from complex communities into population genetics.\kobe{slightly changed wordings throughout to save words}
\end{abstract}

\maketitle

\section*{Introduction}

A major goal of biology is to develop a quantitative theory of evolution. The mathematical framework of population genetics was a foundational step toward this goal. A cornerstone of population genetics is Kimura's diffusion model (also called Wright-Fisher diffusion) \cite{kimura1962probability,kimura1971theoretical,kimura1994stochastic,ewens2004mathematical,good2018effective}, which conceptualizes the evolutionary dynamics of a mutant's frequency as a stochastic process, subject to both diffusion (random genetic drift) and deterministic selection. Using this theory, one can predict two fundamental aspects of a mutant's fate: its fixation probability and the time it takes to either fix in the population or go extinct.

Since then, it has been shown that a large class of evolutionary dynamics map onto Kimura's diffusion model. Examples range from Wright-Fisher and Moran processes \cite{ewens2004mathematical,mohle2001classification} to single-species serial    dilution experiments like Lenski's long-term evolution \cite{good2017dynamics,wahl2002evaluating}. These developments have led to the widespread view that the results of classical population genetics\kobe{got rid of ``universal''} apply broadly to evolving populations \cite{ewens2004mathematical,crow2017introduction,dykhuizen1983selection}. Indeed, subsequent studies have used Kimura's basic framework as a starting point while extending evolution to other biological contexts, e.g., in fluctuating environments \cite{cvijovic2015fate}, and with spatial structure \cite{whitlock2003fixation,hallatschek2008gene}.

While successful, these models make a simplifying assumption: that the evolutionary dynamics of mutants are mainly determined by their interaction with their parents. In doing so, they incorporate ecological context implicitly rather than explicitly. This makes it hard to understand how ecological interactions will alter evolutionary trajectories~\cite{feng2025theory,mcenany2024predicting,good2023eco}, which is important since almost all evolving populations are embedded in complex ecological communities with which they interact \cite{fussmann2007eco,hendry2017eco,govaert2019eco}. 
Any comprehensive theory of evolution must therefore explain how ecological interactions with a complex community will affect mutation outcomes. Without such an eco-evolutionary perspective, our understanding of evolution in natural contexts will remain incomplete \cite{holt2005integration,mittelbach2015ecological,govaert2021integrating}.

\kobe{rewritten this paragraph}
Numerous works have sought to bridge this gap. These include population genetics models with frequency-dependent selection that capture more realistic evolutionary scenarios \kobe{more citations}\cite{robertson1962selection,antal2006fixation,altrock2010stochastic,assaf2010large,connallon2012general,assaf2015obstruction}. However, since these models assume communities with one to two species \cite{blount2008historical, herron2013parallel,frenkel2015crowded}, it is unclear whether these apply to more complex ecological settings~\cite{marquet2017proportional,overcast2019integrated}. Other evolutionary theories seek to integrate ecological context \cite{foster2017evolution,ferreiro2018multiscale} but many of these efforts remain qualitative in nature and give rise to very different predictions \cite{lawrence2012species,calcagno2017diversity,venkataram2023mutualism,schluter2017speciation,hall2018competitive,scheuerl2020bacterial}. \kobe{added example} \akshit{reworded slightly} For example, experimental studies have shown that communities may promote \cite{lawrence2012species,calcagno2017diversity} or constrain \cite{hall2018competitive,scheuerl2020bacterial} adaptation depending on the specific ecological setting. 

Over the past decade, quantitative approaches have made significant progress understanding the role of ecological interactions in shaping mutant evolutionary trajectories. However, several of these works often make specific assumptions that make it hard to assess their general applicability, such as low community diversity \cite{yamamichi2022integrating} and specific fitness functions \cite{sireci2024statistical}.
\kobe{new introduction for these papers}
Some of these studies have used models of trait evolution \cite{scheffer2006self, doebeli2010complexity} and pathogen competition \cite{zinder2013roles,azarian2020frequency,he2018networks,liu2025eco}, and found that ecological interactions can generate effective frequency-dependent selection. \akshit{streamlined sentence}
Others have used consumer-resource models to explain evolutionary outcomes such as diversification \cite{good2018adaptation} and functional robustness \cite{fant2021eco}. 
Yet others \cite{mcenany2024predicting,feng2025theory} have focused on predicting mutation outcomes in complex communities. However, these studies have largely focused on deterministic dynamics and neglected stochastic dynamics, which are crucial for new mutant lineages which start at extremely low population sizes (often a single individual). 
Hence, despite these advances, a major outstanding challenge remains to develop a theoretical framework to predict the stochastic dynamics of mutants in highly-diverse communities that is general enough to apply across a wide variety of ecological models.

At first, incorporating all the complex details of community ecological interactions might seem intractable due to the large number of unknown parameters. However, recent approaches from statistical physics and random matrix theory have revealed that for a sufficiently complex ecological community, it is possible to gain useful theoretical insights\kobe{changed wording} about ecological properties of an ecosystem such as its diversity or stability \cite{roy2019numerical,pearce2020stabilization,BuninGLV,MayStability,mahadevan2023spatiotemporal,advani2018statistical,altieri2021properties,mahadevan2025continual}. More recent works have started to adapt these methods to ask about short-term evolutionary processes and ecological invasions \cite{mcenany2024predicting,feng2025theory}. A key insight from these works has been the realization that at the community level, the effects of ecological interactions can often be summarized in terms of a few emergent parameters that relate to susceptibilities that measure the strength of ecological feedback \cite{goyal2025universal}. These efforts suggest it should be possible to use tools from statistical physics to incorporate between ecological interactions into classic population genetics models.

\kobe{added assumptions throughout this paragraph} \akshit{edited in various places}
Inspired by these works, here we derive an effective theory of population genetics in complex ecological communities near a steady state. We use dynamical mean-field theory (DMFT) to coarse-grain the effect of a highly-diverse, randomly assembled community on a mutant's frequency dynamics. \kobe{slight rewordings}
Our central result is that community-mediated feedbacks generically lead to emergent frequency-dependent selection across a variety of ecological models. Importantly, a community's combined effect can be captured by only a single new parameter beyond the mutant's fitness \kobe{beyond fitness}that quantifies the strength of community-mediated feedbacks on mutant dynamics. This parameter is tightly related to the community's statistical properties such as its niche-packing. 
We find that community effects dramatically alter the fixation probabilities of mutants compared to Kimura's classical predictions derived in the absence of frequency dependent selection. Further, in diverse communities, mutants can often coexist with their parents for extremely long periods, which drastically alters their fixation or extinction times. Our results highlight that the fate of a mutant can be quantitatively predicted by population genetics models even in complex ecological settings, and provide a path to estimate ecological feedbacks in a complex community from mutant frequency dynamics.\kobe{weaken the statement}

\begin{figure*}[]
    \centering
    \includegraphics[width=0.75\linewidth]{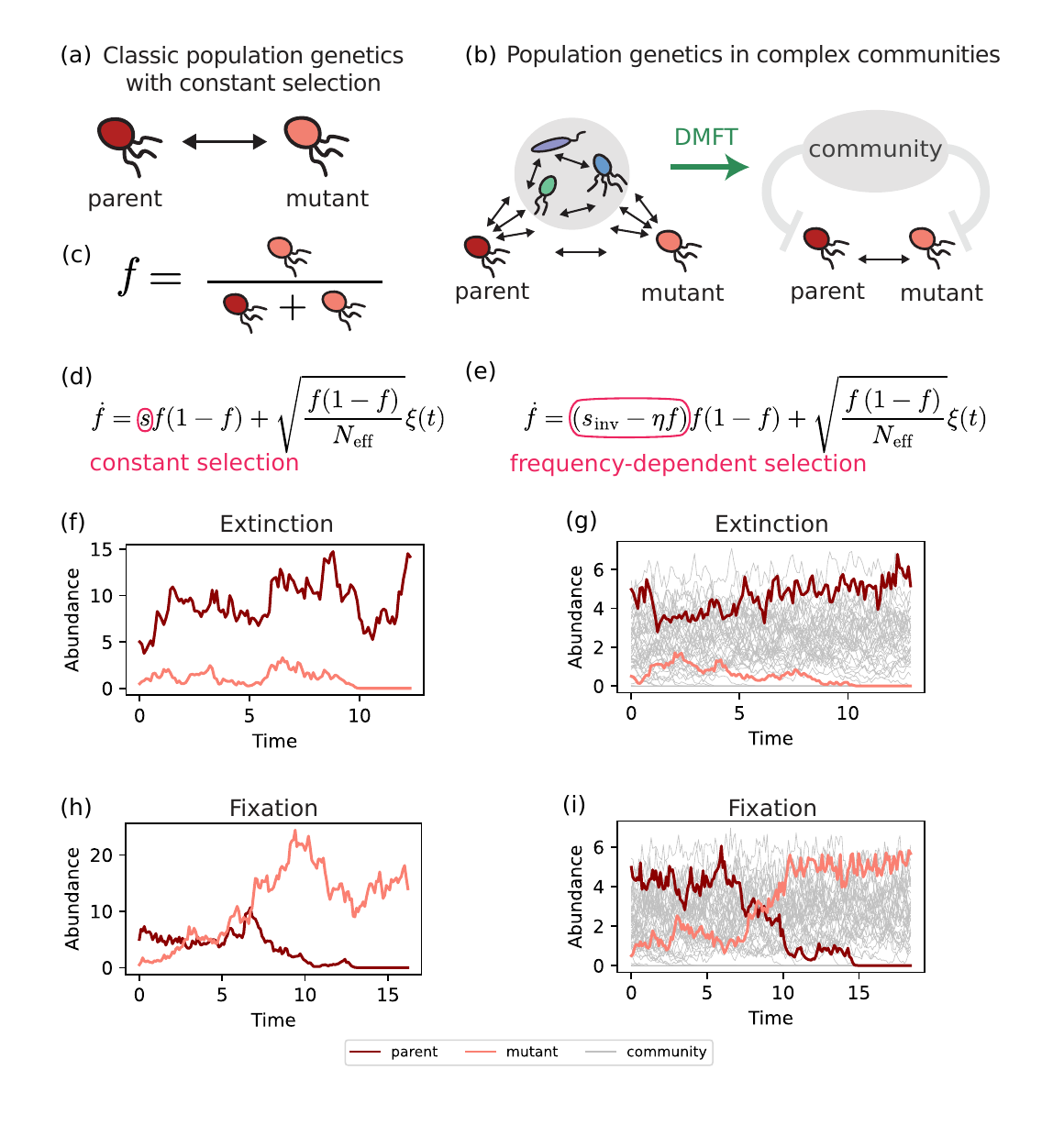}
    \caption{\justifying
\textsf{\textbf{Population genetics and parent-mutant dynamics in the absence and presence of complex communities.}
(a) Classic population genetics, with constant selection, typically models dynamics only for the parent and the mutant, independent of community context.\kobe{added constant selection to figure and caption}
(b) In complex communities, ecological interactions among species collectively influence the parent-mutant dynamics. As shown using dynamical mean-field theory, the ecological community mediates an emergent frequency-depdendent selection between the parent and the mutant.
(c) Parent-mutant dynamics can be expressed in terms of mutant frequency $f$.
(d) In classical population genetics, the dynamics of $f$ are driven by constant selection and stochastic drift.
(e) In complex communities, additional frequency-dependent selection arises, with an emergent parameter $\eta$ characterizing the strength of ecological feedbacks.
(f-i) Examples of dynamics of species abundances in classical population genetics (left) versus in complex communities (other species' dynamics in grey). After a long time, the mutant eventually reaches either extinction (f--g) or fixation (h--i).}
}
\label{fig:dynamics}
\end{figure*}

\section*{Results}
\textbf{\textsf{Ecological feedbacks manifest as emergent frequency-dependent selection.}} 
Population genetics aims to describe how mutations spread through a species population. To do so, it follows the dynamics of a mutant's frequency $f(t)$ in the population given by
\begin{equation}
f(t) = {N_m(t)\over N_p(t) + N_m(t)},
\end{equation} 
where $N_p(t)$ and $N_m(t)$ are the abundances of parent and mutant strains at time $t$, respectively (Fig.~\ref{fig:dynamics}a and c). For convenience, throughout this work, we use the term ``classical population genetics'' synonymously with the single-locus diffusion limit of population genetics in haploid populations with constant selection \cite{kimura1962probability, kimura1971theoretical, ewens2004mathematical}\kobe{moved constant selection to here}. Such models assume that mutant dynamics are shaped by two major evolutionary forces:  (i) selection, which is characterized by a selection coefficient $s$ that measures the fitness difference between the mutant and parent, and (ii) stochastic drift, which is parameterized in terms of an effective population size $N_\mathrm{eff}$ (Fig.~\ref{fig:dynamics}d). Stochastic effects play an important role in mutational dynamics because mutants typically emerge at low frequencies $f(0)\ll 1$, making their early dynamics inherently noisy, e.g., due to randomness in birth and death events \cite{ewens2004mathematical,kimura1971book}.  At long times, mutants either fix ($f=1$) or go extinct ($f=0$, Fig.~\ref{fig:dynamics}f and h).

\kobe{moved constant selection to the above} \akshit{rewrote this topic sentence}
Nevertheless, by treating the selection coefficient $s$ as a constant (Fig.~\ref{fig:dynamics}d), 
these models implicitly assume that ecological interactions do not change the selective landscape over the course of mutant dynamics. \akshit{removed next commented sentence since it now sounded redundant}
For this reason, they often neglect the dynamic environmental feedbacks through which ecological communities can potentially alter mutant evolutionary trajectories (Fig.~\ref{fig:dynamics}b) \cite{fussmann2007eco,hendry2017eco,govaert2019eco}. This assumption becomes especially hard to justify when communities are highly-diverse, from rainforests to microbiomes.

In this work we develop a general framework to understand how ecology affects mutant dynamics. \kobe{added citations}Consistent with Refs.~\cite{good2018adaptation,mcenany2024predicting}, we find that when mutations arise in a complex ecological community, the selection coefficient $s$ can no longer be treated as a constant or simple time-dependent function~\cite{cvijovic2015fate}. Instead, for communities near a steady state, selection becomes frequency-dependent.\kobe{steady} Importantly, this frequency dependence emerges as a consequence\kobe{removed ``generic''} of community-mediated ecological feedbacks on the parent and mutant (Fig.~\ref{fig:dynamics}b), which can be captured with just a few parameters.\kobe{added about parameters} 

To illustrate these ideas, we begin by analyzing the evolutionary dynamics of a mutant in an ecosystem modeled using a generalized Lotka-Volterra model (GLV) with demographic noise. The GLV model describes a complex, highly-diverse community of $S \gg 1$ species with abundances $N_i$ ($i=1,\ldots,S$) whose dynamics are governed by stochastic differential equations of the form
\begin{equation}
\label{eq:GLVEq}
    \frac{dN_i}{dt} = N_i\left(r_i - N_i - \sum_{j \neq i,j=1}^{S} A_{ij}N_j\right) + \sqrt{2DN_i}\xi_i(t),
\end{equation}
with the $\{ \xi_i(t)\}$ independent normal random variables with $\langle \xi_i(t)\rangle=0$ and $\langle \xi_i(t)\xi_j(t^\prime)\rangle=\delta_{ij}\delta(t-t^\prime)$.
In this expression, $r_i$ is the intrinsic growth rates\kobe{removed ``capacity''} of species $i$; $A_{ij}$ captures ecological interactions between species and encodes how species $j$ affects the abundance of species $i$; and $\sqrt{2DN_i}\xi_i(t)$ represents demographic noise. The parameter $D$ sets the overall scale of demographic noise and serves as an analog of a temperature or diffusion coefficient. Following recent work, we model parameters $r_i$ and $A_{ij}$ as random variables \cite{BuninGLV, advani2018statistical, fisher2014transition} drawn from Gaussian distributions (Methods), although our results remain valid for a wide variety of other distributions, as long as they have finite mean and variance \cite{barbier2017cavity,cui2024houches} and negligible higher cumulants \cite{azaele2024generalized} (SI Sec.~\ref{sec:gLVderivation}).\kobe{clarified about higher cumulants} \akshit{minor rewording}

To study population genetics in the GLV, we follow \cite{mcenany2024predicting,mahadevan2023spatiotemporal,mahadevan2025continual} and first initialize a community with $S$ randomly chosen species at steady-state (Methods).\kobe{added citations} We then pick one of these species at random to be the parent $p$. We denote the abundance of the parent species in the community by $N_p(t)$. At time $t=0$, we introduce a mutant $m$ to this community at a small abundance $N_m(0) \ll N_p(0)$. This is equivalent to assuming that the mutant starts at low frequency $f_0 \ll 1$. To model the ecological similarity of parent and mutant, we assume that the mutant's interaction parameters, $A_{mj}$, with other species in the community are correlated with those of the parent, $A_{pj}$. We denote the strength of this correlation as $\rho$, which is independent of any difference in intrinsic growth rates\kobe{removed ``capacity''} between parent and mutant.

Crucially, we assume that mutants and parents are highly correlated ($\rho\approx 1$) in their interactions with the community. With a high $\rho$, the total parent and mutant population behaves effectively as a single species from the community perspective, which greatly simplifies the mutant dynamics by suppressing various community effects (SI Sec.~\ref{sec:gLVderivation}, \ref{sec:parentMutantCorrelation}).\kobe{stated this assumption more clearly} Moreover, mutants and parents interact much more strongly with each other than with any other species in the community. To model this, we set parent-mutant interactions $A_{mp},A_{pm}$ to be close to but weaker than the intraspecific competition ($\approx 1$, SI Sec. \ref{sec:gLVderivation}), while interactions with all other species in the community are assumed to be much weaker ($\approx 1/S$, SI Sec. \ref{sec:gLVderivation}). However, as we will show below, even though mutants interact weakly with any individual species in the community, these weak effects add up due to the large number of species ($\approx S$) present in the community. Thus as a collective, community interactions can no longer be neglected and can strongly influence mutant dynamics. Throughout this work, we focus on the strong-selection-weak-mutation regime \cite{gillespie1983some}: no subsequent mutations of the same species\kobe{same species} occur before a mutant fixes or goes extinct (Discussion).\kobe{referred to discussion, otherwise too much detail here} For this reason, we focus on a single mutation and ignore the scenarios of multiple mutants of the same species\kobe{same species} (known as clonal interference) or the role of mutation rates. 
We also assume that the community is at an ecological steady state when the mutant arises.

Our goal is to derive the effective dynamics of the mutant frequency $f(t)$ in a complex steady \kobe{steady}community described by the GLV. To do so, we use dynamical mean-field theory (DMFT) \cite{pearce2020stabilization,mahadevan2023spatiotemporal,roy2019numerical,sompolinsky1988statistical}. The key idea behind DMFT is that in large, diverse communities the net effect of the community can be coarse-grained into an effective feedback on parent-mutant dynamics which is encoded in a single emergent parameter $\eta$ beyond fitness\kobe{beyond fitness} that summarizes the strength of ecological feedbacks (Fig.~\ref{fig:dynamics}b). Using this technique, we find that mutant frequency can
be described using the equation
\begin{equation}
\label{eq:frequencyDynamics}
    \frac{df}{dt} = \left(s_\mathrm{inv} - \eta f\right)f(1-f) + \sqrt{\frac{f(1-f)}{N_\mathrm{eff}}}\xi(t),
\end{equation}
where $s_\mathrm{inv}$ measures the invasion fitness of the mutant in the community. Since $s_\mathrm{inv}$ contains both the intrinsic selection coefficient $s$ (e.g., differences in intrinsic growth rates\kobe{removed ``capacity''}) and community effects (SI Sec. \ref{sec:gLVderivation}), it plays a role analogous to the mutant's selective coefficient in classical population genetics. $N_{\text{eff}}=\left<N_p(0)+N_m(0)\right>/2D$ measures the effective population size of the combined parent-mutant population, which is approximately constant on average, especially for mutants where $\rho\approx 1$ (SI Sec. \ref{sec:gLVderivation}). $\xi(t)$ is a random normal variable with $\langle\xi(t)\rangle=0$ and $\langle \xi(t) \xi(t')\rangle = \delta(t-t')$. 

This equation results from two distinct eco-evolutionary processes: deterministic selection -- the term proportional to $f(1-f)$ -- and stochastic drift -- the term proportional to $\xi(t)$. Remarkably, once the (invasion) fitness is known, just introducing a single new parameter $\eta$ suffices to capture the collective effect of the entire community on parent-mutant dynamics.\kobe{clarified on invasion fitness}  Eq.~\eqref{eq:frequencyDynamics} shows that selection is now frequency-dependent since the selection coefficient $s_\mathrm{inv} - \eta f$ changes with mutant frequency. When $\eta = 0$, e.g., in the absence of a community, we recover classical population genetics with a constant selection coefficient. However, when  $\eta \neq 0$, ecological feedbacks fundamentally alter evolutionary dynamics.

We show that community-mediated feedbacks can\kobe{removed ``generically''} manifest as frequency-dependent selection across a wide variety of ecological models. In the SI, we derive expressions for the effective parameters---$s_\mathrm{inv}$, $N_{\text{eff}}$ and $\eta$---for MacArthur Consumer-Resource Models (CRMs) \cite{Levins1962I,MacArthurLimitingSimilarity} and generalized CRMs with nonlinear per-capita growth rates, which includes models with cross-feeding \cite{marsland2019available} (SI Sec. \ref{sec:CRM} and \ref{sec:gCRM}). Remarkably, despite the diverse mathematical structures of these models, they all reduce to the same form of frequency-dependent selection near ecological steady states.\kobe{steady} 
\kobe{removed sentence that claims too much; see commented content}
\kobe{rewrote this to make central result clearer}
While it has been well known that ecological interactions with a few other species can lead to frequency-dependent selection (Introduction), 
our results show that the same type of frequency-dependent selection in Eq.~\eqref{eq:frequencyDynamics} also emerges in communities with many species near a steady state. As a result, mutation outcomes can be predicted using simple models despite the ecological complexity of communities. Our results also generalize previous work~\cite{mcenany2024predicting} and predict the full stochastic trajectories of mutants, in addition to being applicable across a wide variety of ecological models.\kobe{added this for Ben} \akshit{slight rewording}

\begin{figure*}[]
    \centering
    \includegraphics[width=0.77\linewidth]{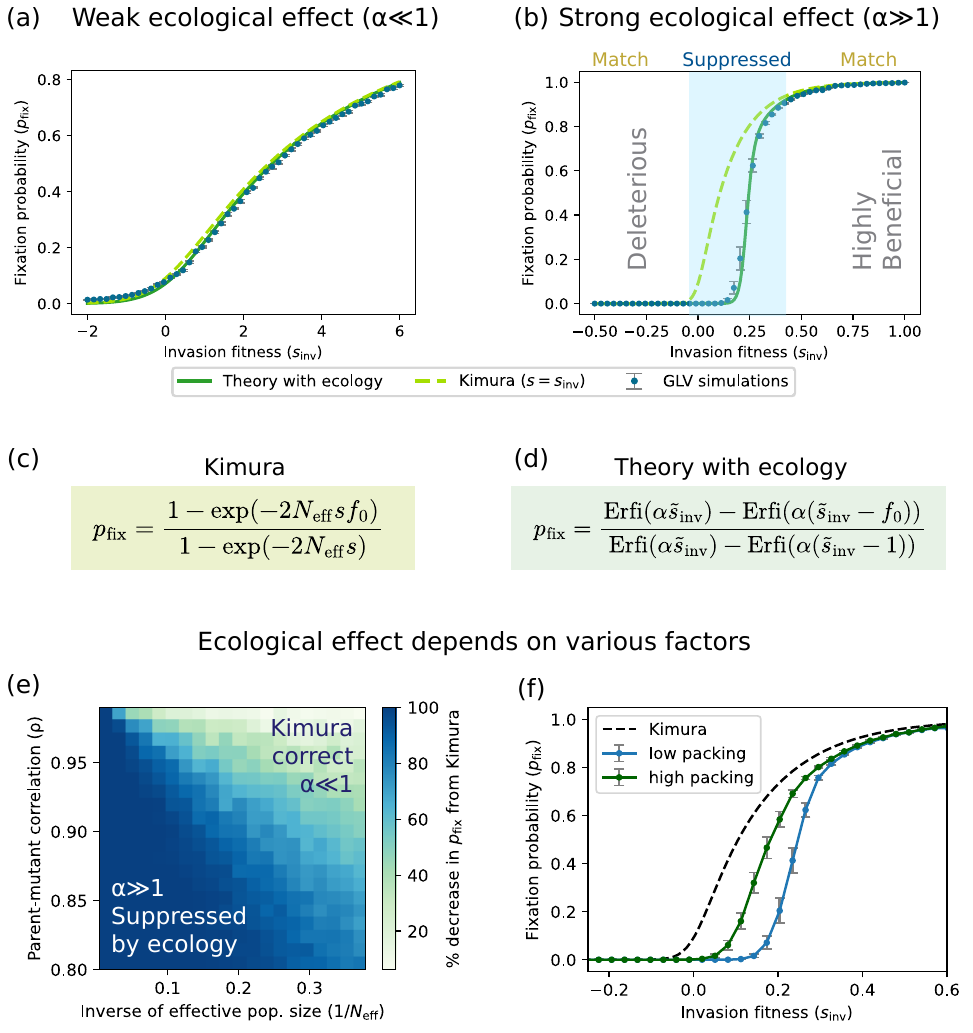}
    \caption{\justifying
\textsf{\textbf{Fixation probabilities suppressed by strong ecological effects.}
We compute fixation probabilities $p_\mathrm{fix}$ at various invasion fitness 
$s_\mathrm{inv}$ by simulating mutations within generalized Lotka-Volterra models.
Error bars denote standard errors from multiple instances of demographic noise and mutants.
(a) When ecological effects are weak relative to stochastic drift ($\alpha \ll 1$), our theory matches  Kimura's formula with $s=s_\mathrm{inv}$ for fixation probabilities and predicts them accurately in simulations.
(b) When ecological effects are strong ($\alpha \gg 1$), Kimura's formula applies only for deleterious or highly beneficial mutants. For moderately beneficial mutants with $s_\mathrm{inv}\lesssim \eta/2$, fixation probabilities are strongly suppressed relative to Kimura's prediction. Our theory with ecology can capture such suppression and matches simulations.
We give analytic expressions for fixation probability in (c) Kimura's theory and (d) our theory with ecology.
Ecological suppression of mutants depends on several factors:
(e) For mutations with $s_\mathrm{inv}=0$, we quantify the suppression by the percentage decrease in fixation probability from Kimura's prediction ($1-p_\mathrm{fix}/f_0$). Suppression increases with larger effective population size $N_{\text{eff}}$ and lower parent-mutant correlation $\rho$.
(f) Suppression is also stronger when the community is less packed (has more ``open niches''), i.e., the number of surviving species is low relative to the maximum number of species allowed.\kobe{changed description since this is GLV, not competitive exclusion} The extent of packing is controlled by different $\sigma^2$, the variance of the interactions (SI Sec.~\ref{sec:simulationParameters}).\kobe{clarified about how we get different packing}}
}
    \label{fig:fixationProbability}
\end{figure*}

\vskip 10pt
\textbf{\textsf{Generalization of Kimura's formula for fixation probabilities.}} \kobe{rewritten the opening so that it is clear why we organize the main text this way} To test whether Eq.~\eqref{eq:frequencyDynamics} indeed captures ecological effects, we now explore the consequences of community interactions on the fate of a mutant. To do so, we draw on existing results in population genetics on both constant and frequency-dependent selection~\cite{kimura1962probability,robertson1962selection,nei1973probability,antal2006fixation,altrock2010stochastic,assaf2010large,connallon2012general,assaf2015obstruction,ruzicka2026century}). \akshit{reworded here}

One of the celebrated results of classical population genetics is Kimura's formula (with constant selection)\kobe{added constant selection} for fixation probability $p_\mathrm{fix}$ \cite{kimura1962probability}, which describes how likely a mutant is to replace its parent and fix in a population (Fig. \ref{fig:fixationProbability}c). Kimura's formula is given by
\begin{equation}
\label{eq:Kimura}
    p_\mathrm{fix}= \frac{1-\exp(-2N_\mathrm{eff}sf_0)}{1-\exp(-2N_\mathrm{eff}s)}\,.
\end{equation}
This formula shows that as the fitness difference between mutant and parent $s$ increases, mutants become more likely to fix. Specifically, neutral mutants with $s=0$ fix with a probability equal to their initial frequency $p_\mathrm{fix}=f_0$. When selection is weak, i.e., $N_{\text{eff}} |s| \ll 1$, mutants are ``effectively neutral'' and fix with the same probability $f_0$. When selection is strong, i.e., $N_{\text{eff}} |s| \gg 1$, the formula reduces to $p_\mathrm{fix} = 1-e^{-2N_{\text{eff}}sf_0}$ for positive $s$ and $p_\mathrm{fix} = e^{-2N_{\text{eff}}|s|}$ for negative $s$. Deleterious mutants with $s < 0$ thus have a negligible fixation probability, while for beneficial mutants with $s>0$ the fixation probability increases linearly $p_\mathrm{fix} \propto s$. Ultimately, once a mutation is strongly beneficial ($s\gg1/N_{\text{eff}}f_0$) the mutant will almost surely fix since $p_\mathrm{fix} \approx 1$.

Starting with Eq.~\eqref{eq:frequencyDynamics}, one can solve the corresponding backward equation to obtain an analytic formula for the fixation probability $p_\mathrm{fix}$ in complex communities \cite{kimura1962probability,nei1973probability} (see also SI Sec.~\ref{sec:fixProbAnalysis}):\kobe{weaken the ``derive'' statement as Ben didn't like it}
\begin{equation}
\label{eq:ecoPfix}
    p_\mathrm{fix}=\frac{\mathrm{Erfi}\left(\alpha\tilde s_\mathrm{inv}\right)-\mathrm{Erfi}\left(\alpha(\tilde s_\mathrm{inv}-f_0)\right)}{\mathrm{Erfi}\left(\alpha\tilde s_\mathrm{inv}\right)-\mathrm{Erfi}\left(\alpha(\tilde s_\mathrm{inv}-1)\right)}\,,
\end{equation}
 where
\begin{equation}
    \mathrm{Erfi}(x)=\frac{2}{\sqrt{\pi}}\int_0^x dy\,e^{y^2}\,,
\end{equation}
is the imaginary error function,
\begin{equation}
\label{eq:alpha}
    \alpha=\sqrt{N_\mathrm{eff}\eta}\,,
\end{equation}
is the ratio of the strength of ecological feedbacks and stochastic drift
and
\begin{equation}
    \tilde s_\mathrm{inv}=\frac{s_\mathrm{inv}}{\eta}\,,
\end{equation}
is the mutant invasion fitness normalized by the strength of the ecological feedback. The quantity $\tilde s_\mathrm{inv}$ is closely related to the ``dressed invasion fitness'' introduced in \cite{feng2025theory} for predicting the outcomes of ecological invasions. In SI Sec.~\ref{sec:fixProbAnalysis}, we elaborate on the similarities and differences between the two quantities.\kobe{added more explanation on relationship and referred to SI} \akshit{rewrote to keep it simple}
Eq.~\eqref{eq:ecoPfix} serves as the generalization of Eq.~\eqref{eq:Kimura} in the context of complex ecological communities.\kobe{we are not generalizing ``Kimura's formula'' according to Ben}

\kobe{replaced ``our formula'' with Eq. 5 since this is not ``our formula''}
To gain intuition for this formula, it is useful to look at various limits of this expression. When ecological effects are weak, i.e., $\alpha\ll 1$, Eq.~\eqref{eq:ecoPfix} reduces back to Kimura's formula in Eq.~\eqref{eq:Kimura} (SI Sec. \ref{sec:fixProbAnalysis}) with $s=s_\mathrm{inv}$. To numerically confirm this, we simulated the dynamics of complex communities governed by the GLV model into each of which we introduced a mutant of a randomly chosen parent strain. We made ecological effects weak simply by setting a large value of $D$, increasing the strength of stochastic drift. We repeated simulations for mutants with a given invasion fitness $s_\mathrm{inv}$ and measured the fraction of simulations in which mutants fixed (Methods). As shown in Fig.~\ref{fig:fixationProbability}a, the simulated fixation probabilities match Eq.~\eqref{eq:ecoPfix} remarkably well across a range of $s_\mathrm{inv}$. Further, in this limit, Eq.~\eqref{eq:ecoPfix} virtually overlaps Kimura's formula. In the next section, we will look at the opposite case of strong ecological effects, i.e., $\alpha \gg 1$, and show that we get a qualitatively different picture of the fate of a mutant compared to Kimura's formula.

\vskip 10pt
\textbf{\textsf{Ecological interactions strongly suppress mutant fixation.}} In the presence of strong ecological effects, i.e., when $\alpha\gg 1$, our numerical simulations for the fixation probability show a stark deviation from Kimura's predictions (Fig.~\ref{fig:fixationProbability}b), with an almost switch-like behavior with increasing $s_\mathrm{inv}$. This is in contrast with Kimura's formula which shows a gradual, linear increase in fixation probability over a range of $s$. The deviation from Kimura is most pronounced for moderately beneficial mutants where ecological effects strongly suppress the fixation probability (Fig.~\ref{fig:fixationProbability}b, blue region). The origin of this suppression is easiest to understand for mutants with $s_\mathrm{inv}=0$ where Eq.~\eqref{eq:ecoPfix} reduces to (SI Sec.~\ref{sec:fixProbAnalysis})\kobe{replaced ``neutral'' to $s_\mathrm{inv}=0$ throughout the paper}
\begin{equation}
    p_\mathrm{fix}(s_\mathrm{inv}=0)=\begin{cases}
        f_0 & \alpha\ll 1\\
        e^{-\alpha^2} & \alpha\gg 1
    \end{cases}\,.
\end{equation} 
This formula agrees with Kimura's results when ecological effects are weak ($\alpha\ll 1$) and shows that ecological effects exponentially suppress mutations with $s_\mathrm{inv}=0$ when ecological effects are large ($\alpha\gg 1$). 

\begin{figure*}[]
    \centering
    \includegraphics[width=0.9\linewidth]{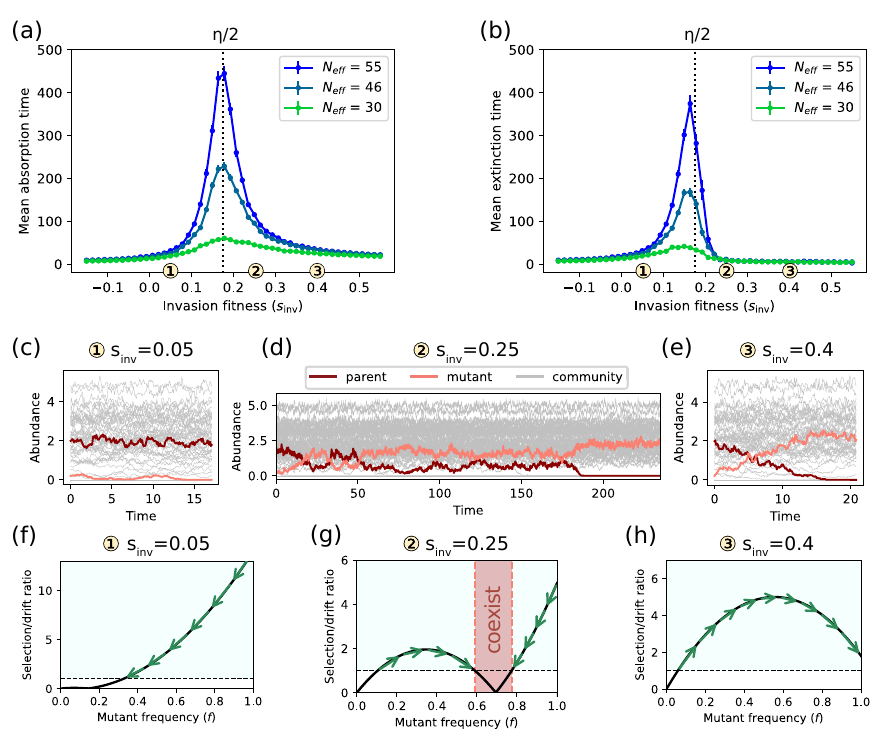}
    \caption{\justifying
\textsf{\textbf{Parent-mutant coexistence in complex communities.}
We compute mean absorption and extinction times at various invasion fitness 
$s_\mathrm{inv}$ by simulating mutations within generalized Lotka-Volterra models.
Error bars denote standard errors from multiple instances of demographic noise.
(a, b) The time scales of parent-mutant dynamics at various invasion fitness $s_\mathrm{inv}$ reveal the possibility of coexistence. Under strong ecological effects, the mean times to absorption (either extinction or fixation) (a) and to extinction (b) are exponentially longer for moderately beneficial mutants with $0\lesssim s_\mathrm{inv}\lesssim \eta$, peaking at $s_\mathrm{inv}\simeq \eta/2$ (dashed lines) as predicted by our theory. The maximum mean times also grow exponentially with the effective population size $N_\mathrm{eff}$.
(c-h) Representative trajectories and phase portraits (in terms of the ratio between selection and drift) show distinct behaviors across different $s_\mathrm{inv}$. For low $s_\mathrm{inv}$ (f) or high $s_\mathrm{inv}$ (h), there is a single crossover (dashed lines) from drift to selection dominated regimes. Selection drives (green arrows) the mutant to extinction (c) or fixation (e). However, there are additional crossovers (g) when $s_\mathrm{inv}$ is moderate with $s_\mathrm{inv}\lesssim\eta$. Selection instead drives the mutant into a coexistence region centered at frequency $f^*$ (dark red), resulting in exponentially long times for parent-mutant coexistence (d).}
}
\label{fig:coexistence}
\end{figure*}

Using Eq.~\eqref{eq:ecoPfix}, we can also compute the full range of invasion fitness $s_\mathrm{inv}$ for which mutants are suppressed when ecological effects are strong (SI Sec.~\ref{sec:fixProbAnalysis}). We find that suppression occurs when
\begin{equation}
\label{eq:criticalFitnessDiff}
    0\lesssim s_\mathrm{inv}\lesssim s_c:= \frac{\eta}{2}\,,
\end{equation}
where $s_c=\eta/2$ is the critical invasion fitness at which fixation probability switches from roughly 0 to almost 1. As we show in SI Sec.~\ref{sec:fixProbAnalysis}, in this case the dominant contribution to ecological suppression in the range $0\lesssim s_\mathrm{inv}\lesssim s_c$ comes from the last term in the denominator of Eq.~\eqref{eq:ecoPfix}.

The ecological suppression of mutants is an emergent community-mediated phenomenon whose strength depends on $\alpha=\sqrt{N_{\text{eff}}\eta}$. To understand how different ecological properties of the underlying community control the strength of suppression, we varied $\alpha$ by scanning over ranges of effective population sizes $N_\mathrm{eff}$ and parent-mutant correlation $\rho$\kobe{clarified about what we exactly did}, and plotted the corresponding deviation of the simulated fixation probability from Kimura's predictions for a mutation with $s_\mathrm{inv}=0$ (Fig.~\ref{fig:fixationProbability}e).  Increasing 
$N_\mathrm{eff}$ further suppressed mutant fixation by increasing $\alpha$\kobe{added $\alpha$} \akshit{edited slightly} (Fig.~\ref{fig:fixationProbability}e). This can be understood by noting that increasing $N_\mathrm{eff}$ weakens stochastic drift, and hence increases the importance of ecological feedbacks. Suppression also increased with decreasing $\rho$ (Fig.~\ref{fig:fixationProbability}e). The reason for this is that the ecological feedback parameter $\eta$\kobe{removed $\alpha$} is proportional to $1-\rho$ (SI Sec.~\ref{sec:gLVderivation}--\ref{sec:gCRM}). Intuitively, as parent and mutant become less similar (i.e., $\rho$ decreases), mutants interact less strongly with their parents, allowing community feedbacks to play a stronger role in the dynamics. 

Most interestingly\kobe{changed wording so that it sounds less paradoxical}, suppression was stronger in less ``niche-packed'' communities (Fig.~\ref{fig:fixationProbability}f), where we have defined packing as the ratio of surviving species $S^*$ compared to largest number of species that can stably coexist in the community (often referred to as the species packing bound). \kobe{added explanation for GLV and CRM}In GLV, the maximum comes from May's stability bound $S^*\lesssim 1/\sigma^2$ \cite{MayStability}, where $\sigma^2$ is the variance of the interactions $A_{ij}$ (SI Sec.~\ref{sec:gLVderivation}); in CRM, it comes from competitive exclusion that $S^*\lesssim M^*$, the number of non-depleted resources (SI Sec.~\ref{sec:CRM}). This suppression suggests that ecological feedbacks become weaker as communities get closer to a fully packed regime where all niches are occupied, consistent with \cite{mcenany2024predicting} (SI Sec.~\ref{sec:fixProbCRM}).\kobe{added citation, also pointed to SI where we discuss fully packed case}
To understand this effect, we draw on an analogy between ecological communities and mechanical jamming \cite{cui2020effect}. Mechanical systems become harder to deform as they are packed with more particles, ultimately becoming rigid when they reach the jamming limit. Similarly, ecological communities become more ``rigid'' and less ``deformable'' as vacant niches get occupied due to a change in their susceptibility to perturbations. (Note that in consumer-resource models, the relevant susceptibility is in resource (niche) space, rather than species abundance space.) \kobe{clarified about Ben's confusion on jamming} \akshit{edited previous sentences} In support of this analogy, we have verified that it is really packing ratio that contributes to $\eta$ by affecting community susceptibility, rather than the biomass or species richness\kobe{made clear about $\eta$} (SI Sec.~\ref{sec:gLVderivation}, \ref{sec:CRM}, \ref{sec:fixProbCRM}).\kobe{clarified about dependence on other things}

Finally, we show in the SI that the results in Fig.~\ref{fig:fixationProbability} generalize beyond the GLV model to many variants of consumer resource models. This leads us to conclude that ecological interactions likely generically suppress fixation of moderately beneficial mutants, strongly deviating from the predictions of classical population genetics.

\begin{figure}[]
    \centering
    \includegraphics[width=0.8\linewidth]{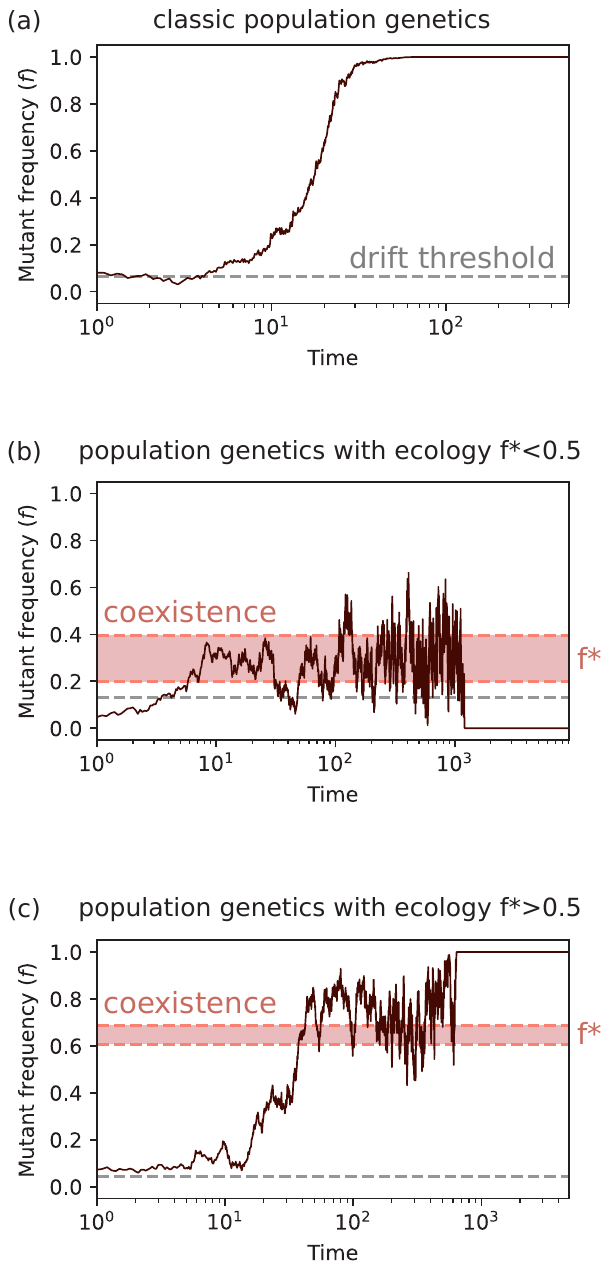}
    \caption{\justifying
\textsf{\textbf{Suppression of fixation probabilities explained by parent-mutant coexistence.}
(a) In classic population genetics, mutants are very likely to reach fixation after crossing the drift threshold (dashed gray line) at low frequencies, a process known as establishment.
(b, c) With ecological effects, the coexistence region (dark red, Fig.~\ref{fig:coexistence}(g)) modifies the fate of mutants. When the coexistence region is closer to extinction (b), i.e., $f^*<1/2$, the region acts as an additional barrier to fixation, causing extinction even after crossing the drift threshold. This mechanism accounts for the suppression of fixation probabilities (Fig.~\ref{fig:coexistence}b). In contrast, the coexistence region no longer prevents mutants from fixation when it becomes closer to fixation (c), i.e., $f^*>1/2$.}
}
\label{fig:fixProbExplanation}
\end{figure}
\vskip 10pt
\textsf{\textbf{Parent-mutant coexistence alters fixation and extinction times.}} In addition to fixation probability, the second important quantity in population genetics is the time it takes for a mutant to reach fixation or extinction (which we refer to as absorption time). This quantity characterizes the timescale of adaptation in evolutionary dynamics.
It is also informative to study the establishment time or equivalently the time to extinction. 
These two time scales are intimately related since extinction probability becomes negligible as soon as the mutant becomes established in the population, long before fixation. The mean extinction time can also be theoretically estimated (SI Sec.~\ref{sec:extinctionTime}).
These times are relatively short in classic population genetics. Namely, they reach a maximum of order $N_\mathrm{eff}f_0|\log f_0|$ \cite{kimura1969average} for neutral mutations ($s=0$), and decay with the magnitude of the fitness difference $|s|$.

On the other hand, we find that when ecological effects are strong ($\alpha\gg 1$), the mean times to absorption 
(Fig.~\ref{fig:coexistence}a) and extinction (Fig.~\ref{fig:coexistence}b)
become exponentially large for moderately beneficial mutants (i.e., $s_\mathrm{inv}\lesssim \eta$).  Both of these times
show a sharp peak at the critical fitness difference $s_\mathrm{inv}\simeq \eta/2$ 
(Eq.~\eqref{eq:criticalFitnessDiff}). This increase in absorption time is so dramatic that it can even be directly observed in simulation trajectories of mutant and parent dynamics (see Fig.~\ref{fig:coexistence}c--e). We further find that the maximum of both the absorption and extinction times grows
exponentially with the effective population size $N_\mathrm{eff}$. This is in contrast with Kimura's theory where the maximum time grows linearly with $N_\mathrm{eff}$.
The presence of these exponentially long time scales raises the possibility that parents and mutants can transiently coexist for long periods in complex communities \cite{mcenany2024predicting,feng2025theory}, highlighting a crucial difference between classical population genetics with constant selection and our eco-evolutionary approach.\kobe{constant selection, and included citations}

Such coexistence is a direct consequence of frequency-dependent selection (Eq.~\eqref{eq:frequencyDynamics}) \cite{gillespie1983some}.\kobe{cited for Ben}
To see this, it is helpful to think about the infinite effective population size limit of Eq.~\eqref{eq:frequencyDynamics}. In this limit, we can ignore stochasticity and focus on the fixed points of the deterministic dynamics which are given by the solutions to the equation $0= \left(s_\mathrm{inv} - \eta f\right)f(1-f)$. When the fitness satisfies $s_\mathrm{inv}>\eta$ or $s_\mathrm{inv}<0$, this equation has two fixed points at $f^*=0$ and $f^*=1$. On the other hand, when $0 \le s_\mathrm{inv} \le \eta$, this equation has an additional fixed point at $f^*=\tilde s_\mathrm{inv}=s_\mathrm{inv}/\eta$, implying that the parent and mutant can potentially coexist in the absence of stochastic drift.

In the presence of drift, for small and large $\tilde{s}_\mathrm{inv}$ --- i.e., for deleterious and strongly beneficial mutants --- selection almost always drives mutants to extinction (Fig.~\ref{fig:coexistence}c) and fixation (Fig.~\ref{fig:coexistence}e) respectively. On the other hand, for moderately beneficial mutants the presence of the additional fixed point in the deterministic limit gives rise to additional crossovers (Fig.~\ref{fig:coexistence}g) which
create a ``coexistence region'' between extinction and fixation, centered at frequency
$f^*\simeq \tilde s_\mathrm{inv}$. Selection drives the
mutant into this region instead of extinction or fixation, hence the parent and mutant coexist for exponentially long
times (Fig.~\ref{fig:coexistence}d) before escaping the region (SI Sec.~\ref{sec:extinctionTime}). Such a region is absent in classical population genetics with constant selection.\kobe{constant selection}

Parent-mutant coexistence also qualitatively explains the suppression of fixation
probabilities for moderately beneficial mutants with $0<s_\mathrm{inv}<s_c$ 
(Fig.~\ref{fig:fixationProbability}b). Recall that in classical population genetics,
beneficial mutants reach fixation through establishment, i.e., crossing the drift threshold from
drift to selection dominated regimes (Fig.~\ref{fig:fixProbExplanation}a, see also
Fig.~\ref{fig:coexistence}f and h). The process of establishment is central to Kimura's
theory, since after this point, mutants grow almost deterministically towards fixation. However, this qualitative picture is modified by the coexistence region under strong ecological effects, as selection now drives mutants into this region after establishment (Fig.~\ref{fig:coexistence}g). 
When the coexistence region is closer to extinction (Fig.~\ref{fig:fixProbExplanation}b), i.e., $f^*<1/2$ or $s_\mathrm{inv}<s_c$, mutants that escape this region due to stochastic drift are more likely to go extinct than fix. Since such extinction occurs even after crossing the drift threshold, it suppresses fixation probabilities relative to Kimura's prediction. On the other hand, when the coexistence region is closer to 
fixation (Fig.~\ref{fig:fixProbExplanation}c), i.e., $f^*>1/2$ or $s_\mathrm{inv}>s_c$, it is more
likely for mutants inside the region to drift to fixation instead, hence in this case coexistence
does not suppress fixation.

\begin{figure}[]
    \centering
    \includegraphics[width=0.9\linewidth]{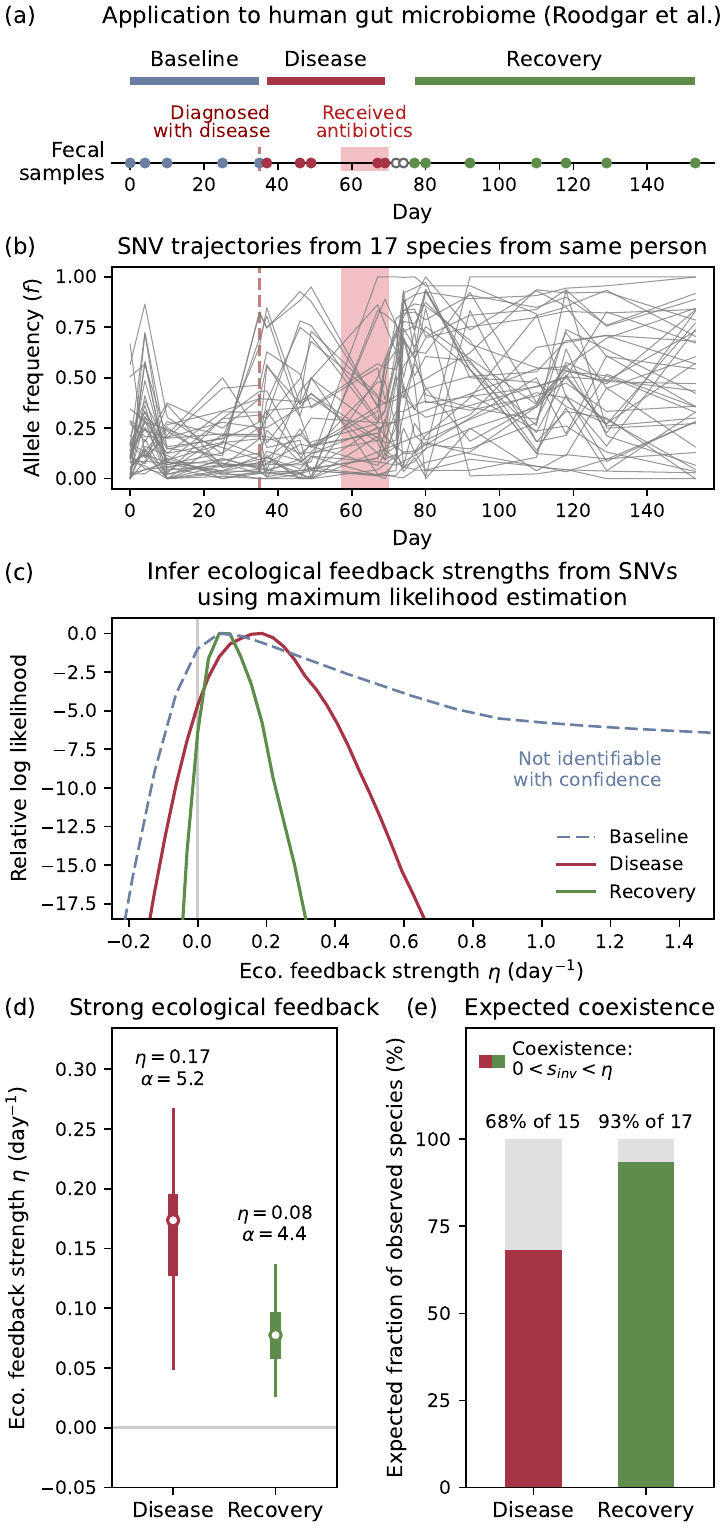}
    \caption{\justifying
\textsf{\textbf{Mutant frequency dynamics recapitulate strong ecological feedbacks and parent-mutant coexistence in the human gut microbiome.}
(a) We analyze longitudinal sequencing data from the gut microbiome of a single patient across 19 fecal samples collected over 5 months~\cite{roodgar2021longitudinal}. The patient was diagnosed with disease at the end of the first month and received antibiotics for 2 weeks around day 60 (pink). We split the timeline into baseline, disease, and recovery periods, excluding the period shortly after antibiotic treatment when the community underwent substantial compositional shifts.
(b) Frequency trajectories of sweeping single nucleotide variant (SNV) clusters~\cite{roodgar2021longitudinal} from 17 selected species within the same microbiome.
(c) We infer the strength of ecological feedback $\eta$ in different periods by fitting the trajectories to the model of frequency-dependent selection (Eq.~\eqref{eq:frequencyDynamics}, Methods). Maximum likelihood estimation yields confident estimates of $\eta$ in the disease and recovery periods, but not in the baseline period.
(d) The inferred $\eta$ in disease and recovery phases are both positive. Thick and thin lines denote 50\% and 95\% confidence intervals obtained by bootstrapping across species. Combined with the inferred effective population sizes $N_\mathrm{eff}$ (SI Sec.~\ref{sec:dataAnalysis}), these estimates indicate strong ecological effects in both periods, with $\alpha=\sqrt{N_\mathrm{eff}\eta}$ around $4$ to $5$ (Eq.~\eqref{eq:alpha}, SI Sec.~\ref{sec:dataAnalysis}).
(e) We also infer the distribution of invasion fitness $s_\mathrm{inv}$ across species (SI Sec.~\ref{sec:dataAnalysis}). In both periods, parent-mutant coexistence is expected for majority of the selected species, based on the fixed-point condition $0<s_\mathrm{inv}<\eta$ (Fig.~\ref{fig:coexistence}).}
}
\label{fig:dataAnalysis}
\end{figure}

\vskip 10pt
\textsf{\textbf{Mutant dynamics reveal ecological feedbacks in the human gut microbiome.}}
A key prediction of our theory is that the strength of frequency-dependent selection $\eta$ encodes the ecological feedbacks acting on a mutant. In practice, we can directly infer $\eta$ from data by fitting Eq.~\eqref{eq:frequencyDynamics} to a set of mutant frequency trajectories from the same community, similar to previous approaches~\cite{wolff2023ecological}.\kobe{added citation} \akshit{moving assumption of a common pooled $\eta$ to later in the text}

To explore this application, we analyzed longitudinal metagenomic sequencing data from the gut microbiome of a single patient, sampled across 19 fecal samples over 5 months~\cite{roodgar2021longitudinal}.
Deep sequencing allows us to infer frequency trajectories of single nucleotide variants (SNVs), which serve as proxies for mutant dynamics (SI Sec.~\ref{sec:dataAnalysis}). These SNVs span 17 species from the same community (Fig.~\ref{fig:dataAnalysis}b), allowing us to estimate $\eta$
by fitting Eq.~\eqref{eq:frequencyDynamics} (Methods). During the 5-month period, the patient was diagnosed with disease at the end of the first month and received a two-week course of antibiotics at the end of the second month (Fig.~\ref{fig:dataAnalysis}). These perturbations have the potential to change community composition, resulting in an ecological feedback strength that shifts over time. To account for this possibility, we split the data into baseline, disease, and recovery periods (Fig.~\ref{fig:dataAnalysis}a). Further, we excluded the period shortly after the antibiotic treatment, during which the community underwent rapid compositional shifts, violating our assumption of ecological steady state. Note that our model does not require mutants to differ from parents by a single point mutation; they may differ by several SNVs, as is the case here. For simplicity, in our analysis we assumed that all mutants in the same community experience a common average ecological feedback strength $\eta$. In reality, different mutants may experience slightly different ecological feedbacks due to, e.g., differences in their phenotypic correlation $\rho$ with their parents \akshit{added assumption about common $\eta$ here.} Taken together, this dataset is well suited for applying our model: it tracks the frequency dynamics of several mutants in a diverse community that appear to coexist for a long period of time (in fact, some of these were previously interpreted as supported by spatial or metabolic niches~\cite{roodgar2021longitudinal}).

We fit Eq.~\eqref{eq:frequencyDynamics} to this dataset using a maximum likelihood estimation (MLE) method we developed (SI Sec.~\ref{sec:dataAnalysis}). This method allowed us to infer the average strength of ecological feedbacks $\eta$, the average effective population size $N_{\mathrm{eff}}$, and the distribution of invasion fitness $s_{\mathrm{inv}}$ across different mutants (SI Sec.~\ref{sec:dataAnalysis}). We then used the best-fit parameters to: (1) test whether the data are consistent with widespread negative frequency-dependent selection, i.e., $\eta>0$ predicted by our theory; (2) estimate the strength of ecological effects $\alpha=\sqrt{\eta \cdot N_{\mathrm{eff}}}$; and (3) estimate the fraction of mutants that could coexist with their parents due to community ecological feedbacks.
For simplicity, we also assumed there was only one mutant and one parent strain per species. This assumption is consistent with previous studies where two strains per species are commonly observed~\cite{garud2019evolutionary,roodgar2021longitudinal,zhao2019adaptive,good2017dynamics,goyal2022interactions} 
Our MLE procedure led to confident estimates of $\eta$ in the disease and recovery periods, but not in the baseline period. We confirmed that the best-fit models in these two periods were well-calibrated for one-step-ahead predictions ($p=0.60$ and $p=0.15$ respectively; Cram\'er-von Mises criterion, SI Sec.~\ref{sec:dataAnalysis}).

Encouragingly, we found that the inferred $\eta$ were positive during both the disease and recovery periods
(Fig.~\ref{fig:dataAnalysis}d). This is consistent with our prediction that communities typically exert negative frequency-dependent selection on mutants (SI Sec.~\ref{sec:gLVderivation}--\ref{sec:gCRM}). Combined with the estimates $N_\mathrm{eff}\simeq 160$ and $N_\mathrm{eff}\simeq 250$ (normalized by the number of generations within a day)\kobe{added normalization since Ben may complain $N_\mathrm{eff}$ looks too small} for the two periods respectively (SI Sec.~\ref{sec:dataAnalysis}), these results suggest that mutants in the gut microbiome are subject to strong ecological effects $\alpha>1$ in both periods, with $\alpha\approx$4-5 (Eq.~\eqref{eq:alpha}). Although the inferred $\eta$ is slightly lower in the recovery period, more data are required to statistically resolve the difference between the two periods. 

Our theory also predicts that strong ecological effects promote parent-mutant coexistence when $0<s_\mathrm{inv}<\eta$ (Fig.~\ref{fig:coexistence}). To examine this prediction, we used the inferred distribution of $s_\mathrm{inv}$ across species to estimate the expected prevalence of parent-mutant coexistence (SI Sec.~\ref{sec:dataAnalysis}). We found that the inferred parameters are consistent with parent-mutant coexistence for most of the observed species in the community (Fig.~\ref{fig:dataAnalysis}e).
Together, these results 
demonstrate how mutant frequency dynamics can be used to infer strong ecological effects and their implications across the community.

\vskip -10pt
\section*{Discussion}

 In this study, we developed an analytic framework for integrating ecological interactions with a complex community into population genetics models. Our calculations show that interactions with a highly-diverse community can dramatically affect evolutionary outcomes. Despite their complexity, ecological feedbacks give rise to a remarkably simple form of
frequency-dependent selection between ecologically similar \kobe{similar}parent and mutant strains when the community is near its ecological steady state.\kobe{steady} Surprisingly, the frequency dependence from the entire community\kobe{fixed wording} can be captured by a single parameter $\eta$ which can be easily calculated from the community's statistics\kobe{statistics} for a wide variety of ecological models. Thus our results reveal a principled method to apply existing population genetics models with frequency-dependent selection even in the presence of complex communities. 

\kobe{added citations for this paragraph}
We find that when ecological feedbacks are strong, the fixation probabilities for 
moderately beneficial mutants are suppressed compared to Kimura's predictions for constant selection\kobe{constant selection} (Fig.~\ref{fig:fixationProbability}b) \cite{antal2006fixation}. The suppression increases with effective population size (Fig.~\ref{fig:fixationProbability}e) \cite{antal2006fixation}, weaker
parent-mutant correlations (Fig.~\ref{fig:fixationProbability}e), and decreases as ecosystems become more packed (Fig.~\ref{fig:fixationProbability}f). We further show that the  
deviations from Kimura's formula arise from prolonged parent-mutant coexistence arising from ecological interactions
(Fig.~\ref{fig:fixProbExplanation}b and c) \cite{assaf2015obstruction}. This coexistence results in exponentially 
long absorption (Fig.~\ref{fig:coexistence}a) and extinction times (Fig.~\ref{fig:coexistence}b) for moderately beneficial mutants \cite{antal2006fixation,assaf2010large,assaf2015obstruction}, as well as
coexistence regions in the phase portraits of parent-mutant dynamics (Fig.~\ref{fig:coexistence}g).

Our theory is robust to the details of the exact ecological model used to mathematically represent the
community dynamics. We show that our framework
applies to various ecological models including generalized Lotka-Volterra models and multiple variants of the
consumer-resource model (SI Sec.~\ref{sec:gLVderivation}--\ref{sec:gCRM}), \kobe{fixed wording starting here}which all support regimes with parent-mutant coexistence. Therefore, consistent with \cite{mcenany2024predicting} for deterministic equilibria\kobe{added citation},
our analysis suggests that the possibility of parent-mutant coexistence is a hallmark of population genetics in complex communities,
and we expect 
the framework developed here to be widely applicable. As an exploration\kobe{slightly weaken}, we
applied our framework to understand ecological feedbacks in data from the
human gut microbiome \cite{roodgar2021longitudinal} (Fig.~\ref{fig:dataAnalysis}).


Our theory makes testable predictions for high-resolution strain tracking in natural microbiomes~\cite{goyal2022interactions,roodgar2021longitudinal}. With high enough temporal resolution, strain dynamics should exhibit intermittent stochastic fluctuations in a range of frequencies \textit{above} the drift barrier (Fig.~\ref{fig:fixProbExplanation}b--c). These dynamics should be most pronounced for moderately beneficial mutants with $s_\mathrm{inv} \lesssim \eta/2$. Importantly, estimating both $f^*$ and $s_\mathrm{inv}$ from observed trajectories could enable direct inference of the ecological feedback strength $\eta$ from strain dynamics alone. Together with the analysis in Fig.~\ref{fig:dataAnalysis}, our theory suggests a new empirical technique to estimate average \kobe{average}ecological feedbacks on evolutionary dynamics \cite{ascensao2025frequency}.

Our findings also have broader implications for inference using population genetics models. Current methods largely neglect ecology and for this reason may yield biased estimates of effective population sizes, demographic histories, and selection coefficients. Designing predictors that incorporate $\eta$ into coalescent models and demographic inference frameworks has the potential to improve the accuracy of statistical methods, especially in settings where ecology plays an important role in the evolutionary dynamics. Our results also have important implications for conservation genetics. They suggest that traditional metrics that ignore ecology may under- or over-estimate extinction risks.
 
In the future, it will be useful to extend our theory to a wider range of ecological and 
evolutionary settings. On the ecological side, while our theory has so far focused on communities near steady state, ecosystems can exhibit more
complex dynamical behavior such as limit cycles, multistability \cite{goyal2018multiple,lopes2024cooperative}, and chaos \cite{pearce2020stabilization,mahadevan2023spatiotemporal,blumenthal2024phase,arnoulx2024many,hu2022emergent}. Extending our theory to these settings will enable us to understand how the fate of a mutant depends on the underlying dynamical behavior of a community. 

\kobe{new discussion on strong-selection-weak-mutation, I know this may be too much but Ben asked for specific parameter combinations}
On the evolutionary side,
for many species, especially microbes, mutation rates are sufficiently large so that multiple mutants of the same species can emerge before previous ones fix or go extinct \cite{desai2007beneficial}. This phenomenon challenges our assumption of the strong-selection-weak-mutation regime, since we have let a mutant to coexist transiently with its parent without new mutants. Mathematically, our assumption holds when the stochastic extinction time of a coexisting mutant is faster than the time for new mutations (see SI Sec.~\ref{sec:extinctionTime} for precise mathematical condition).
This condition is typically violated in many microbial communities due to their large population sizes and mutation rates. In this case, multiple mutant lineages compete with each other simultaneously, a phenomenon known as clonal interference~\cite{yoshida2007cryptic,good2017dynamics,desai2007beneficial}. 
In the future, it would be interesting to extend DMFT approaches like the one we developed here to understand the effect of clonal interference on eco-evolutionary dynamics in complex communities.

\kobe{new discussion on small $1-\rho$}
On the modeling front, our model relies on the assumption that mutant strains have high phenotypic correlation $\rho\approx 1$ with their parents. While this assumption might be reasonable for mutations with small effect sizes,  the lack of genotype-phenotype maps in ecological contexts makes it challenging to estimate $\rho$ from community data.
Future work on better understanding how genotype-phenotype maps~\cite{liu2026ecology} affect ecological dynamics could help test the applicability of our approximation \cite{kinsler2020fitness}.

\kobe{new discussion on random assembly}
Finally, our theory assumes random and dense ecological interactions, and mutations that weakly alter many interactions, allowing feedbacks to be summarized using community-level statistics. However, ecological networks may become more nonrandom through ecological assembly~\cite{maynard2018network,pearl2025structured} or longer-term eco-evolution~\cite{feng2025theory,mcenany2024predicting,good2018adaptation,mahadevan2025continual}. Mutational effects may also be structured and context-dependent~\cite{kinsler2020fitness,hart2019disentangling,ascensao2023quantifying}, affecting only a few partners or changing multiple interactions coherently. In the future, it will be useful to extend our framework to these scenarios.



\section*{Methods}
\vskip -10pt

The details of all the analytical and numerical methods used can be found in the Supplementary Information. Briefly, we started with defining the equations for the generalized Lotka-Volterra model and various consumer-resource models, designating two of the species to 
be a parent and mutant. We analyzed these equations using dynamical 
mean-field theory and other approximations to derive Eq. \eqref{eq:frequencyDynamics} and the value of $\eta$. We then
solved the resulting stochastic differential equation to obtain fixation probabilities (Eq. \eqref{eq:ecoPfix}),
selection-drift ratio, and mean extinction time (Fig. \ref{fig:coexistence}).
We performed numerical simulations by sampling random 
ecosystems at steady states, then explicitly solved the equations in
the ecological models using the Euler method with demographic noise \cite{dornic2005integration,altieri2021properties,garcia2024interactions}. We repeated
the same simulation with multiple instances of demographic noise to obtain the
statistical quantities such as fixation probabilities and mean 
absorption/extinction times. Finally, we fitted the SNV trajectories 
(Fig.~\ref{fig:dataAnalysis}) to Eq. \eqref{eq:frequencyDynamics}, accompanied 
with a model of measurement noise, using maximum likelihood estimation. We 
provide details on the data analysis including the fitted model, the 
algorithms for maximum likelihood estimation, the full set of
inferred values with bootstrapping, and the procedures for testing the goodness of fit on one-step-ahead predictions.\kobe{more data methods}
This analysis was assisted by Claude Opus 5.5 for consultation of statistical approaches and production of the analysis code.\kobe{AI disclosure}

\vskip 10pt
\textsf{\textbf{Acknowledgements.}} We would like to thank Michael M. Desai and Benjamin H. Good\kobe{added Ben} for helpful discussions. This work was
funded by NIH NIGMS R35GM119461 to PM and
Chan-Zuckerberg Initiative Investigator grant to PM. SYL acknowledges support from the National Science Foundation through the Physics Frontier Center for Living Systems (PHY-2317138).
ZF acknowledges support by the NSF-Simons National Institute for Theory
and Mathematics in Biology (NITMB) Fellowship supported via grants from the NSF (DMS-2235451) and
Simons Foundation (MPS-NITMB-00005320).
AG acknowledges support from the Ashok and Gita
Vaish Junior Researcher Award, the DST-SERB Ramanujan Fellowship, as well the DAE, Govt. of India,
under project no. RTI4001.
This work was completed in part with resources provided by the University
of Chicago’s Research Computing Center.
Claude Opus 5.5 and ChatGPT-6 Astra were used to assist with minor editing of the manuscript, writing the simulation code, and checking the theoretical calculations.\kobe{AI disclosure}

\bibliography{references}

\appendix

\onecolumngrid

\newpage

\section*{Supplementary Information}
\renewcommand{\thefigure}{S\arabic{figure}}
\setcounter{figure}{0} 
\renewcommand{\thetable}{S\arabic{table}}
\setcounter{table}{0} 

\section{Lotka-Volterra model}
\label{sec:gLVderivation}

We begin with the details of our eco-evolution models. Consider an ecological
community of $S$ competing species, with the abundance of each species $N_i$ 
(where $1\leq i\leq S$) described by the Lotka-Volterra dynamics:
\begin{equation}
    \frac{dN_i}{dt}=N_i\left(r_i-N_i-\sum_{j\neq i}A_{ij}N_j\right)+\sqrt{2DN_i}\xi_i(t)\,,
\end{equation}
where $r_i$ is the intrinsic growth rate\kobe{renamed} of species $i$, and $A_{ij}$ is the 
competition coefficient between species $i$ and $j$. We have also introduced a 
white demographic noise $\xi_i(t)$ for each species with diffusion coefficient $D$, satisfying
\begin{equation}
    \left<\xi_i(t)\right>=0\,,\quad\left<\xi_i(t)\xi_j(t')\right>=\delta_{ij}\delta(t-t')\,.
\end{equation}
To model the diversity of this
community, we work in the limit $S\gg 1$ and assume that $r_i$ and $A_{ij}$ 
are randomly drawn from Gaussian distributions, such that
\begin{equation}
\label{eq:GaussianSampling}
    r_i=\mu_r+\sigma_r \delta r_i\,,\quad A_{ij}=\frac{\mu}{S}+\frac{\sigma}{\sqrt{S}}a_{ij}\,,
\end{equation}
where $\mu_r,\sigma_r,\mu,\sigma$ are the distribution parameters, and $\delta r_i,a_{ij}$ are zero-mean Gaussian random variables with
\begin{equation}
    \left[\delta r_i \delta r_j\right]=\delta_{ij}\,,\quad \left[a_{ij}a_{kl}\right]=\delta_{ik}\delta_{jl} + \gamma\delta_{il}\delta_{jk}\,,
\end{equation}
where the square brackets denote averages over random parameters instead of the demographic noise.
The reciprocity $-1\leq\gamma\leq 1$ is the correlation between the 
off-diagonal entries $A_{ij}$ and $A_{ji}$. The results below also apply to other distributions with mean and variance of order $1/S$, and higher order in $1/S$ for higher cumulants.\kobe{clarified higher cumulants}

Suppose at time $t=0$, a mutation of a parent strain $p$ occurs and a new mutant
strain $m$ with small abundance $N_m(0)\ll N_p(0)$ invades the community. Biologically, the mutant should be highly similar to its parent; we assume the parent-mutant correlation of ecological interactions $\rho$ to
be close to $1$. We sample the competition coefficients for the mutant as
\begin{align}
    a_{mi}&=\rho a_{pi}+\sqrt{1-\rho^2}z_i\,,\\
    a_{im}&=\rho a_{ip}+\sqrt{1-\rho^2}\left(\gamma z_i+\sqrt{1-\gamma^2}z_i'\right)\,,
\end{align}
where $z_i,z_i'$ are independent zero mean, unit variance Gaussian random
variables, and indices $i,j$ now denotes the rest of the community without the
parent or the mutant. These choices of coefficients ensure the right correlations
$\mathrm{Corr}\left(A_{pi},A_{mi}\right)=\mathrm{Corr}\left(A_{ip},A_{im}\right)=\rho$ and $\mathrm{Corr}\left(A_{mi},A_{im}\right)=\gamma$. We also have
$\mathrm{Corr}\left(A_{pi},A_{im}\right)=\mathrm{Corr}\left(A_{mi},A_{ip}\right)=\gamma\rho$. 
On the other hand, the competition between the parent and the mutant should be 
much stronger than that with other species, 
controlled by their niche overlap. 
In the main text, we assume
\begin{equation}
    A_{pm}=A_{mp}=\rho\,,
\end{equation}
but here we leave them to be arbitrary for generality as long as $1-A_{pm}$ and $1-A_{mp}$ are at order $1-\rho$.\kobe{$1-\rho$ expansion} As defined in 
Eq. \eqref{eq:GaussianSampling}, the other interaction coefficients 
$A_{pi},A_{mi},A_{ip},A_{im}$ (where $i\neq p,m$) are of order $1/S$
and much smaller than $A_{pm},A_{mp}$.
Finally, we do not specify any certain values for the intrinsic growth rate\kobe{renamed} $r_m$, in order
to obtain the functional dependence of the fixation probability and to control the order of $1-\rho$ in selection. If we sample $r_m$ as how we sample $A_{mi}$, the difference $r_m-r_p$ would be at order $\sqrt{1-\rho^2}\sim\sqrt{1-\rho}$. Throughout the analysis below, we make use of the small $1-\rho$ expansion to simplify our results; see also Appendix \ref{sec:parentMutantCorrelation} for more validation on results.\kobe{$1-\rho$ expansion}

Including the parent-mutant dynamics, the full set of Lotka-Volterra equations are
\begin{align}
\label{eq:GLVNi}
    \frac{dN_i}{dt}&=N_i\left(r_i-N_i-\sum_{j\neq i}A_{ij}N_j-A_{ip}N_p-A_{im}N_m\right)+\sqrt{2DN_i}\xi_i(t)\,,\\
\label{eq:GLVNp}
    \frac{dN_p}{dt}&=N_p\left(r_p-N_p-A_{pm} N_m-\sum_i A_{pi}N_i\right)+\sqrt{2DN_p}\xi_p(t)\,,\\
\label{eq:GLVNm}
    \frac{dN_m}{dt}&=N_m\left(r_m-N_m-A_{mp} N_p-\sum_i A_{mi}N_i\right)+\sqrt{2DN_m}\xi_m(t)\,.
\end{align}
We focus on the ``strong-selection-weak-mutation'' limit: we assume an ecological steady state and let the mutant
invade when the community is close to the steady state, then we let the community dynamics run till the parent or the mutant becomes extinct, even though the parent and mutant may transiently coexist for a long time. This approximation is one of the main
assumptions of our analysis, which ignores the possibility of having multiple
mutations of the same species in the community simultaneously due to nonzero mutation rates. By assuming an ecological steady state, we also require sufficiently small $D$ such that demographic fluctuations around the deterministic steady state are small over the timescale we study, and that other species in the community do not go extinct during mutant dynamics. This is a self-consistent requirement: we show in Appendix \ref{sec:extinctionTime} that under small $1-\rho$, the stochastic extinction time of a species is indeed much longer than that of the mutant. Furthermore, since  selection (terms such as $r_m-r_p$) is suppressed by $1-\rho$, $D$ should also be suppressed by $1-\rho$ accordingly in the strong-selection regime.\kobe{added assumption on $D$}

\subsection{Dynamical mean-field theory (DMFT)}
\kobe{added subsections for reviewer 3 since they care too much about what DMFT should be}

We now simplify the above set of Lotka-Volterra equations using DMFT. Instead of the abundances themselves, we will focus on the dynamics
of the mutant frequency, as is typical in this literature:
\begin{equation}
    f=\frac{N_m}{N_p+N_m}\,,
\end{equation}
which changes dynamically and stochastically till reaching the 
point $f=0$ or $f=1$. We denote the initial frequency as $f_0\ll 1$. Using
the Lotka-Volterra equation, we have
\begin{align}
\label{eq:rawFrequencyDynamicsGLV}
    \frac{df}{dt}&=\frac{1}{(N_p+N_m)^2}\left(N_p\frac{dN_m}{dt}-N_m\frac{dN_p}{dt}\right)\\
    &=f(1-f)\left(r_m-r_p-(1-A_{pm})N_m+(1-A_{mp})N_p-\sum_i (A_{mi}-A_{pi})N_i\right)+\sqrt{\frac{2Df(1-f)}{N_p+N_m}}\xi(t)\,.
\end{align}
Here we have combined the noise terms of the parent and the mutant into a 
single white noise by adding their variances. The ecological feedback from the community on parent-mutant dynamics is determined by how the parent and the mutant
impact other species in the community. We see that in Eq. \eqref{eq:GLVNi},
the abundances $N_p,N_m$ serve only as a perturbation of order $1/S$. We can
then model their influence to $N_i$ as a linear response. We write
\begin{equation}
    N_i(t)=N_{i/pm}(t)-\sum_j\int_0^t dt'\,\nu_{ij}(t,t')\left(A_{jp}N_p(t')+A_{jm}N_m(t')\right)\,,
\end{equation}
where $N_{i/pm}$ is the abundance of species $i$ if \emph{both} the parent and the mutant 
are excluded from the community, and
\begin{equation}
\label{eq:susceptibilityDef}
    \nu_{ij}(t,t')=\frac{\partial N_i(t)}{\partial r_j(t')}\,,
\end{equation}
is the susceptibility 
kernel. 
Using the linear response, we can now write the community contribution in 
Eq. \eqref{eq:rawFrequencyDynamicsGLV} as
\begin{equation}
\label{eq:DMFTresponse}
    \sum_i(A_{mi}-A_{pi})N_i(t)=\sum_i(A_{mi}-A_{pi})N_{i/pm}(t)-\sum_{ij}\int_0^t dt'\,\nu_{ij}(t,t')(A_{mi}-A_{pi})\left(A_{jp}N_p(t')+A_{jm}N_m(t')\right)\,.
\end{equation}
\kobe{new treatment}
For the first term in Eq.~\eqref{eq:DMFTresponse}, we use the assumption that the community without the parent or the mutant fluctuates around its ecological steady state, hence we decompose
\begin{equation}
\label{eq:coloredNoiseDef}
    N_{i/pm}(t)=\left<N_{i/pm}(0)\right>+\delta N_{i/pm}(t)\,,
\end{equation}
where we recall that the angle
bracket denotes the average over demographic fluctuations, and $\delta N_{i/pm}(t)$ is an effective noise term.
For the last term in Eq.~\eqref{eq:DMFTresponse}, it can be simplified at leading order with the self-averaging property of the community:
\begin{align}
    &\sum_{ij}\int_0^t dt'\,\nu_{ij}(t,t')(A_{mi}-A_{pi})\left(A_{jp}N_p(t')+A_{jm}N_m(t')\right)\\
    \simeq\, &\sum_{ij}\int_0^t dt'\,\nu_{ij}(t,t')\left[(A_{mi}-A_{pi})\left(A_{jp}N_p(t')+A_{jm}N_m(t')\right)\right] \\
    \simeq\, &\frac{\gamma\sigma^2 (1-\rho)}{S}\sum_i\int_0^t dt'\,\nu_{ii}(t,t')\left(N_m(t')-N_p(t')\right)\\
    =\, &\gamma\sigma^2(1-\rho)\int_0^t dt'\,\nu(t,t')\left(N_m(t')-N_p(t')\right)\,,
\end{align}
where we have introduced the susceptibility
\begin{equation}
    \nu(t,t')=\frac{1}{S}\sum_i\nu_{ii}(t,t')\,.
\end{equation}
Therefore, the mutant frequency satisfies
\begin{align}
\label{eq:frequencyDynamicsGLVAfterDMFT}
    \frac{df}{dt}=\,& f(1-f)\left(r_m-r_p-(1-A_{pm})N_m+(1-A_{mp})N_p-\sum_i (A_{mi}-A_{pi})\left(\left<N_{i/pm}(0)\right>+\delta N_{i/pm}(t)\right)\right.\\
    &\left.+\gamma\sigma^2(1-\rho)\int_0^t dt'\,\nu(t,t')\left(N_m(t')-N_p(t')\right)\right)+\sqrt{\frac{2Df(1-f)}{N_p+N_m}}\xi(t)\,.
\end{align}
\kobe{moved the comment to here}This is the DMFT equation for the mutant frequency with effective colored community noise and response, which in principle should be determined self-consistently together with the community dynamics. Note that in the conventional DMFT of the GLV, the term similar to the $\left<N_{i/pm}(0)\right>$ term in Eq.~\eqref{eq:frequencyDynamicsGLVAfterDMFT} also contributes to the community noise, even when the full dynamics is deterministic. This is because the cavity species can be chosen to be any species in the community, hence there is noise from heterogeneity in interactions across different species. Here, the cavity species is instead a specific pair of parent and mutant with quenched interactions, hence the $\left<N_{i/pm}(0)\right>$ term is deterministic. Nevertheless, there is still community noise from demographic fluctuations in $\delta N_{i/pm}(t)$.

\subsection{Small $1-\rho$ expansion}

\begin{figure}
    \centering
    \includegraphics[width=0.8\linewidth]{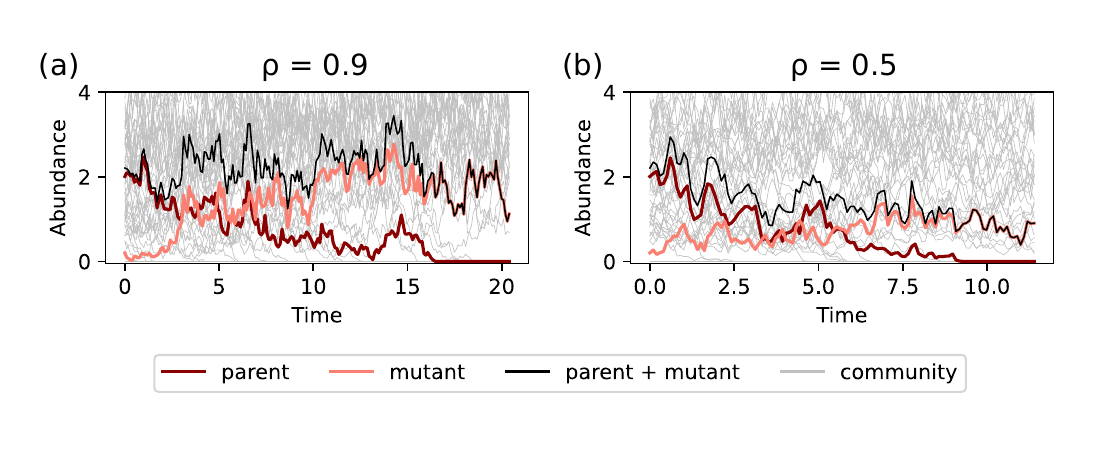}
    \caption{\justifying\textbf{Dynamics of parent and mutant abundances during mutant fixation at various parent-mutant correlation.} Fixation of a mutation generally requires deviation from the deterministic ecological steady state. At high $\rho=0.9$ (a), the total parent-mutant abundance remains constant on average throughout the full dynamics. At low $\rho=0.5$ (b), the total parent-mutant abundance drops significantly towards the final mutant abundance.\kobe{also got rid of ``neutral'' mutation since Ben didn't like it}}
    \label{fig:fig_rho_1}
\end{figure}

We now show that the DMFT equation implies that the dynamics of $f$ are slower than those of the community. This is because all the terms in Eq.~\eqref{eq:frequencyDynamicsGLVAfterDMFT} are suppressed by small $1-\rho$ to various extents.
To obtain an equation involving $f$ only, we then make further approximations using the small $1-\rho$ expansion.\kobe{$1-\rho$ expansion, also moved some of the other Appendix to here}

Qualitatively, when the parent and the mutant are highly similar in their 
ecological interactions with the community, the total parent-mutant population 
behaves effectively as a single species from the community perspective. Therefore,
the community as a whole stays near the ecological steady state. This is true even when the 
parent and the mutant frequencies are rather different from their steady state frequencies during the dynamics.
As a result, the total parent-mutant abundance $N_p+N_m$ always stays constant on 
average, with fluctuations coming from small demographic noise only. This is also consistent with Kimura's theory, which assumes a constant effective population size. In contrast, when 
$\rho$ is sufficiently less than 1, $N_p+N_m$ may not be stationary on average when a mutant  fixes, as shown in Fig. \ref{fig:fig_rho_1}. In conclusion, under the small $1-\rho$ expansion, we can replace
\begin{equation}
\label{eq:totalPopulationConstant}
    N_p(t)+N_m(t)\simeq \left<N_p(0)+N_m(0)\right>:=N_0\,,
\end{equation}
and the effective population size is $N_\mathrm{eff}=N_0/2D$. Note that we have not specified the physical unit of time for the dynamics; to match with real experiments, $N_\mathrm{eff}$ should be further normalized by the actual generation timescale.\kobe{added this comment since Ben complained our $N_\mathrm{eff}$ is too small}

\kobe{rewrote the whole analysis here}

Next, based on the ecological steady state, we calculate the community noise and the response in the DMFT equation and show that they can be greatly simplified under small $1-\rho$. Namely, the community noise is subdominant compared to the demographic noise, and the dynamic susceptibility can be replaced by a static one. 

First, consider the deterministic version of the community without the parent and the mutant. The abundances of surviving species $N_i^*=\left<N_{i/pm}(0)\right>$ satisfies the fixed point condition
\begin{equation}
    r_i-N_i^*-\sum_{j\neq i}A_{ij}N_j^*=0\,.
\end{equation}
Next, we add two sources of perturbations: the demographic noise, and the parent-mutant dynamics as an effective $\delta r_i(t)$. Accordingly, we decompose $N_i(t)=N_i^*+\delta N_i(t)$ for the linear response. The GLV dynamics becomes
\begin{equation}
    \frac{d(\delta N_i(t))}{dt}=N_i^*\left(\delta r_i(t)-\sum_j(I+A)_{ij}^*\delta N_j(t)\right)+\sqrt{2DN_i^*}\xi_i(t)\,,
\end{equation}
where the ``star'' on the matrices means taking the components of surviving species only. We can solve this ordinary differential equation using standard methods to obtain
\begin{equation}
    \delta N_i(t)=\sum_j\int_0^t dt'\,\exp\left(-\mathrm{diag}(N_i^*)(I+A)^*(t-t')\right)_{ij}\left(N_j^*\delta r_j(t')+\sqrt{2DN_j^*}\xi_j(t')\right)\,.
\end{equation}
Using the definitions in Eqs.~\eqref{eq:susceptibilityDef} and \eqref{eq:coloredNoiseDef}, we can then read off the susceptibility from the first term
\begin{equation}
\label{eq:susceptibilityExact}
    \nu_{ij}(t,t')=\nu_{ij}(t-t')=\exp\left(-\mathrm{diag}(N_i^*)(I+A)^*(t-t')\right)_{ij}N_j^*\,.
\end{equation}
and the community noise from the second term
\begin{equation}
    \delta N_{i/pm}(t)=\sqrt{2D}\sum_j\int_0^t dt'\,\nu_{ij}(t,t')(N_j^*)^{-1/2}\xi_j(t')\,.
\end{equation}

\begin{figure}
    \centering
    \includegraphics[width=0.8\linewidth]{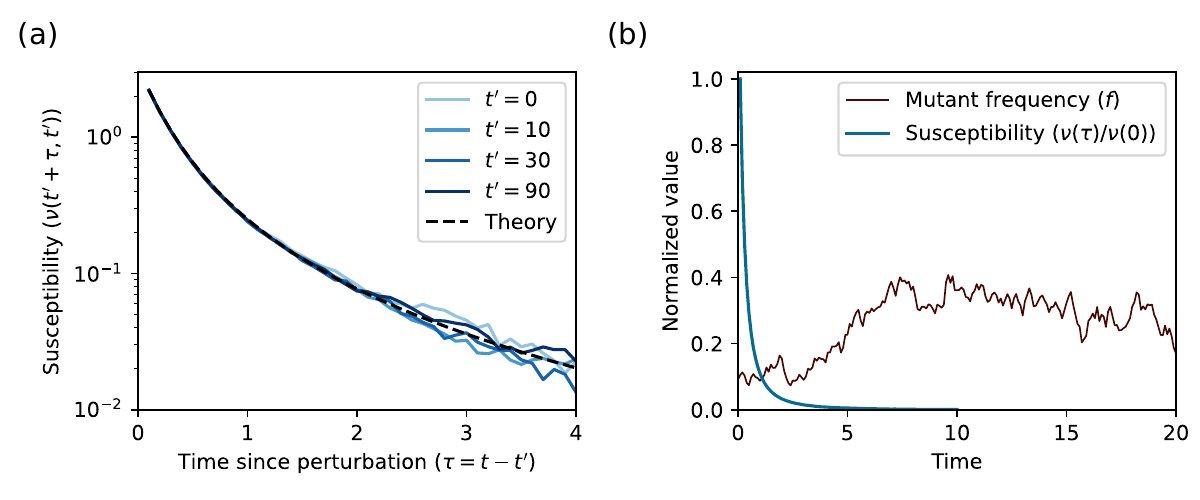}
    \caption{\textbf{Dynamical susceptibility and comparison with the dynamics of mutant frequencies.} (a) The susceptibility function is time-translational invariant and exponentially decaying, matching with the prediction of our theory (Eq.~\eqref{eq:susceptibilityExact}). (b) The susceptibility function decays much faster compared to the dynamics of mutant frequencies.}
    \label{fig:susceptibility}
\end{figure}

\begin{figure}
    \centering
    \includegraphics[width=0.8\linewidth]{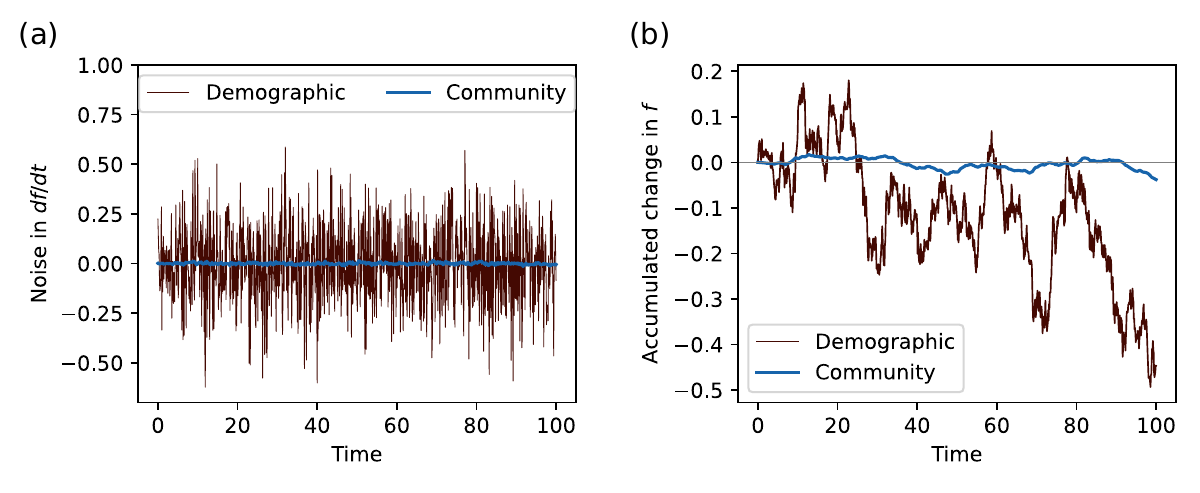}
    \caption{\textbf{Community noise and comparison with demographic noise of the parent and the mutant.} Given a mutation and its frequency trajectory, we re-simulate the community dynamics without the parent and the mutant. The resulting community noise is negligible compared to the demographic noise (or genetic drift) of the parent and the mutant, in terms of (a) the rates of change in $df/dt$ and (b) the accumulated changes in $f$.}
    \label{fig:communityNoise}
\end{figure}

We now tackle each of the quantities separately. First, the susceptibility is time-translational invariant can be decomposed into a sum of exponential decays in time difference $\tau=t-t'$. Notice that all the decay rates, given by the eigenvalues of $\mathrm{diag}(N_i^*)(I+A)^*$, are not suppressed by $1-\rho$. Therefore, since the derivatives of $\delta r_i(t)$ are suppressed by $1-\rho$ as in the frequency dynamics, we can approximate the response as
\begin{equation}
    \sum_j\int_0^t d\tau\,\nu_{ij}(\tau)\delta r_j(t-\tau)\simeq \sum_j\int_0^t d\tau\,\nu_{ij}(\tau)\delta r_j(t)\simeq\sum_j\nu_{ij}^0\delta r_j(t)\,,
\end{equation}
where
\begin{equation}
\label{eq:staticSusceptibilityDef}
    \nu_{ij}^0=\int_0^\infty d\tau\,\nu_{ij}(\tau)=(I+A)^{*-1}_{ij}\,,
\end{equation}
is the static susceptibility. Equivalently, we have approximated the dynamical susceptibility with a delta function $\nu_{ij}(t-t')\simeq \nu_{ij}^0 \delta(t-t')$. To numerically verify these approximations, we also compute the susceptibility function directly from the GLV simulation; see Appendix \ref{sec:simulationMethod} for methods. The results in Fig.~\ref{fig:susceptibility} indeed match with our theory.

Next, for the community noise term in Eq.~\eqref{eq:frequencyDynamicsGLVAfterDMFT}, we can now calculate its covariance using the self-averaging property.
\begin{align}
    &\,\left<\sum_i (A_{mi}-A_{pi})\delta N_{i/pm}(t)\sum_j (A_{mj}-A_{pj})\delta N_{j/pm}(t')\right>\\
    \simeq&\,\sum_{ij}\left[(A_{mi}-A_{pi})(A_{mj}-A_{pj})\right]\left<\delta N_{i/pm}(t)\delta N_{j/pm}(t')\right>\\
    =&\,\frac{2\sigma^2(1-\rho)}{S}\sum_i\left<\delta N_{i/pm}(t)\delta N_{i/pm}(t')\right>\\
    =&\,\frac{4D\sigma^2(1-\rho)}{S}\sum_{ijk}\int_0^t dt_1\int_0^{t'}dt_2\,\nu_{ij}(t-t_1)\nu_{ik}(t'-t_2)(N_j^*N_k^*)^{-1/2}\left<\xi_j(t_1)\xi_k(t_2)\right>\\
    =&\,\frac{4D\sigma^2(1-\rho)}{S}\sum_{ij}\int_0^{\min(t,t')} dt_1\,\nu_{ij}(t-t_1)\nu_{ij}(t'-t_1)(N_j^*)^{-1}\,.
\end{align}
Note that the covariance decays in time in the same way as the susceptibility, hence the decay is faster than the frequency dynamics. We can then compare the community noise with the demographic noise in the following way. Suppose the frequency $f$ evolves for a time $\Delta t$ that is longer than the timescale of covariance decay, but shorter than the timescale of frequency dynamics. In this regime, $f$ is unchanged on average and we can focus on its variance. Since the demographic noise is given by a white noise $\xi(t)$, its contribution to the variance is simply
\begin{equation}
    \mathrm{Var}(f)_\mathrm{demographic}\simeq \frac{2Df(1-f)}{N_0}\Delta t\,.
\end{equation}
On the other hand, the variance from the community noise is
\begin{align}
    \mathrm{Var}(f)_\mathrm{community}&\simeq f^2(1-f)^2\Delta t\int_{t'}^\infty dt\,\left<\sum_i (A_{mi}-A_{pi})\delta N_{i/pm}(t)\sum_j (A_{mj}-A_{pj})\delta N_{j/pm}(t')\right>\\
    &\simeq \frac{4D\sigma^2(1-\rho)f^2(1-f)^2\Delta t}{S}\sum_{ij}\int_0^{t'} dt_1\int_{t'}^\infty dt\, \nu_{ij}(t-t_1)\nu_{ij}(t'-t_1)(N_j^*)^{-1}\\
    &=\frac{4D\sigma^2(1-\rho)f^2(1-f)^2\Delta t}{S}\sum_{ij}\int_0^{t'} d\tau_1\int_{\tau_1}^\infty d\tau_2\, \nu_{ij}(\tau_1)\nu_{ij}(\tau_2)(N_j^*)^{-1}\\
    &\simeq \frac{2D\sigma^2(1-\rho)f^2(1-f)^2\Delta t}{S}\sum_{ij}\frac{(\nu_{ij}^0)^2}{N_j^*}\,.
\end{align}
In other words,
\begin{equation}
    \frac{\mathrm{Var}(f)_\mathrm{community}}{\mathrm{Var}(f)_{\mathrm{demographic}}}\simeq (1-\rho)\cdot \sigma^2 f(1-f)\cdot \frac{1}{S}\sum_{ij}\frac{N_0}{N_j^*}(\nu_{ij}^0)^2\,.
\end{equation}
Since all the factors except $1-\rho$ are $O(1)$, we conclude that the demographic noise is always dominant over the community noise, hence we omit the community noise from now on. We also verify this approximation in the GLV simulation in Fig.~\ref{fig:communityNoise}.

\subsection{Frequency-dependent selection}

With all these approximations, we arrive
at
\begin{align}
    \frac{df}{dt}=&\,f(1-f)\left(r_m-r_p-(1-A_{pm})N_0 f+(1-A_{mp})N_0(1-f)-\sum_i(A_{mi}-A_{pi})\left<N_{i/pm}(0)\right>\right.\\
    &\,\left.+2N_0\gamma\sigma^2\nu(1-\rho)(f-1/2)\right)+\sqrt{\frac{f(1-f)}{N_\mathrm{eff}}}\xi(t)\\ 
    =&\,f(1-f)s(f)+\sqrt{\frac{f(1-f)}{N_\mathrm{eff}}}\xi(t)\,,
\end{align}
where $s(f)$ represents frequency-dependent selection and the static susceptibility is
\begin{equation}
    \nu=\frac{1}{S}\sum_i\nu_{ii}^0\,.
\end{equation}
We see that $s(f)$ is 
linear in $f$. To better interpret the above result, we note that $s(f)$ is in
fact the difference between the growth rates between the mutant and the parent
at frequency $f$. The difference $s(f=0)$ is also the invasion
fitness of the mutant $s_\mathrm{inv}$ and plays the same role as the selection
coefficient $s$ in classic population genetics. Therefore, it is useful to
rewrite $s(f)$ as a linear function in $f$ involving $s_\mathrm{inv}$:
\begin{equation}
    s(f)=s_\mathrm{inv}-\eta f\,,
\end{equation}
where
\begin{equation}
    s_\mathrm{inv}=r_m-r_p+(1-A_{mp})N_0-\sum_i(A_{mi}-A_{pi})\left<N_{i/pm}(0)\right>-N_0\gamma\sigma^2\nu(1-\rho)\,,
\end{equation}
and
\begin{equation}
    \eta=2N_0\left(1-\frac{A_{pm}+A_{mp}}{2}-\gamma\sigma^2\nu(1-\rho)\right)\,,
\end{equation}
parametrizes the strength of the overall ecological effects. Note that the 
ecological feedback proportional to $\nu$ also contributes to the invasion fitness since the parent was already in
the community before $t=0$. In conclusion, we get
\begin{equation}
\label{eq:frequencyDynamicsSI}
    \frac{df}{dt}=f(1-f)\left(s_\mathrm{inv}-\eta f\right)+\sqrt{\frac{f(1-f)}{N_\mathrm{eff}}}\xi(t)\,,
\end{equation}
which is the same dynamics as in the main text.\kobe{comment on $1-\rho$ expansion} Note that all the three terms in the dynamics are suppressed by $1-\rho$ to different but independent extents. Therefore, one term may be dominant over the others depending on the parameter regime. Nevertheless, it is mathematically consistent in the sense that the effects we ignore in the previous section are all subleading compared to the dominant terms here.

To better 
understand the impact of the community, we can estimate the susceptibility using Eq.~\eqref{eq:staticSusceptibilityDef}, implying
\begin{equation}
    \nu=\frac{1}{S}\mathrm{Tr}(I+A)^{*-1}\,.
\end{equation}
The trace is simply the resolvent of $A^*$, which can be calculated using random
matrix theory. The result is \cite{cui2024elementary}
\begin{equation}
    \nu=\frac{1-\sqrt{1-4\gamma\sigma^2\phi}}{2\gamma\sigma^2}\,,
\end{equation}
where $\phi$ is the fraction of surviving species.
Since we have started with a near steady state of the community for the mutation, the above 
expression for $\nu$ should also hold approximately after the mutation. 
Therefore, we also have
\begin{equation}
    \eta = N_0\left(2-A_{pm}-A_{mp}-(1-\rho)\left(1-\sqrt{1-4\gamma\sigma^2\phi}\right)\right)\,.
\end{equation}
For the main text, we set $A_{pm}=A_{mp}=\rho$ and the above nicely
factorizes into
\begin{equation}
\label{eq:etaGLV}
    \eta = N_0(1-\rho)\left(1+\sqrt{1-4\gamma\sigma^2\phi}\right)\,.
\end{equation}
Note that this expression is valid only when $\phi<1/4\gamma\sigma^2$, which is
an upper bound on the number of surviving species similar to May's stability
bound for ecological communities \cite{MayStability}, since we have assumed that the community is near a stable state. This is the species packing bound referred
in the main text.

\section{MacArthur consumer-resource model}
\label{sec:CRM}

Following the same procedure as in Appendix \ref{sec:gLVderivation}, it is
straightforward to extend the results to consumer-resource models. We will see
that the mutant frequency follows the same dynamics as before, only with a 
different $\eta$.

We first study the MacArthur consumer-resource model with self-renewing resources.
We consider an ecological community of $S$ species and $M$ resources. The species
abundances $N_i$ (where $1\leq i\leq S$) and resource abundances $R_\alpha$
(where $1\leq \alpha\leq M$) follow
\begin{align}
\label{eq:CRM-N}
    \frac{dN_i}{dt}&=N_i\left(\sum_\beta C_{i\beta}R_\beta-m_i\right)+\sqrt{2DN_i}\xi_i(t)\,,\\
\label{eq:CRM-R}
    \frac{dR_\alpha}{dt}&=R_\alpha\left(K_\alpha-R_\alpha-\sum_j E_{j\alpha} N_j \right)\,,
\end{align}
where $K_\alpha$ is the carrying capacity of resource $\alpha$, $m_i$ is the 
intrinsic mortality rate for species $i$, $C_{i\alpha}$
represents the consumption preferences of species $i$ for resource $\alpha$, and
$E_{i\alpha}$ is the corresponding impact of species $i$ on resource $\alpha$. 
The demographic noise $\xi_i(t)$ is the same as in the previous section.
For a diverse community, we work in the limit $S,M\gg 1$ but finite $S/M$. We
further assume that the parameters are randomly drawn from Gaussian 
distributions, such that
\begin{gather}
    K_\alpha=\mu_K+\sigma_K\delta K_\alpha\,,\quad m_i=\mu_m+\sigma_m \delta m_i\,,\\
    C_{i\alpha}=\frac{\mu}{M}+\frac{\sigma}{\sqrt M}c_{i\alpha}\,,\quad E_{i\alpha}=\frac{\mu}{M}+\frac{\sigma}{\sqrt M}e_{i\alpha}\,,
\end{gather}
where $\mu_K,\sigma_K,\mu_m,\sigma_m,\mu,\sigma$ are the distribution parameters,
and $\delta K_\alpha,\delta m_i,c_{i\alpha},e_{i\alpha}$ are zero-mean Gaussian
random variables with
\begin{gather}
    \left[\delta K_\alpha\delta K_\beta\right]=\delta_{\alpha\beta}\,,\quad\left[\delta m_i\delta m_j\right]=\delta_{ij}\,,\\
    \left[c_{i\alpha}c_{j\beta}\right]=\left[e_{i\alpha}e_{j\beta}\right]=\delta_{ij}\delta_{\alpha\beta}\,,\quad \left[c_{i\alpha}e_{j\beta}\right]=\kappa \delta_{ij}\delta_{\alpha\beta}\,.
\end{gather}
The reciprocity $0\leq\kappa\leq 1$ of species-resource interaction is the
correlation between the matrices $C_{i\alpha}$ and $E_{i\alpha}$.

Suppose at time $t=0$, a mutation of a parent strain $p$ occurs and a new mutant
strain $m$ with small abundance $N_m(0)\ll N_p(0)$ invades the community.
Let the parent-mutant correlation be $\rho$. We sample the new interaction coefficients as
\begin{align}
    c_{m\alpha}&=\rho c_{p\alpha}+\sqrt{1-\rho^2}z_\alpha\,,\\
    e_{m\alpha}&=\rho e_{p\alpha}+\sqrt{1-\rho^2}\left(\kappa z_\alpha+\sqrt{1-\kappa^2}z_\alpha'\right)\,,
\end{align}
where $z_\alpha,z_\alpha'$ are independent zero mean, unit variance Gaussian
random variables. These choices of coefficients ensure the right correlation
$\mathrm{Corr}(C_{p\alpha},C_{m\alpha})=\mathrm{Corr}(E_{p\alpha},E_{m\alpha})=\rho$ and $\mathrm{Corr}(C_{m\alpha},E_{m\alpha})=\kappa$. We leave the mortality rate $m_m$ to be arbitrary. Note that we do
not introduce any new resources.

After introducing the mutant, the consumer-resource dynamics becomes
\begin{align}
    \frac{dN_p}{dt}&=N_p\left(\sum_\beta C_{p\beta}R_\beta-m_p\right)+\sqrt{2DN_p}\xi_p(t)\,,\\
    \frac{dN_m}{dt}&=N_m\left(\sum_\beta C_{m\beta}R_\beta-m_m\right)+\sqrt{2DN_m}\xi_m(t)\,,\\
    \frac{dR_\alpha}{dt}&=R_\alpha\left(K_\alpha-R_\alpha-\sum_j E_{j\alpha}N_j-E_{p\alpha}N_p-E_{m\alpha}N_m\right)\,.
\end{align}
In terms of the mutant frequency $f=N_m/(N_p+N_m)$, the above becomes
\begin{align}
\label{eq:rawCRMFrequencydynamics}
    \frac{df}{dt}&=\frac{1}{(N_p+N_m)^2}\left(N_p\frac{dN_m}{dt}-N_m\frac{dN_p}{dt}\right)\\
    &=f(1-f)\left(\sum_\beta (C_{m\beta}-C_{p\beta})R_\beta-m_m+m_p\right)+\sqrt{\frac{2Df(1-f)}{N_p+N_m}}\xi(t)\,.
\end{align}
The dynamics of the other species $N_i$ will not be relevant in the following
calculation. Now we adopt the same set of approximations as before. We can
model the changes of $R_\alpha$ as linear responses:
\begin{equation}
    R_\alpha=R_{\alpha /pm}-\sum_\beta\int_0^t dt'\, \chi_{\alpha\beta}(t,t')\left(E_{p\beta}N_p(t')+E_{m\beta}N_m(t')\right)\,,
\end{equation}
where
\begin{equation}
    \chi_{\alpha\beta}(t,t')=\frac{\partial R_\alpha(t)}{\partial K_\beta(t')}\,,
\end{equation}
is the susceptibility kernel. Now we can write the resource contribution in 
Eq. \eqref{eq:rawCRMFrequencydynamics} as
\begin{equation}
    \sum_\alpha (C_{m\alpha}-C_{p\alpha})R_\alpha(t)=\sum_\alpha (C_{m\alpha}-C_{p\alpha})R_{\alpha/pm}(t)-\sum_{\alpha\beta}\int_0^t dt'\,\chi_{\alpha\beta}(t,t')(C_{m\alpha}-C_{p\alpha})\left(E_{p\beta}N_p(t')+E_{m\beta}N_m(t')\right)
\end{equation}
We can then apply self-averaging to the products 
between $C$ and $E$ and approximate the last term as
\begin{align}
    &\sum_{\alpha\beta}\int_0^t dt'\,\chi_{\alpha\beta}(t,t')(C_{m\alpha}-C_{p\alpha})\left(E_{p\beta}N_p(t')+E_{m\beta}N_m(t')\right)\\
    \simeq\, &\sum_{\alpha\beta}\int_0^t dt'\,\chi_{\alpha\beta}(t,t')\left[(C_{m\alpha}-C_{p\alpha})\left(E_{p\beta}N_p(t')+E_{m\beta}N_m(t')\right)\right]\\
    \simeq\, &\frac{\kappa\sigma^2 (1-\rho)}{M}\sum_\alpha\int_0^t dt'\,\chi_{\alpha\alpha}(t,t')\left(N_m(t')-N_p(t')\right)\\
    =\, &\kappa\sigma^2(1-\rho)\int_0^t dt'\,\chi(t,t')\left(N_m(t')-N_p(t')\right)\,,
\end{align}
where we have introduced the susceptibility
\begin{equation}
    \chi(t,t')=\frac{1}{M}\sum_\alpha\chi_{\alpha\alpha}(t,t')\,.
\end{equation}
Therefore, the mutant frequency satisfies
\begin{align}
    \frac{df}{dt}=\,& f(1-f)\left(\sum_\beta (C_{m\beta}-C_{p\beta})R_{\beta/pm}-m_m+m_p\right.\\
    &\left.-\kappa\sigma^2(1-\rho)\int_0^t dt'\,\chi(t,t')\left(N_m(t')-N_p(t')\right)\right)+\sqrt{\frac{2Df(1-f)}{N_p+N_m}}\xi(t)\,.
\end{align}
As before, we then further approximate
\begin{equation}
    \sum_\beta (C_{m\beta}-C_{p\beta})R_{\alpha /pm}(t)\simeq \sum_\beta (C_{m\beta}-C_{p\beta})\left<R_{\alpha /pm}(0)\right>\,,\quad N_p(t)+N_m(t)\simeq N_0\,,\quad \chi(t,t')\simeq \chi\delta(t-t')\,.
\end{equation}
We then arrive at:
\begin{align}
    \frac{df}{dt}=&\,f(1-f)\left(\sum_\beta (C_{m\beta}-C_{p\beta})R_{\beta/pm}(0)-m_m+m_p\right.\\
    &\,\left.-2\kappa\sigma^2\chi(1-\rho)N_0(f-1/2)\right)+\sqrt{\frac{f(1-f)}{N_\mathrm{eff}}}\xi(t)\\
    =&\,f(1-f)s(f)+\sqrt{\frac{f(1-f)}{N_\mathrm{eff}}}\xi(t)\,.
\end{align}
As before, we can separate the invasion fitness $s_\mathrm{inv}=s(f=0)$ from $s(f)$ and 
rewrite the dynamics in the form of Eq. \eqref{eq:frequencyDynamicsSI} as
\begin{equation}
    \frac{df}{dt}=f(1-f)\left(s_\mathrm{inv}-\eta f\right)+\sqrt{\frac{f(1-f)}{N_\mathrm{eff}}}\xi(t)\,,
\end{equation}
where the invasion fitness is
\begin{equation}
    s_\mathrm{inv}=\sum_\beta \left(C_{m\beta}-C_{p\beta}\right)\left<R_{\beta /pm}(0)\right>-m_m+m_p+\frac{\eta}{2}\,,
\end{equation}
and the strength of ecological effect is
\begin{equation}
\label{eq:etaCRM}
    \eta=2\kappa\sigma^2\chi(1-\rho) N_0\,.
\end{equation}

Now, we can estimate the susceptibility by the steady state conditions for 
surviving species and non-depleted resources (denoted by asterisks):
\begin{equation}
    \sum_\beta C^*_{i\beta}R_\beta-m_i=K_\alpha-R_\alpha-\sum_jE^*_{j\alpha}N_j=0\,.
\end{equation}
The variations satisfy
\begin{equation}
    \left(\begin{matrix}
        0 & C^*\\
        E^{*T} & I^*
    \end{matrix}\right)
    \left(\begin{matrix}
        \delta N \\ \delta R
    \end{matrix}\right)=
    \left(\begin{matrix}
        0 \\ \delta K
    \end{matrix}\right)\,.
\end{equation}
Using block matrix inversion, we have
\begin{equation}
    \delta R=\left(I^*-E^{*T}\left(C^*E^{*T}\right)^{-1}C^*\right)\delta K\,,
\end{equation}
hence,
\begin{align}
    \chi&=\frac{1}{M}\mathrm{Tr}\left(I^*-E^{*T}\left(C^*E^{*T}\right)^{-1}C^*\right)\\
    &=\frac{1}{M}\mathrm{Tr}\,I^* - \frac{1}{M}\mathrm{Tr}\left(\left(C^*E^{*T}\right)^{-1}C^*E^{*T}\right)\\
    &=\phi_R-\frac{S}{M}\phi_N\,,
\end{align}
where $\phi_R$ is the fraction of non-depleted resources, and $\phi_N$ is the
fraction of surviving species. In other words, $\chi$ measures how far the 
community is from the competitive exclusion limit. This $\chi$ is again the same 
as the corresponding susceptibility appeared in the static cavity method.
We finally have
\begin{equation}
    \eta=2\kappa\sigma^2\left(\phi_R-\frac{S}{M}\phi_N\right)(1-\rho)N_0\,.
\end{equation}

\kobe{new comment for Ben}Note that when the community is fully-packed, i.e., the number of surviving species $S\phi_N$ is the same as the number of non-depleted resources $M\phi_R$, the above expression for $\eta$ vanishes. However, the $\eta$ term is only a linear approximation to the ecological effects. We expect that other nonlinear effects including mutation-induced extinctions \cite{mcenany2024predicting,feng2025theory} and variances from random matrices can still contribute to the ecological suppression of fixation probabilities. These effects should still be suppressed by high $\rho$, and we show in Figs.~\ref{fig:fixProbCRMPacking} and \ref{fig:fig_rho_2} that the Kimura theory (i.e., with $\eta=0$) indeed works well in the fully-packed regime when $\rho$ is high. However, as $\rho$ decreases, these effects can become stronger, and we would need nonlinear corrections to better predict the dynamics.

\section{Generalized consumer-resource models}
\label{sec:gCRM}

We can further extend the above results to more general consumer-resource models
with abiotic resources.
The dynamics take the form \cite{goyal2025universal}
\begin{align}
    \frac{dN_i}{dt}&=N_i\left(g_i(\vec R)-m_i\right)+\sqrt{2DN_i}\xi_i(t)\,,\\
    \frac{dR_\alpha}{dt}&=h_\alpha(\vec K)-q_\alpha(\vec R)-f_\alpha(\vec R,\vec N)\,,
\end{align}
where $g_i(\vec R)$ describes how the growth rate
of species $i$ depends on resource abundances, $h_\alpha(\vec K)$ is the
external supply rate of resource $\alpha$ to the community, $q_\alpha(\vec R)$
represents the resource dynamics in the absence of species, and
$f_\alpha(\vec R,\vec N)$ is the rate at which resource $\alpha$ is produced
or consumed by the species in the community. An example of such a model is
the consumer-resource model with externally supplied resources:
\begin{align}
    \frac{dN_i}{dt}&=N_i\left(\sum_\alpha C_{i\alpha}R_\alpha-m_i\right)+\sqrt{2DN_i}\xi_i(t)\,,\\
    \frac{dR_\alpha}{dt}&=K_\alpha-R_\alpha-\sum_i C_{i\alpha}N_iR_\alpha\,.
\end{align}
The consumer preference matrix $C_{i\alpha}$ matches the definition in 
below. The sampling schemes for $C_{i\alpha},K_\alpha,m_i$ are the same as
in Sec. \ref{sec:CRM}.

Due to the generality, here we will
derive the corresponding equation to Eq. \eqref{eq:frequencyDynamicsSI} in terms of
the functions $g_i,h_\alpha,q_\alpha,f_\alpha$ themselves, but not the random 
matrix statistics. It will be helpful to define the following matrices \cite{goyal2025universal}:
\begin{align}
    C_{i\alpha}&=\left.\frac{\partial g_i}{\partial R_\alpha}\right|_{\vec R^*}\,,\\
    E_{i\alpha}&=\left.\frac{\partial f_\alpha}{\partial N_i}\right|_{\vec R^*,\vec N^*}\,,\\
    Q_{\alpha\beta}&=\left.\frac{\partial f_\alpha}{\partial R_\beta}\right|_{\vec R^*,\vec N^*}+\left.\frac{\partial q_\alpha}{\partial R_\beta}\right|_{\vec R^*}\,,
\end{align}
where $\vec N^*,\vec R^*$ are the abundances at the steady state. 
$C_{i\alpha}$ and $E_{i\alpha}$ play similar roles as in the MacArthur
consumer-resource model; see Eqs. \eqref{eq:CRM-N} and \eqref{eq:CRM-R}. 
$Q_{\alpha\beta}$ represents the interactions between resources.

Suppose at time $t=0$, a mutation of a parent strain $p$ occurs and a new mutant
strain $m$ with small abundance $N_m(0)\ll N_p(0)$ invades the community.
The consumer-resource dynamics becomes
\begin{align}
    \frac{dN_p}{dt}&=N_p\left(g_p(\vec R)-m_p\right)+\sqrt{2DN_p}\xi_p(t)\,,\\
    \frac{dN_m}{dt}&=N_m\left(g_m(\vec R)-m_m\right)+\sqrt{2DN_m}\xi_m(t)\,,\\
    \frac{dR_\alpha}{dt}&=h_\alpha(\vec K)-q_\alpha(\vec R)-f_\alpha(\vec R,N_i,N_p,N_m)\\
    &\simeq h_\alpha(\vec K)-q_\alpha(\vec R)-f_\alpha(\vec R,N_i)-E_{p\alpha}N_p-E_{m\alpha}N_m\,.
\end{align}
In terms of the mutant frequency $f=N_m/(N_p+N_m)$, the above becomes
\begin{align}
    \frac{df}{dt}&=\frac{1}{(N_p+N_m)^2}\left(N_p\frac{dN_m}{dt}-N_m\frac{dN_p}{dt}\right)\\
    &=f(1-f)\left(g_m(\vec R)-g_p(\vec R)-m_m+m_p\right)+\sqrt{\frac{2Df(1-f)}{N_p+N_m}}\xi(t)\,.
\end{align}
The dynamics of the other species $N_i$ will not be relevant in the following
calculation. Now we adopt the same set of approximations as before. We can
model the changes of $R_\alpha$ as linear responses, which can be approximated by
the variations of the steady state conditions:
\begin{equation}
    \left(\begin{matrix}
        0 & C^*\\
        E^{*T} & Q
    \end{matrix}\right)
    \left(\begin{matrix}
        \delta N \\ \delta R
    \end{matrix}\right)=
    \left(\begin{matrix}
        0 \\ \delta h
    \end{matrix}\right)\,,
\end{equation}
where
\begin{equation}
    \delta h_\alpha=-E_{p\alpha}N_p-E_{m\alpha}N_m\,.
\end{equation}
Using block matrix inversion, we have
\begin{equation}
    R_\alpha=R_{\alpha/pm}-\sum_\beta\left(\left(I-P\right)Q^{-1}\right)_{\alpha\beta}\left(E_{p\beta}N_p+E_{m\beta}N_m\right)\,,
\end{equation}
where
\begin{equation}
    P=Q^{-1}E^{*T}\left(C^*Q^{-1}E^{*T}\right)^{-1}C^*\,.
\end{equation}
is a projector in resource space satisfying $P^2=P$ \cite{goyal2025universal}.
We see that the matrix $(I-P)Q^{-1}$ generalizes the susceptibility 
$\chi_{\alpha\beta}$ in the MacArthur consumer-resource model.
Now, the mutant frequency satisfies
\begin{align}
    \frac{df}{dt}=\,& f(1-f)\left(g_m\left(\vec R_{/pm}\right)-g_p\left(\vec R_{/pm}\right)-m_m+m_p\right.\\
    &\left.-\sum_{\alpha\beta}(C_{m\alpha}-C_{p\alpha})\left(\left(I-P\right)Q^{-1}\right)_{\alpha\beta}\left(E_{p\beta}N_p+E_{m\beta}N_m\right)\right)+\sqrt{\frac{2Df(1-f)}{N_p+N_m}}\xi(t)\,.
\end{align}
As before, we then approximate
\begin{equation}
    g_m\left(\vec R_{/pm}(t)\right)-g_p\left(\vec R_{/pm}(t)\right)\simeq \left<g_m\left(\vec R_{/pm}(0)\right)-g_p\left(\vec R_{/pm}(0)\right)\right>\,,\quad N_p(t)+N_m(t)\simeq N_0\,.
\end{equation}
We then arrive at the same equation as Eq. \eqref{eq:frequencyDynamicsSI}:
\begin{align}
    \frac{df}{dt}=\,& f(1-f)\left(\left<g_m\left(\vec R_{/pm}(0)\right)-g_p\left(\vec R_{/pm}(0)\right)\right>-m_m+m_p\right.\\
    &\left.-N_0\sum_{\alpha\beta}(C_{m\alpha}-C_{p\alpha})\left(\left(I-P\right)Q^{-1}\right)_{\alpha\beta}\left(E_{p\beta}(1-f)+E_{m\beta}f\right)\right)+\sqrt{\frac{f(1-f)}{N_\mathrm{eff}}}\xi(t)\,.\\
    =&\,f(1-f)\left(s_\mathrm{inv}-\eta f\right)+\sqrt{\frac{f(1-f)}{N_\mathrm{eff}}}\xi(t)\,,
\end{align}
where the invasion fitness is
\begin{align}
    s_\mathrm{inv}=&\,\left<g_m\left(\vec R_{/pm}(0)\right)-g_p\left(\vec R_{/pm}(0)\right)\right>-m_m+m_p\\
    &\,-N_0\sum_{\alpha\beta}(C_{m\alpha}-C_{p\alpha})\left(\left(I-P\right)Q^{-1}\right)_{\alpha\beta}E_{p\beta}
\end{align}
and the strength of ecological effect is
\begin{equation}
    \eta=N_0\sum_{\alpha\beta}(C_{m\alpha}-C_{p\alpha})\left(\left(I-P\right)Q^{-1}\right)_{\alpha\beta}\left(E_{m\beta}-E_{p\beta}\right)\,.
\end{equation}
Here we do not specify the statistics of the matrices $C,E,Q$. Nevertheless, we expect
that upon self-averaging, the expression of $\eta$ involves only the trace of $(I-P)Q^{-1}$ instead of the full matrix, corresponding to the scalar susceptibility 
$\chi$ in the MacArthur consumer-resource model.
We then conclude that generalized consumer-resource models lead to the same
dynamics of mutant frequency as in the MacArthur consumer-resource model.

\section{Derivation and analysis of fixation probability}
\label{sec:fixProbAnalysis}

From the previous sections, we have seen that all the ecological models
give rise to the same dynamics of mutant frequency $f$, 
satisfying the stochastic differential equation
\begin{equation}
    \frac{df}{dt}=f(1-f)\left(s_\mathrm{inv}-\eta f\right)+\sqrt{\frac{f(1-f)}{N_\mathrm{eff}}}\xi(t)\,,
\end{equation}
where $s_\mathrm{inv}$ is the 
invasion fitness, $\eta$ is the strength of ecological effects
depending on details of the ecological model, and $N_\mathrm{eff}$ is the initial effective population size of the parent and the mutant.

The fixation probability can be obtained by solving the stochastic
differential equation. Let $p(f,t|x)$ be the probability that
the frequency is between $f$ and $f+df$ at time $t$ given that $f(0)=x$. 
The corresponding Kolmogorov 
backward equation is
\begin{equation}
    \frac{\partial p}{\partial t}=x(1-x)(s_\mathrm{inv}-\eta x)\frac{\partial p}{\partial x}+\frac{x(1-x)}{2N_\mathrm{eff}}\frac{\partial^2 p}{\partial x^2}\,.
\end{equation}
Since $f$ stops at either $0$ or $1$ after sufficiently long time, $p$ must
approach the steady state $p^*(f,x)=p(f,\infty|x)$ given by
\begin{gather}
    0=x(1-x)(s_\mathrm{inv}-\eta x)\frac{\partial p^*}{\partial x}+\frac{x(1-x)}{2N_\mathrm{eff}}\frac{\partial^2 p^*}{\partial x^2}\,,\\
    \Rightarrow \frac{\partial p^*}{\partial x}\propto\exp\left[N_\mathrm{eff}\left(\eta x^2-2s_\mathrm{inv}x\right)\right]\,.
\end{gather}
Using the stopping condition $p(1,0)=0$ and $p(1,1)=1$, we arrive at the
fixation probability
\begin{align}
\label{eq:fixationProbSI}
    p_\mathrm{fix}&:= p(1,f_0)=\frac{\int_0^{f_0}dx\,\partial p^*/\partial x}{\int_0^1 dx\,\partial p^*/\partial x}\\
    &=\frac{\mathrm{Erfi}\left(\alpha\tilde s_\mathrm{inv}\right)-\mathrm{Erfi}\left(\alpha(\tilde s_\mathrm{inv}-f_0)\right)}{\mathrm{Erfi}\left(\alpha\tilde s_\mathrm{inv}\right)-\mathrm{Erfi}\left(\alpha(\tilde s_\mathrm{inv}-1)\right)}\,,
\end{align}
where
\begin{equation}
    \mathrm{Erfi}(x)=\frac{2}{\sqrt{\pi}}\int_0^x dy\,e^{y^2}\,,
\end{equation}
is the imaginary error function. We have denoted $\alpha=\sqrt{N_\mathrm{eff}\eta}$ as 
the ratio of strength between ecological
effects and demographic noise. \kobe{$1-\rho$ expansion}Although $\eta$ is suppressed by $1-\rho$, we have also assumed a large $N_\mathrm{eff}$ and it is possible that $N_\mathrm{eff}^{-1}\sim O(1-\rho)$ or lower, so that $\alpha$ can be large. Next, $\tilde s_\mathrm{inv}=s_\mathrm{inv}/\eta$ is closely related to the 
``dressed invasion fitness'' $g_\mathrm{dressed}$ in \cite{feng2025theory}. \kobe{simple justification for mathematical relationship} $\tilde s_\mathrm{inv}$ predicts the frequency for the coexistence region when the parent and mutant can coexist. Meanwhile, $g_\mathrm{dressed}$ predicts the invader abundance when there is no invasion-induced extinction. Therefore, in the regime of parent-mutant coexistence, we have
\begin{equation}
    g_\mathrm{dressed}\simeq N_0\tilde s_\mathrm{inv}\,,
\end{equation}
but there is no definite relationship between $g_\mathrm{dressed}$ and $\tilde s_\mathrm{inv}$ outside this regime, especially when mutant invaders displace their parents.

We can better understand the formula for $p_\mathrm{fix}$ using the asymptotic
forms
\begin{equation}
    \mathrm{Erfi}(x)=\begin{cases}
        2x/\sqrt{\pi} & |x|\ll 1\\
        e^{x^2}/x\sqrt{\pi} & |x|\gg 1\,.
    \end{cases}
\end{equation}
For example, when $s_\mathrm{inv}=0$,
\begin{equation}
    p_\mathrm{fix}(s_\mathrm{inv}=0)=\frac{\mathrm{Erfi}\left(\alpha f_0\right)}{\mathrm{Erfi}\left(\alpha\right)}\,.
\end{equation}
For $\alpha\ll 1$ (and $f_0\ll 1$ as we have assumed), we can use the linear approximation for all the terms, thus
\begin{equation}
    p_\mathrm{fix}(s_\mathrm{inv}=0,\alpha\ll 1)\simeq \frac{\alpha f_0}{\alpha}=f_0\,.
\end{equation}
For $\alpha\gg 1$, the denominator is always exponentially large and dominating.
Ignoring the subdominating factors, we have
\begin{equation}
    p_\mathrm{fix}(s_\mathrm{inv}=0,\alpha\gg 1)\sim e^{-\alpha^2}\,.
\end{equation}
We see that $p_\mathrm{fix}(s_\mathrm{inv}=0)$ matches Kimura's prediction when 
$\alpha\ll 1$, but exponentially suppressed by $\alpha$ otherwise.

The fixation probability for mutations with $s_\mathrm{inv}=0$ suggests that there is a crossover
between the regimes of weak and strong ecological effects, characterized by small and large $\eta$ respectively.
First, for small $\eta$ with
$\alpha\ll 1$, we have $|\alpha\tilde s_\mathrm{inv}|\gg 1$ for any finite $s_\mathrm{inv}$ since 
$\alpha\tilde s_\mathrm{inv}$ scales as $\eta^{-1/2}$. Therefore, 
we can use the exponential approximation for all the terms and $p_\mathrm{fix}$ becomes
\begin{align}
\label{eq:reduceToKimura}
    p_\mathrm{fix}&\simeq \frac{\exp\left(\alpha^2 \tilde s_\mathrm{inv}^2\right)-\exp\left(\alpha^2 (\tilde s_\mathrm{inv}-f_0)^2\right)}{\exp\left(\alpha^2 \tilde s_\mathrm{inv}^2\right)-\exp\left(\alpha^2 (\tilde s_\mathrm{inv}-1)^2\right)}\\
    &\simeq\frac{1-\exp(-2N_\mathrm{eff} s_\mathrm{inv} f_0)}{1-\exp(-2N_\mathrm{eff} s_\mathrm{inv})}\,,
\end{align}
which is exactly Kimura's formula for fixation probability with $s=s_\mathrm{inv}$. Combining the result for $s_\mathrm{inv}=0$, we see that Kimura's 
formula works equally well in
ecological context as long as $\alpha\ll 1$. On the other hand, for large $\eta$
with $\alpha\gg 1$, the behavior of $p_\mathrm{fix}$ is divided into different 
regimes according to $s_\mathrm{inv}$. Now for $|\alpha\tilde s_\mathrm{inv}|\ll 1$, we again have
\begin{equation}
    p_\mathrm{fix}\simeq \frac{\mathrm{Erfi}\left(\alpha f_0\right)}{\mathrm{Erfi}\left(\alpha\right)}\sim e^{-\alpha^2}\,.
\end{equation}
For $|\alpha\tilde s_\mathrm{inv}|\gg 1$, we have
\begin{align}
    p_\mathrm{fix}&\simeq\frac{\exp\left(\alpha^2 \tilde s_\mathrm{inv}^2\right)-\mathrm{sgn}(\tilde s_\mathrm{inv}-f_0)\exp\left(\alpha^2 (\tilde s_\mathrm{inv}-f_0)^2\right)}{\exp\left(\alpha^2 \tilde s_\mathrm{inv}^2\right)-\mathrm{sgn}(\tilde s_\mathrm{inv}-1)\exp\left(\alpha^2 (\tilde s_\mathrm{inv}-1)^2\right)}\\
    &= \frac{1-\mathrm{sgn}(\tilde s_\mathrm{inv}-f_0)\exp\left(-\alpha^2f_0(2\tilde s_\mathrm{inv}-f_0)\right)}{1-\mathrm{sgn}(\tilde s_\mathrm{inv}-1)\exp\left(-\alpha^2(2\tilde s_\mathrm{inv}-1)\right)}\,.
\end{align}
The behavior of the above expression is further divided into two regimes. When $\tilde s_\mathrm{inv}\ll 1/2$, the denominator is dominating and $p_\mathrm{fix}$ is 
still exponentially suppressed:
\begin{equation}
    p_\mathrm{fix}\sim \exp\left(-2N_\mathrm{eff}(s_\mathrm{inv}-\eta/2)\right)\,.
\end{equation}
When $\tilde s_\mathrm{inv} \gg 1/2$, the denominator becomes very close to $1$ and the
numerator becomes dominating instead. We then have
\begin{equation}
    p_\mathrm{fix}\simeq\frac{1 - \exp(-2N_\mathrm{eff}s_\mathrm{inv}f_0)}{1-\mathrm{sgn}(\tilde s_\mathrm{inv}-1)\exp(-2N_\mathrm{eff}s_\mathrm{inv})}\,,
\end{equation}
which matches Kimura's formula approximately for $\tilde s_\mathrm{inv}<1$ and exactly for
$\tilde s_\mathrm{inv}>1$.

In conclusion, while
$p_\mathrm{fix}$ for negative $s_\mathrm{inv}$ is always exponentially suppressed regardless
of ecology, the presence of a large $\eta$ now also exponentially suppresses
$p_\mathrm{fix}$ with positive $s_\mathrm{inv}$ up to
\begin{equation}
    s_\mathrm{inv}\simeq s_c:=\frac{\eta}{2}\,.
\end{equation}
We can then interpret $s_c$ as a new
threshold in $s$ for beneficial mutations that overcome the competitive 
ecological effects. In particular, the result for $s_\mathrm{inv}=0$ can now be
understood as
\begin{equation}
    p_\mathrm{fix}(s_\mathrm{inv}=0)\sim \exp(-2N_\mathrm{eff}s_c)\,,
\end{equation}
which is the same as Kimura's prediction but for $s=-s_c$.

\section{Fixation probabilities in consumer-resource models}
\label{sec:fixProbCRM}

\begin{figure*}[]
    \centering
    \includegraphics[width=0.8\linewidth]{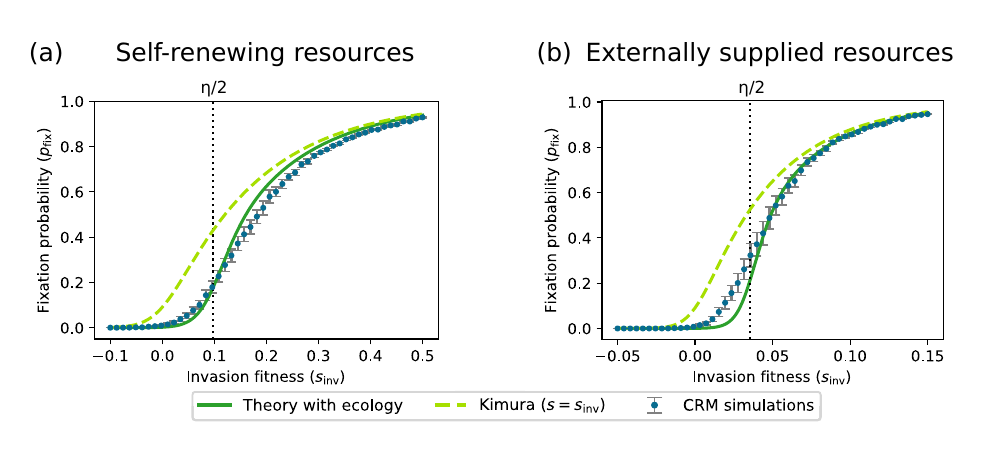}
    \caption{\justifying
\textbf{Theoretical and simulated fixation probabilities in consumer-resource models.}
We compute fixation probabilities $p_\mathrm{fix}$ at various invasion fitness $s_\mathrm{inv}$ by simulating mutations within consumer-resource models
with (a) self-renewing resources (MacArthur consumer-resource model) and
(b) externally supplied resources. In both cases, our theoretical predictions are more accurate than Kimura's predictions, but do not fully
fit with the simulation results.
Error bars denote standard errors from multiple instances of demographic noise and mutants.
}
    \label{fig:fixProbCRM}
\end{figure*}

\begin{figure*}[]
    \centering
    \includegraphics[width=0.4\linewidth]{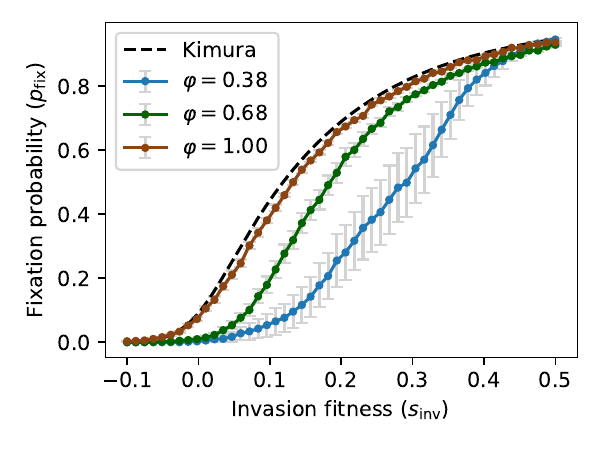}
    \caption{\justifying
\textbf{Ecological suppression of fixation probabilities controlled by species packing in consumer-resource models.}
We compute fixation probabilities $p_\mathrm{fix}$ at various invasion fitness 
$s_\mathrm{inv}$ by simulating mutations within MacArthur consumer-resource models with different packing fractions $\varphi=S^*/M^*$. The ecological 
suppression of fixation probabilities is stronger when the packing fraction is lower, and there is nearly no suppression when the community is fully packed.
Error bars denote standard errors from multiple instances of demographic noise and mutants.\kobe{new figure including fully packed case}
}
    \label{fig:fixProbCRMPacking}
\end{figure*}

In the main text, we have compared our predictions on $p_\mathrm{fix}$ with
the simulation results from generalized Lotka-Volterra model. The comparison
can similarly be done for various consumer-resource models as in Figure 
\ref{fig:fixProbCRM}. We find that our predictions are in general more
accurate than Kimura's predictions. 
On the other hand, there are some deviations for $0<s_\mathrm{inv}<\eta$, where the parent and the mutant can transiently coexist, hence
the deterministic ecological dynamics affect the fixation probabilities more
significantly. Compared to generalized Lotka-Volterra model, here
the species interactions are mediated through resource dynamics. For self-renewing
resources, there are additional complications since resources can also go extinct. 
We suspect that these additional complexities cause our DMFT approximations
in Sec. \ref{sec:CRM} and \ref{sec:gCRM} to be less accurate.

We have also demonstrated in the main text that the ecological suppression
of fixation probabilities is affected by species packing. While the packing
is related to May's stability bound for generalized Lotka-Volterra model,
for consumer-resource models the packing is related to the competitive
exclusion principle. The packing fraction is given by $S^*/M^*$, where $S^*$
is the number of surviving species and $M^*$ is the number of non-depleted
resources. For externally supplied resources we have $M^*=M$. The packing
bound is $S^*/M^*\leq 1$ for self-renewing resources and interestingly 
$S^*/M^*\leq 1/2$ for externally supplied resources \cite{cui2020effect}. In Fig. \ref{fig:fixProbCRMPacking}, we indeed see that the fixation probabilities
are more suppressed for a less packed community with lower packing fraction. \kobe{new comments for Ben} In the other direction, we also see that the fixation probabilities nearly reduce to the results of Kimura when the community is fully packed. This observation apparently contradicts \cite{mcenany2024predicting}, which observed a sudden increase in certain coexistence probabilities near full packing in simulations (Fig. 3b in Ref.~\cite{mcenany2024predicting}). This might arise from minor differences in modeling assumptions which result in communities behaving differently when their niches are fully packed.  These differences may affect the behavior of species and resource susceptibilities, with concomitant effects on invasion probabilities. An additional reason that the difference in behaviors between our models may arise is due to differences in how parameters are sampled, e.g., in Fig.~\ref{fig:fixProbCRMPacking}, we have  imposed specific values of $s_\mathrm{inv}$, in contrast with Ref.~\cite{mcenany2024predicting}.

Note also that the numbers of surviving species in Fig.~\ref{fig:fixProbCRMPacking} are non-monotonic, namely $15,26,22$ from low to high packing, showing that the number of surviving species is indeed not a good predictor for the ecological suppression, because of differences in the number of surviving resources $M^\ast$. The same non-monotonicity is also present in the amount of biomass. \kobe{comment for reviewer 2}

In conclusion, we see that the results for different consumer-resource models
are qualitatively the same as in generalized Lotka-Volterra model. Such 
agreement suggests that our theory is indeed applicable to various ecological models.

\section{Effects of parent-mutant correlation on fixation probabilities}
\label{sec:parentMutantCorrelation}

\begin{figure}
    \centering
    \includegraphics[width=0.8\linewidth]{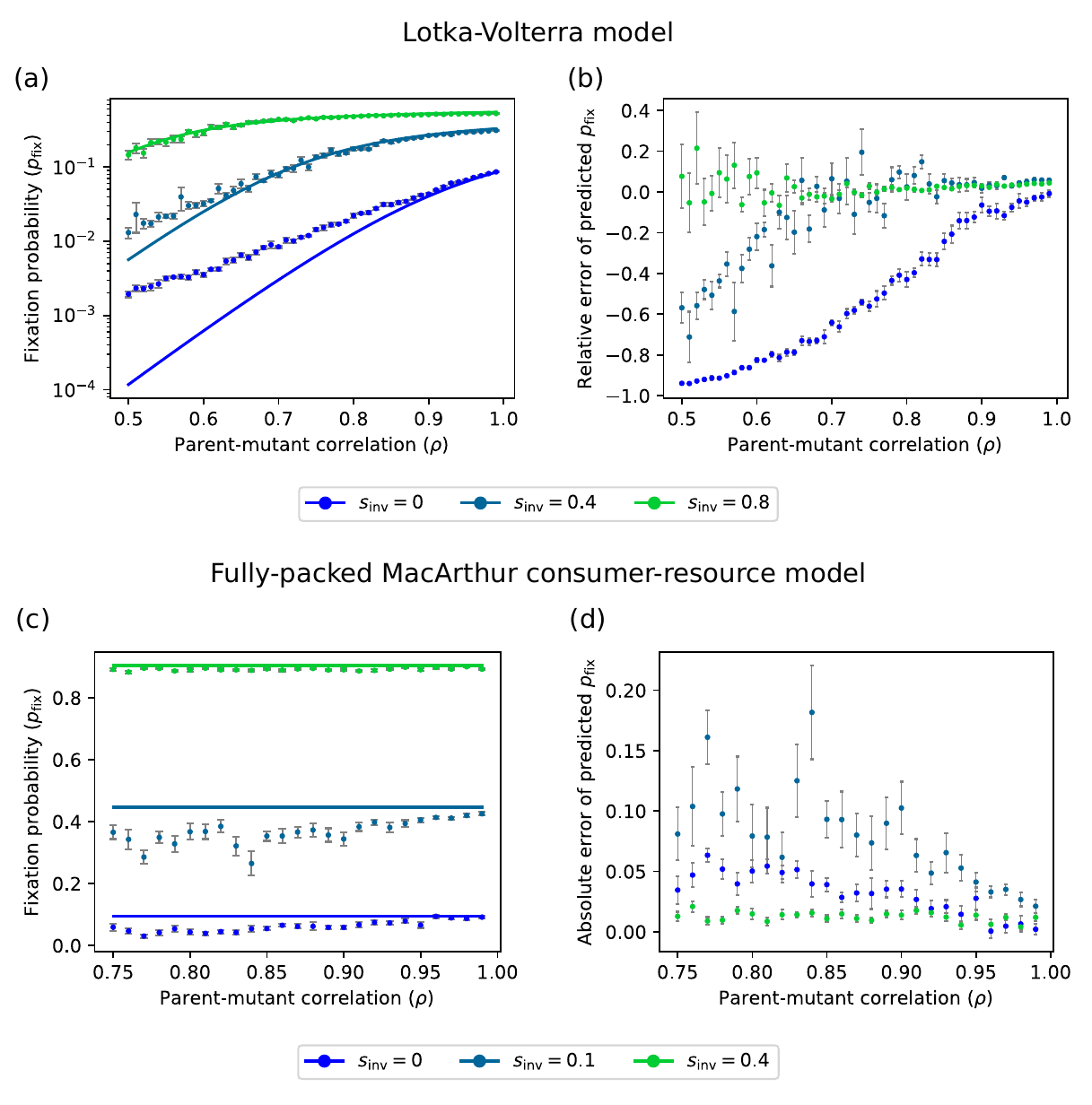}
    \caption{\justifying\textbf{Theoretical and simulated fixation probabilities in generalized Lotka-Volterra model and MacArthur consumer-resource model at various parent-mutant correlation.} 
    We compute fixation probabilities $p_\mathrm{fix}$ at various invasion fitness $s_\mathrm{inv}$ by simulating mutations within generalized Lotka-Volterra model (a--b) and fully-packed MacArthur consumer-resource model (c--d). Our theory (solid curves) matches the simulations (dots) for high $\rho$, but deviates from the simulations for low $\rho$ (a, c). The error is more significant for lower $s_\mathrm{inv}$ (b, d). Error bars denote standard errors from multiple instances of demographic noise and mutants.\kobe{new figure as requested by Ben}}
    \label{fig:fig_rho_2}
\end{figure}

\kobe{simplified this appendix since most of the stuff are in the GLV appendix now}

Throughout the paper, we have assumed high parent-mutant correlation $\rho\approx 1$. As shown in Appendix \ref{sec:gLVderivation}, this is one of the crucial assumptions that allows us to simplify the
mutation dynamics into the form in Eq. \eqref{eq:frequencyDynamicsSI}. Apart from the justifications made in Appendix \ref{sec:gLVderivation}, here we also validate the small $1-\rho$ expansion using the downstream quantities, namely the fixation probabilities. In Fig.~\ref{fig:fig_rho_2}, we see that there are indeed ranges of small $1-\rho$ in which our approximated theory in Eq.~\eqref{eq:fixationProbSI} agrees well with full simulations. That is, the fixation probabilities decrease with smaller $\rho$ due to frequency-dependent selection. In particular, both theory and simulations reduce back to Kimura's results as $\rho$ approaches $1$. The agreement remains true even for fully-packed communities (Fig.~\ref{fig:fig_rho_2}c, d) in which nonlinear effects become important (see the end of Appendix \ref{sec:CRM}). We also see that our theory starts to deviate from the simulations when $\rho$ is no longer large, though the specific value of $\rho$ at which the deviation rises depends on other properties of the community.


\section{Comparison between selection and drift dynamics}
\label{sec:selectionDriftRatio}

In this appendix, we obtain a qualitative understanding of both the results
in classic population genetics and our new results, by comparing the 
relative sizes of selection and drift terms in the differential equation 
for $f$. This analysis helps us understand how parent-mutant coexistence
arises in the presence of ecological effects.

\subsection{Without ecological effects}

For pedagogical purposes, it is helpful to start by re-deriving some classic results from classical population genetics with constant selection (in the absence of many-species ecology).
As explained in the main text, a mutant starts to invade at a very 
small frequency where genetic drift dominates. The mutant can either 
become extinct or grow to a larger frequency due to drift.
This process continues until the surviving mutants reach a frequency 
$f\sim 1/N_\mathrm{eff}|s|$ , where selection becomes dominant instead. This critical frequency is the drift
threshold as shown in Fig. 4a in the main text.

We now obtain a heuristic analytical estimate of drift threshold, following the approach
in \cite{fisher2007course,good2016molecular}. Recall that the dynamics of 
$f$ in classic population genetics is given by
\begin{equation}
    \frac{df}{dt}=f(1-f)s+\sqrt{\frac{f(1-f)}{N_\mathrm{eff}}}\xi(t)\,.
\end{equation}
The exact solution to this differential equation is in general 
complicated as both the growth rate and diffusion rate depends on $f$.
On the other hand, analogous to the Euler method for numerically solving differential equations,
we can approximate the dynamics of $f$ by splitting it into infinitesimal
time intervals with width $\Delta t$. Within a time interval around some time $t$, the changes
in both the growth rate and the diffusion rate can be neglected. Therefore,
the change in frequency $\Delta f$ within the time interval can be easily solved and is given by
\begin{equation}
    \Delta f:=f(t+\Delta t)-f(t)\simeq f(t)(1-f(t))s\Delta t+\sqrt{\frac{f(t)(1-f(t))\Delta t}{N_\mathrm{eff}}}z\,,
\end{equation}
where $z$ is a zero mean, unit variance Gaussian random variable. Below,
we can simply use $f$ to denote $f(t)$.

Let us focus on small frequencies $f\ll1$ where drift dominates over
selection. In this case, $\Delta f$ can be either positive or negative due to randomness, and 
its magnitude can be estimated by the variance
\begin{equation}
    |\Delta f|\sim \sqrt{\frac{f(1-f)\Delta t}{N_\mathrm{eff}}}\,.
\end{equation}
The above approximation of constant growth and diffusion rates remains valid until $\Delta f$ is comparable to $f$, that is
\begin{equation}
\label{eq:goodApproximation}
    |\Delta f|\sim \sqrt{\frac{f(1-f)\Delta t}{N_\mathrm{eff}}}\sim f\Rightarrow \Delta t\sim \frac{N_\mathrm{eff} f}{1-f}\,.
\end{equation}
Within this time interval, the contribution to $\Delta f$ from selection is
\begin{equation}
    |\Delta f|_\mathrm{sel}\sim f(1-f)|s|\Delta t\sim |s|N_\mathrm{eff}f^2\,.
\end{equation}
Note that the absolute value is needed since $\Delta f$ due to selection 
can also be either positive or negative depending on the sign of $s$.
To ensure that the approximation is self-consistent, $\Delta f$ due to 
selection must be negligible, i.e., $|\Delta f|_\mathrm{sel}\ll|\Delta f|$. Therefore, we require
\begin{equation}
    |s|N_\mathrm{eff}f^2\ll f\Rightarrow f\ll \frac{1}{N_\mathrm{eff}|s|}\,.
\end{equation}
Indeed, we see that drift dominates the dynamics when $f$ is below drift
threshold. 

\subsection{With ecological effects}


We can repeat the above analysis under the presence of ecological effects.
As explained in the main text, when $0\lesssim s_\mathrm{inv}\lesssim \eta$, there is
an additional deterministic fixed point $f^*\simeq \tilde s_\mathrm{inv}$. We then expect
a new region around both sides of $f^*$ where drift dominates the dynamics, apart from the one near $f=0$. This is because at every fixed point, by definition, the deterministic contribution to the dynamics vanishes.

Suppose the mutant reaches frequency $f\gg f_0$ at time $t$. Again, we focus
on an infinitesimal time interval with width $\Delta t$.
From Eq. \eqref{eq:frequencyDynamicsSI}, we can approximate $\Delta f$ 
within this time interval as
\begin{equation}
    \Delta f\simeq f(1-f)(s_\mathrm{inv}-\eta f)\Delta t+\sqrt{\frac{f(1-f)\Delta t}{N_\mathrm{eff}}}z\,,
\end{equation}
where $z$ is a zero mean, unit variance Gaussian random variable. Note that
the variance is the same as in the case without ecological effects.
Therefore, if drift dominates the dynamics, the above expression
of $\Delta f$ remains a good approximation until
\begin{equation}
    \Delta t\sim \frac{N_\mathrm{eff}f}{1-f}\,,
\end{equation}
which is the same as Eq. \eqref{eq:goodApproximation}.
To ensure that this approximation is self-consistent, the contribution to $\Delta f$ from selection must be negligible. Therefore, we require
\begin{equation}
    \left|f(1-f)(s_\mathrm{inv}-\eta f)\Delta t\right|\sim |s_\mathrm{inv}-\eta f|N_\mathrm{eff} f^2\ll f\,.
\end{equation}
Note again that the absolute value is needed since frequency can either increase or decrease, i.e., $\Delta f$ due to 
selection can be either positive or negative, particularly when $f$ is 
around the new fixed point $f^*\simeq \tilde s_\mathrm{inv}$, which is between $f=0$ and $f=1$.
Based on the above condition, we can define the selection-drift ratio
\begin{equation}
\label{eq:selectionDriftRatio}
    R=|s_\mathrm{inv}-\eta f|N_\mathrm{eff}f\,,
\end{equation}
such that drift (selection) dominates the dynamics when $R<1$ ($R>1$), 
and $R=1$ is the crossover boundary between selection and drift
dominated regimes. This is the ratio as shown in Fig. 3f-h in the main text.


\begin{figure*}[]
    \centering
    \includegraphics[width=1\linewidth]{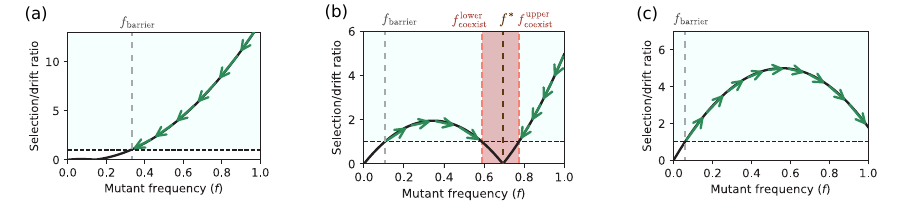}
    \caption{\justifying
\textbf{Phase portraits of frequency dynamics in terms of selection-drift ratio.}
The selection-drift ratios $R$ (Eq. \eqref{eq:selectionDriftRatio}) are calculated in different regimes of invasion fitness $s_\mathrm{inv}$, namely (a) $s_\mathrm{inv}<\sqrt{4\eta/N_\mathrm{eff}}$, (b) $\sqrt{4\eta/N_\mathrm{eff}}<s_\mathrm{inv}<\eta$, and (c) $s_\mathrm{inv}>\eta$. When $s_\mathrm{inv}$ is in the intermediate range, there are more crossovers between selection and drift dominated regimes, hence a coexistence region centered at $f^*$ emerges. The selection-drift ratios are identical to the ones in Fig. 3 in the main text.
}
    \label{fig:phasePortrait}
\end{figure*}

Since we already know that our results reduce back to those in population 
genetics when $\alpha\ll 1$ ($\eta\approx0)$, we now focus on the case of large $\eta$ and 
$\alpha\gg 1$. In this regime, due to the absolute value in $R$, 
the solutions to $R=1$ have different types of behavior depending on the value
of $s$. Recall that we require $0\leq f\leq 1$. When $s_\mathrm{inv}<\sqrt{4\eta/N_\mathrm{eff}}$ (or $\tilde s_\mathrm{inv}<2/\alpha$), there is only one solution with $s_\mathrm{inv}-\eta f<0$: (Fig. \ref{fig:phasePortrait}a)
\begin{equation}
\label{eq:barrierSmallS}
    f_\mathrm{barrier}\sim \frac{2}{N_\mathrm{eff}|s_\mathrm{inv}|}\left(\sqrt{1+\frac{4\eta}{N_\mathrm{eff}s_\mathrm{inv}^2}}-\mathrm{sgn}(s_\mathrm{inv})\right)^{-1}\,.
\end{equation}
When $s_\mathrm{inv}>\eta$, there is also only one solution but with $s_\mathrm{inv}-\eta f>0$: (Fig. \ref{fig:phasePortrait}c)
\begin{equation}
\label{eq:barrierLargeS}
    f_\mathrm{barrier}\sim \frac{2}{N_\mathrm{eff}s_\mathrm{inv}}\left(\sqrt{1-\frac{4\eta}{N_\mathrm{eff}s_\mathrm{inv}^2}}+1\right)^{-1}\,.
\end{equation}
Therefore, the dynamics of $f$ is qualitatively the same as the one in classic
population genetics for $s_\mathrm{inv}<\sqrt{4\eta/N_\mathrm{eff}}$ or $s_\mathrm{inv}>\eta$, with
$f_\mathrm{barrier}$ being the drift threshold.

In contrast, the dynamics of $f$ is qualitatively distinct for $\sqrt{4\eta/N_\mathrm{eff}}<s_\mathrm{inv}\lesssim\eta$. We see that there are three solutions to $R=1$: (Fig. \ref{fig:phasePortrait}b)
\begin{align}
    f_\mathrm{barrier}&=\frac{2}{N_\mathrm{eff}s_\mathrm{inv}}\left(1+\sqrt{1-\frac{4\eta}{N_\mathrm{eff}s_\mathrm{inv}^2}}\right)^{-1},\\
    f_\mathrm{coexist}^\mathrm{lower}&=\frac{2}{N_\mathrm{eff}s_\mathrm{inv}}\left(1-\sqrt{1-\frac{4\eta}{N_\mathrm{eff}s_\mathrm{inv}^2}}\right)^{-1},\\
    f_\mathrm{coexist}^\mathrm{upper}&=\frac{2}{N_\mathrm{eff}s_\mathrm{inv}}\left(\sqrt{1+\frac{4\eta}{N_\mathrm{eff}s_\mathrm{inv}^2}}-1\right)^{-1}\,,
\end{align}
with $f_\mathrm{barrier}<f_\mathrm{coexist}^\mathrm{lower}<f_\mathrm{coexist}^\mathrm{upper}$. Note that we have $f_\mathrm{coexist}^\mathrm{lower}\simeq f_\mathrm{coexist}^\mathrm{upper}\simeq f^*\simeq \tilde s_\mathrm{inv}$,
hence these two frequencies correspond to the bounds of the coexistence region centered at $f^*$.

As demonstrated in the main text, our results in fixation probability and mean absorption/extinction time can be qualitatively
understood using the nonlinear behavior of $R$ (Fig. \ref{fig:phasePortrait}b). Suppose a mutant starts with 
$f=f_0\ll 1$ at the beginning. It must overcome the first barrier at  $f=f_\mathrm{barrier}$ or it
becomes extinct. After that, selection brings the mutant to the
region $f_\mathrm{coexist}^\mathrm{lower}<f<f_\mathrm{coexist}^\mathrm{upper}$, which is the coexistence region. Note that the 
deterministic fixed point $f^*\simeq \tilde s_\mathrm{inv}$ is within this region. Since 
$\left<df/dt\right>>0$ when $f<f_\mathrm{coexist}^\mathrm{lower}$ and $\left<df/dt\right><0$ when $f>f_\mathrm{coexist}^\mathrm{upper}$, 
both boundaries of the coexistence region are reflecting and the mutant is trapped 
inside this region. It can escape this region and reach near the absorbing
boundaries $f=0,1$
only when the demographic noise reaches the tail of its distribution with 
exponentially low probability.

Now we see that if $s_\mathrm{inv}<s_c= \eta/2$, the 
coexistence region is closer to $f=0$ and it is exponentially more likely for the
demographic noise to drive the mutant to extinction. Therefore, $p_\mathrm{fix}$
is exponentially suppressed for $s_\mathrm{inv}<s_c$. In contrast, when $s_\mathrm{inv}>s_c$, the 
coexistence region is closer to $f=1$ and it is exponentially more likely for the
demographic noise to drive the mutant to fixation. The mutant
becomes extinct mostly because it cannot overcome the first barrier at $f=f_\mathrm{barrier}$, hence 
$p_\mathrm{fix}$ agrees with Kimura's formula for $s_\mathrm{inv}>s_c$.

The above argument also explains the behavior of mean absorption/extinction time. It grows
and decays exponentially in the region $\sqrt{4\eta/N_\mathrm{eff}}<s_\mathrm{inv}\lesssim\eta$ due to
the required time to escape the coexistence region. More precisely, the time grows 
exponentially as $s_\mathrm{inv}$ increases and the coexistence region becomes farther from $f=0$.
The growth continues till $s_\mathrm{inv}=s_c$. For $s_\mathrm{inv}>s_c$, the coexistence region becomes
closer to fixation, and the mean absorption time decays exponentially as 
fixation becomes more likely. For the mean extinction time, although the 
time for escaping from
the coexistence region to $f=0$ is even longer, it also becomes exponentially unlikely
that the mutant becomes extinct after reaching the coexistence region. 
Namely, a mutant becomes extinct mainly due to the first barrier at $f=f_\mathrm{barrier}$.
As a result,
the mean extinction time reaches maximum at $s_\mathrm{inv}=s_c$ and decays exponentially for $s_\mathrm{inv}>s_c$.


\section{Mean extinction time}
\label{sec:extinctionTime}

Using the above qualitative picture of coexistence region, we can derive analytical
estimates for the mean extinction time of the mutant. It is also similar to the 
establishment time in population genetics, which is the time when selection dominates over drift and drives a mutant with positive $s$
to fixation. As explained in the main text, the extinction or establishment
time is less sensitive to model details and is easier for analytical estimation.

For the regimes without the coexistence region, i.e. $s_\mathrm{inv}<\sqrt{4\eta/N_\mathrm{eff}}$ or $s_\mathrm{inv}>\eta$,
the extinction time can be approximated by the time required to reach the barrier at 
$f=f_{R=1}$; see Eqs. \eqref{eq:barrierSmallS} and \eqref{eq:barrierLargeS}. For 
$s_\mathrm{inv}<\sqrt{4\eta/N_\mathrm{eff}}$, the barrier is reflecting and the mutant becomes extinct in
the same order of magnitude of time after reaching the barrier. For $s_\mathrm{inv}>\eta$, the 
barrier is absorbing and the mutant can no longer become extinct after passing through
the barrier. Now, the time for reaching the barrier can be estimated using Eq. 
\eqref{eq:goodApproximation}; since $\Delta t$ increases with $f$, the required time to
reach $f=f_{R=1}$ is dominated by $\Delta t(f_{R=1})$. Therefore, for 
$s_\mathrm{inv}<\sqrt{4\eta/N_\mathrm{eff}}$, the mean extinction time $T$ is
\begin{equation}
    T\sim \frac{2}{|s_\mathrm{inv}|}\left(\sqrt{1+\frac{4\eta}{N_\mathrm{eff}s_\mathrm{inv}^2}}-\mathrm{sgn}(s_\mathrm{inv})\right)^{-1}\,,
\end{equation}
and for $s_\mathrm{inv}>\eta$,
\begin{equation}
    T\sim \frac{2}{s_\mathrm{inv}}\left(\sqrt{1-\frac{4\eta}{N_\mathrm{eff}s_\mathrm{inv}^2}}+1\right)^{-1}\,.
\end{equation}
Both cases match the expectation from population genetics that $T\sim 1/|s_\mathrm{inv}|$ 
asymptotically.

On the other hand, the mean extinction time deviates from the expectation from
population genetics when the coexistence region is present, i.e. when 
$\sqrt{4\eta/N_\mathrm{eff}}<s_\mathrm{inv}<\eta$. In this regime, the mean extinction time is dominated by the
exponentially long time to escape the coexistence region driven by the stochastic 
dynamics. The scale of the escape time can be estimated by translating the stochastic
dynamics into 1D diffusion in the presence of a potential $V(f)$. Since $f$ relaxes to
near equilibrium around the deterministic fixed point $f^*\simeq \tilde s_\mathrm{inv}$ in the long term 
(but before being absorbed at $f=0,1$), the probability distribution $p(f)$ can be
obtained from the Fokker-Planck equation:
\begin{align}
\label{eq:FokkerPlanck}
    \frac{\partial p}{\partial t}&=-\frac{\partial}{\partial f}\left(f(1-f)(s_\mathrm{inv}-\eta f)p\right)+\frac{\partial^2}{\partial^2 f}\left(\frac{f(1-f)}{2N_\mathrm{eff}}p\right)\\
    &=\frac{\partial}{\partial f}\left(\frac{dV}{df}f(1-f)p\right)+\frac{\partial^2}{\partial^2 f}\left(k_BT_\mathrm{eff}f(1-f)p\right)=0\,,
\end{align}
where we have used the language of statistical physics and defined a potential for $f$ (with minimum at $f=f^*$)
\begin{equation}
    V(f)=\frac{1}{2}\eta f^2-s_\mathrm{inv}f\,,
\end{equation}
and an effective temperature
\begin{equation}
    k_BT_\mathrm{eff}=\frac{1}{2N_\mathrm{eff}}\propto D\,.
\end{equation}
As long as $f$ is not absorbed at $0$ or $1$, the solution involves a Boltzmann distribution \cite{hanggi1990reaction}
\begin{equation}
    p(f)\propto \frac{e^{-V(f)/k_BT_\mathrm{eff}}}{f(1-f)}\,,
\end{equation}
For large $\eta$ with $\alpha\gg 1$, we can focus on the exponential factor and 
approximate the escape time as 
$e^{\Delta V/k_BT_\mathrm{eff}}$ according to Kramer's 
formula, where $\Delta V$ is the barrier height in the potential. In particular, the
escape time to extinction is
\begin{equation}
    T_0\sim \exp \left(\frac{V(0)-V(\tilde s_\mathrm{inv})}{k_BT_\mathrm{eff}}\right)=\exp\left(\alpha^2\tilde s_\mathrm{inv}^2\right)\,.
\end{equation}

\kobe{clarified the argument here since reviewer 2 got confused}
However, $T_0$ is conditional on that the mutant reaches near equilibrium around $f^*$ and does not eventually escape to fixation. When $s_\mathrm{inv}>s_c=\eta/2$, this is no longer the main route to extinction since the coexisting region is closer to fixation as in Fig.~\ref{fig:phasePortrait}b. Instead, almost all extinctions occur below the drift threshold, i.e., $f<f_\mathrm{barrier}$ in Fig.~\ref{fig:phasePortrait}, with much shorter extinction time $T_b\sim 1/s_\mathrm{inv}$ according to population genetics. As explained in the main text, the probability for a mutant to pass through the drift threshold is precisely Kimura's fixation probability
\begin{equation}
    p_\mathrm{Kimura}\simeq 1-\exp(-\alpha^2\tilde s_\mathrm{inv}f_0)\,,
\end{equation}
for large $\alpha$ and $s_\mathrm{inv}>s_c$. Note that the exponent is much smaller than the one in $T_0$ since $f_0$ is small. Meanwhile, conditional on passing through the drift threshold, the probability for a mutant to reach $f=0$ instead of $f=1$ is
\begin{equation}
    p_0\sim\frac{p(0)}{p(0)+p(1)}\sim \begin{cases}
        1&s<s_c\\
        \frac{p(0)}{p(1)}=\exp\left(\alpha^2(1-\tilde s_\mathrm{inv})^2-\alpha^2\tilde s_\mathrm{inv}^2\right)&s>s_c\,.
    \end{cases}
\end{equation}
Therefore, the mean extinction time is
\begin{equation}
    T\simeq \frac{(1-p_\mathrm{Kimura})T_b+p_\mathrm{Kimura}p_0T_0}{1-p_\mathrm{Kimura}+p_\mathrm{Kimura}p_0}\simeq \frac{p_\mathrm{Kimura}p_0T_0}{1-p_\mathrm{Kimura}}\,,
\end{equation}
since $p_0T_0$ is still exponentially larger than $T_b$, while $1-p_\mathrm{Kimura}$ is less suppressed than $p_0$. Taking the leading exponents, we arrive at
\begin{equation}
    T\sim p_0T_0\sim\begin{cases}
        \exp\left(\alpha^2\tilde s_\mathrm{inv}^2\right)&s_\mathrm{inv}<s_c\\
        \exp\left(\alpha^2(1-\tilde s_\mathrm{inv})^2\right)&s_\mathrm{inv}>s_c\,.
    \end{cases}
\end{equation}
In particular, the mean extinction time reaches maximum when $s_\mathrm{inv}=s_c\simeq \eta/2$, i.e.
\begin{equation}
    T_\mathrm{max}\sim \exp(\alpha^2/4)\,,
\end{equation}
which is determined by $\alpha$ only. \kobe{addressing timescale problem}Importantly, $T$ is still exponentially suppressed by the exponent of $1-\rho$ compared to the stochastic extinction time of a whole species. Therefore, we can indeed observe the stochastic extinction or fixation of mutants well before the community collapses from its steady state.

\kobe{added prefactor for the discussion point}
Apart from the exponential behavior of $T$, it is useful to estimate its prefactor, which has the unit of time. In the ordinary Kramer's theory, where both the potential well (let say at $f=f_a$) and the barrier (at $f=f_b$) contain saddle points, the prefactor is roughly
\begin{equation}
    D^{-1}\left(\frac{V''(f_a)}{k_B T_\mathrm{eff}}\right)^{-1/2}\left(\frac{V''(f_b)}{k_B T_\mathrm{eff}}\right)^{-1/2}\,,
\end{equation}
where the second derivative factors come from the normalization of the Boltzmann distribution.
However, in our case the barrier contains an absorbing boundary at $f=0$ instead of a saddle point. Its second derivative factor is then replaced by a first derivative factor \cite{bender1999advanced}, hence the prefactor of $T$ is
\begin{equation}
    D^{-1}\left(\frac{V''(\tilde s_\mathrm{inv})}{k_B T_\mathrm{eff}}\right)^{-1/2}\left(\frac{V'(0)}{k_B T_\mathrm{eff}}\right)^{-1}\simeq \frac{1}{\eta\alpha}\,,
\end{equation}
where we have used the fact that $s_\mathrm{inv}\sim\eta$ for parent-mutant coexistence. Therefore,
\begin{equation}
    T_\mathrm{max}\simeq \frac{\exp(\alpha^2/4)}{\eta\alpha}\,.
\end{equation}
This expression can be directly compared to the typical time for a new mutation $T_\mathrm{mut}\simeq 1/N_\mathrm{ind}U$, where $N_\mathrm{ind}$ is the number of individuals of a species and $U$ is the mutation rate of an individual. The ``strong-selection-weak-mutation'' limit holds when $T_\mathrm{max}<T_\mathrm{mut}$.

\begin{figure*}[]
    \centering
    \includegraphics[width=0.4\linewidth]{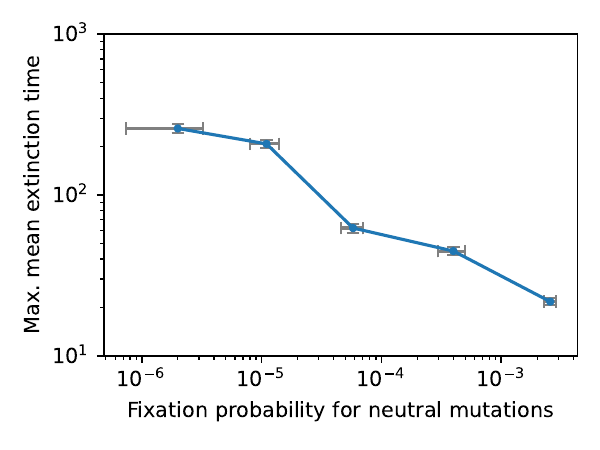}
    \caption{\justifying
\textbf{Correlation between fixation probabilities and mean extinction times.}
In log scales, there is negative correlation between the fixation 
probability for mutations with $s_\mathrm{inv}=0$ and the maximum mean extinction time.
Error bars denote standard errors from multiple instances of demographic noise and mutants.
}
    \label{fig:extinctionTimeVSFixProb}
\end{figure*}

It is interesting to observe that there are two independent quantities
that are related to $e^{\alpha^2}$,
namely the fixation probability for mutations with $s_\mathrm{inv}=0$ and the maximum mean
extinction time. In particular, we expect that the maximum time is 
exponentially longer as the fixation probability is exponentially suppressed,
since both exponential behavior is caused by parent-mutant coexistence as
explained in the main text. Fig. \ref{fig:extinctionTimeVSFixProb} shows
that the expectation is indeed realized in simulations.

\section{Details of simulation methods}
\label{sec:simulationMethod}

We use simulations to compute the fixation probabilities and other 
statistical quantities within the full ecological models. The code for the below 
simulation and the associated figures in this paper can be found on
\url{https://github.com/Emergent-Behaviors-in-Biology/Pop-Gen-Ecology}.

\subsection{Community dynamics}

First, we sample
the ecological properties of the community as
described in Sec. \ref{sec:gLVderivation} and \ref{sec:CRM}. We then evolve
the community towards its unique steady state 
\emph{without demographic noise} by solving the full
differential equations using the \textsc{LSODA} method (with absolute and relative
tolerances $10^{-12}$). The convergence is reached when all the derivatives
$dN_i/dt,dR_\alpha/dt$ are less than $10^{-10}$. We can then use these
results to compute the predicted values of $\eta$.

After that, since we only focus on a single step of mutation, we choose one of the 
surviving species as the parent. Then, we look at multiple instances of the mutation with 
the same parent.
Each time, we introduce a mutant associated with the chosen
parent, sampled as described in Sec. \ref{sec:gLVderivation} and \ref{sec:CRM}.
We then solve the full differential equations \emph{with demographic noise} using a generalized Euler method as explained below.
Due to the multiplicative nature of the noise,
the simulation can become numerically unstable if we sample the noise directly. Instead,
we make use of the result in \cite{dornic2005integration}. Within a time step $dt$, we can approximate that the
total growth rates of the species remain unchanged, since the growth rates involve 
averages over all the species or resource abundances. In other words, the species
dynamics within time $dt$ becomes
\begin{equation}
\label{eq:dtDynamics}
    \frac{dN_i}{dt}=N_ig_i+\sqrt{2DN_i}\xi_i(t)\,,
\end{equation}
where $g_i$ is the momentarily constant growth rate depending on the 
current species or resource abundances. It was noticed that the 
above equation can be
solved exactly, and the solution is given by
\begin{equation}
    N_i(t+dt)\sim \frac{D(e^{g_idt}-1)}{g_i}\cdot\mathrm{Gamma}\left(\mathrm{Poisson}\left(\frac{N_i(t)g_i}{D(1-e^{-g_idt})}\right)\right)\,,
\end{equation}
where $\mathrm{Gamma}(\alpha)$ is the gamma distribution with shape parameter $\alpha$
and unit scale. We define $\mathrm{Gamma}(0)$ as the Dirac delta distribution at $0$.
For each time $t$, we sample $N_i(t+dt)$ according to the above distribution. If 
resources are involved in the model, their abundances are updated at the same time 
using the ordinary Euler method. After that, to ensure uninvadability, we impose a hard wall for
the abundances at $\lambda=10^{-7}$, i.e.
\begin{equation}
    N_i(t+dt)\rightarrow \max(\lambda,N_i(t+dt))\,,
\end{equation}
and similarly for resources if present.
We iterate the above process with a time step $dt=0.1$ till the parent or mutant abundance reaches $\lambda$, which is treated as extinction.

The parent and mutant trajectories in classic population genetics are 
simulated in the same way except that $g_i$ are now truly constants and
the fitness difference is $s=g_m-g_p$.

\subsection{Susceptibility}
\kobe{new subsection for computation of susceptibility}

Recall the definition of susceptibility in the GLV in Eq.~\eqref{eq:susceptibilityDef}. Following \cite{glynn1990likelihood,warren2012malliavin}, we use the likelihood-ratio estimation method to numerically compute the susceptibility function; here we provide the full derivation of the method for completeness.\kobe{added citations} Due to the stochastic nature of our simulation, we need not perturb $r_i$ and rerun the noisy simulation to calculate the derivatives. This is because the rerun would result in the same possible set of trajectories but weighted by a different probability measure based on $\delta r$. Therefore, we instead apply a change of measure to the original unperturbed trajectories:
\begin{equation}
    \left<N_i(t)\right>_{\delta r}=\left<N_i(t)\frac{dP_{\delta r}}{dP_0}\right>_0\,,
\end{equation}
where the subscript $0$ represents the unperturbed measure. By the Girsanov's theorem,\kobe{changed to Girsanov}
\begin{equation}
    \frac{dP_{\delta r}}{dP_0}=\exp\left[\sum_i \int dt'\,b_i\xi_i(t')-\frac{1}{2}\sum_i\int dt'\,b_i^2\right]\,,
\end{equation}
where
\begin{equation}
    b_i=\frac{N_i \delta r_i}{\sqrt{2DN_i}}\,,
\end{equation}
is the ratio between the deterministic perturbation $N_i\delta r_i$ and the stochastic standard deviation $\sqrt{2DN_i}$. The susceptibility can then be obtained by taking the derivative on $\delta r$.
\begin{equation}
    \nu_{ij}(t,t')=\left.\frac{\partial \left<N_i(t)\right>_{\delta r}}{\partial (\delta r_j(t'))}\right|_{\delta r=0}=\left<N_i(t)\sqrt{\frac{N_j(t')}{2D}}\xi_j(t')\right>_0\,.
\end{equation}
In the simulation, we reconstruct $\xi_j(t')$ using Eq.~\eqref{eq:dtDynamics}:
\begin{equation}
    \xi_j(t')\simeq \frac{1}{\sqrt{2DN_j(t')}}\left(\frac{N_j(t'+dt)-N_j(t')}{dt}-N_j(t')g_j(t')\right)\,.
\end{equation}
In conclusion, we can numerically compute the susceptibility using the formula
\begin{equation}
    \nu_{ij}(t,t')\simeq \frac{1}{2D}\left<N_i(t)\left(\frac{N_j(t'+dt)-N_j(t')}{dt}-N_j(t')g_j(t')\right)\right>_0\,,
\end{equation}
applied on the unperturbed trajectories.

\subsection{Fixation probabilities and times}

To obtain fixation probabilities, we sample $10$ mutants with the same chosen
parent. For each mutant, we adjust the value of $r_m$ (for generalized 
Lotka-Volterra model) or $m_m$ (for consumer-resource models) so that the 
invasion fitness is at a given value of $s_\mathrm{inv}$. We then run the above 
simulation $10^3$ times with different instances of demographic noise 
and count the number of fixations. We take the average of the above results 
to find the fixation probability $p_\mathrm{fix}(s_\mathrm{inv})$.

To obtain absoprtion and extinction times, we sample only one mutant for
the chosen parent, then run the simulation $10^3$ times and record the time
to fixation or extinction for each run.

\section{Simulation parameters used in the figures}
\label{sec:simulationParameters}

\kobe{updated for new figures}

Throughout all simulation figures, the initial mutant frequency is $f_0=0.09$. We set $A_{pm}=A_{mp}=\rho$ for all simulations of generalized Lotka-Volterra models.

In Fig. \ref{fig:dynamics}(f) and (h), we simulate the dynamics as in 
classic population genetics with the parent growth rate $g_p=0.1$, the 
parent abundance $N_p=5.0$, the fitness difference $s=0.01$, and the diffusion
coefficient $D=0.5$. In Fig. \ref{fig:dynamics}(g) and (i), we simulate
generalized Lotka-Volterra model with parameters
\begin{equation}
    S=50,\mu=2.0,\sigma=0.4,\mu_r=6.0,\sigma_r=0.6,\gamma=1,\rho=0.99,s_\mathrm{inv}=0.2,N_p=5.0,D=0.2\,.
\end{equation}

In all the subfigures in Fig. \ref{fig:fixationProbability}, we simulate generalized Lotka-Volterra model with parameters
\begin{equation}
    S=50,\mu=2.0,\mu_r=6.0,\sigma_r=0.6,\gamma=1\,.
\end{equation}
In Fig. \ref{fig:fixationProbability}(a) and (b), we further use the parameters
\begin{equation}
    \sigma=0.4,\rho=0.9,N_p=2.6
\end{equation}
We change the control parameter $\alpha$ by using $D=1.0$ in 
Fig. \ref{fig:fixationProbability}(a) and $D=0.04$ in Fig. \ref{fig:fixationProbability}(b). In Fig. \ref{fig:fixationProbability}(e),
we use
\begin{equation}
    \sigma=0.4,s_\mathrm{inv}=0,N_p=4.8\,.
\end{equation}
In Fig. \ref{fig:fixationProbability}(f), the simulation 
parameters for the case of low packing are the same as those for Fig. \ref{fig:fixationProbability}(b). To control the extent of packing, we
keep all the parameters unchanged except using $\sigma=0.7$ for the case
with high packing. While $\sigma$ also controls the ecological contribution in $s_\mathrm{inv}$, it is accordingly balanced by shifting $r_m$, so that the two communities can be compared at given $s_\mathrm{inv}$.\kobe{clarified for reviewer 2}

In Fig. \ref{fig:coexistence}(a) and (b), we simulate generalized Lotka-Volterra model with parameters
\begin{equation}
    S=50,\mu=2.0,\sigma=0.4,\mu_r=6.0,\sigma_r=0.6,\gamma=1,\rho=0.9,N_p=2.0\,,
\end{equation}
and $D=0.020,0.024,0.036$ corresponding to the three values of $N_\mathrm{eff}$. We focus on $D=0.024$ in 
Fig. \ref{fig:coexistence}(c)-(h). The selection-drift ratios in 
Fig.\ref{fig:coexistence} are calculated using Eq. \eqref{eq:selectionDriftRatio} with the predicted values of $\eta$.

In Fig. \ref{fig:fixProbExplanation}(a), we simulate the dynamics as in 
classic population genetics with parameters
\begin{equation}
    s=0.14,N_p=2.0,D=0.01\,.
\end{equation}
In Fig. \ref{fig:fixProbExplanation}(b) and (c), we simulate generalized Lotka-Volterra model with parameters
\begin{equation}
    S=50,\mu=2.0,\sigma=0.4,\mu_r=6.0,\sigma_r=0.6,\gamma=1,\rho=0.9,N_p=2.0,D=0.01\,,
\end{equation}
and $s_\mathrm{inv}=0.12,0.23$ in Fig. \ref{fig:fixProbExplanation}(b) and (c) respectively.

In Fig. \ref{fig:fig_rho_1}, we simulate generalized Lotka-Volterra model with parameters
\begin{equation}
    S=50,\mu=2.0,\sigma=0.4,\mu_r=6.0,\sigma_r=0.6,\gamma=1,N_p=2.0,D=0.2,s_\mathrm{inv}=0\,.
\end{equation}
Figs. \ref{fig:susceptibility} and \ref{fig:communityNoise} use the same parameters with $\rho=0.9$, except that $s_\mathrm{inv}=0.1$.

In Fig. \ref{fig:fixProbCRM}(a), we simulate MacArthur consumer-resource
model with parameters
\begin{gather}
    S=50,M=50,\mu=2.0,\sigma=0.8,\mu_K=5.0,\sigma_K=0.5,\\
    \mu_m=2.0,\sigma_m=0.2,\kappa=1,\rho=0.9,N_p=5.8,D=0.1\,.
\end{gather}
In Fig. \ref{fig:fixProbCRM}(b), we simulate the consumer-resource
model with externally supplied resources with parameters
\begin{gather}
    S=50,M=50,\mu=2.0,\sigma=0.4,\mu_K=10.0,\sigma_K=0,\\
    \mu_m=2.0,\sigma_m=0.2,\rho=0.9,N_p=42.0,D=0.2\,.
\end{gather}
In Fig. \ref{fig:fixProbCRMPacking}, the simulation parameters for the case
of $S^*/M^*=0.68$ are the same as in Fig. \ref{fig:fixProbCRM}.
To control the extent of packing, we
keep all the parameters unchanged except using $S=20$ for the case
of $S^*/M^*=0.38$, and $(\mu_m,\sigma_m)=(0.2,0.02)$ for the case of $S^*/M^*=1$.

Fig. \ref{fig:fig_rho_2}(a) and (b) use the same parameters as in Fig. \ref{fig:fig_rho_1} with $\rho=0.9$, except that $s_\mathrm{inv}=0,0.4,0.8$. To obtain the 
exponentially suppressed fixation probabilities, we repeat the simulation
for each mutant $10^4$ times to count the number of fixations. Fig. \ref{fig:fig_rho_2}(c) and (d) use the same parameters as in the case of $S^*/M^*=1$ in Fig. \ref{fig:fixProbCRMPacking}, with $s_\mathrm{inv}=0,0.1,0.4$.

In Fig. \ref{fig:phasePortrait}, the selection-drift ratios are identical
to the ones in Fig. \ref{fig:coexistence}.

In Fig. \ref{fig:extinctionTimeVSFixProb}, the simulation parameters are
the same as in Fig. \ref{fig:coexistence}(a) and (b), except that we now use $D=0.02,0.024,0.03,0.036,0.05$. To obtain the 
exponentially suppressed fixation probabilities, we repeat the simulation
for each mutant $10^5$ times to count the number of fixations.

\section{Details of experimental data analysis}
\label{sec:dataAnalysis}

\kobe{whole section rewritten}
We analyze the longitudinal sequencing data in \cite{roodgar2021longitudinal} from the 
gut microbiome of a single patient. Here, we 
provide the details including the fitted model, the 
procedures of maximum likelihood estimation, the inference results with bootstrapping, and the procedures for testing the goodness of fit. 
The dataset, the code for data analysis, and the
detailed fit results can be found on
\url{https://github.com/Emergent-Behaviors-in-Biology/Pop-Gen-Ecology}. 
\kobe{should I disclose here?}\akshit{It might be better to have a separate disclosure statement like the data and code availability statement, about the use of AI.}\kobe{ok, will move it to main text}

\subsection{Data and model}

First, we download the dataset from Supplemental Data S1--S3 in 
\cite{roodgar2021longitudinal}. The dataset contains time series data for sweeping SNV 
clusters (with frequencies ranging from $f<0.2$ to $f>0.7$) from 17 abundant species in the same microbiome. For each cluster, the read count of the mutant allele $A_t$ and the
total read count $D_t$ are reported for each of the observed time points among the 19 sampled time points. 
Note that a species may have
multiple SNV clusters, but they may not correspond to different mutant strains. 
We make a simplifying but biologically reasonable assumption \cite{garud2019evolutionary,roodgar2021longitudinal,good2017dynamics,goyal2022interactions}
\kobe{cite two strains paper again}\akshit{we can cite a few different ones}\kobe{cited as in your two strain paper} that different clusters within the same species are proxy for the dynamics of a single mutant.

For each of the periods in Fig.~\ref{fig:dataAnalysis}, we fit the model of 
frequency-dependent selection in Eq.~\eqref{eq:frequencyDynamicsSI} to the mutant
trajectories. However, Eq.~\eqref{eq:frequencyDynamicsSI} only describes the 
dynamics of frequencies $f(t)$, and we need an additional model of measurement
noise to convert $f(t)$ to the read counts $A_t$. First, the read sampling noise in 
$A_t$ given a true $f$ is described by a binomial distribution $A_t\sim B(D_t,f)$.

Moreover, there may be various errors from 
the data source itself during DNA extraction from samples, PCR amplification, and sequencing. In previous work \cite{levy2015quantitative, nguyen2019high}, these errors were typically modeled as random Poisson noise. It was because they largely focused on trajectories with very low frequencies, hence very small populations in which counting errors dominate. However, we specifically focus on sweeping trajectories with intermediate frequencies and large populations, in which multiplicative errors (such as in amplification factors) dominate instead. 
Therefore, we assume log-normal errors for these factors and let the sampled abundances be
\begin{equation}
    \hat N_p=e^{\epsilon_p}N_p\,,\quad \hat N_m=e^{\epsilon_m}N_m\,,
\end{equation}
where $\epsilon_p,\epsilon_m$ are random Gaussian. Note that by applying the logit transform
\begin{equation}
    \mathrm{logit}(x)=\log\frac{x}{1-x}\,,
\end{equation}
the sampled frequency $\hat f$ becomes 
\begin{equation}
    \mathrm{logit}(\hat f)=\log\frac{\hat N_m}{\hat N_p}=\log\frac{N_m}{N_p}+\epsilon_m-\epsilon_p=\mathrm{logit}(f)+\epsilon_m-\epsilon_p\,.
\end{equation}
Therefore, it is equivalent to assume a random Gaussian noise on $\mathrm{logit}(f(t))$ with strength 
$\sigma$. 
\akshit{Here, we should clarify what the error model is essentially assuming a bit more, e.g., like in our Slack conversation, and why we feel it is justified especially for trajectories that sweep through a wide range of frequencies.}\kobe{added} To summarize, we assume that $A_t$ follows the distribution
\begin{equation}
\label{eq:measurementNoise}
    A_t\sim B\left(D_t,\mathrm{expit}\left(\mathrm{logit}(f(t))+\sigma z_t\right)\right)\,,
\end{equation}
where $f(t)$ follows Eq.~\eqref{eq:frequencyDynamicsSI} and $z_t$ are standard Gaussian random variables. The function
\begin{equation}
    \mathrm{expit}(x)=\frac{1}{1+e^{-x}}\,,
\end{equation}
is the inverse of the logit transform.

As described in the main text, we assume shared values of $\eta,N_\mathrm{eff},\sigma$ across species since they belong to the same community, but different invasion fitnesses $s_\mathrm{inv}$ for each species. This is because while we mainly focus on pooled estimates of the average parameters, we expect $s_\mathrm{inv}$ to be more heterogeneous and possibly change sign across species. A distribution of $s_\mathrm{inv}$ is also necessary for predicting the extent of coexistence in the community. 

To model the distribution of $s_\mathrm{inv}$, we make the simplifying assumption that the invasion fitness of each species is drawn from a Gaussian distribution with some mean $\mu$ and variance $\tau^2$, which we must infer from the data. \akshit{edited slightly here.} That is,
\begin{equation}
    s_\mathrm{inv}^i\sim N(\mu,\tau^2)\,,
\end{equation}
where $i$ denotes the species. However, as shown later, our results are largely insensitive to the choice of a Gaussian distribution.\kobe{added comment for later}

\subsection{Likelihood function}

We fit the above model to the trajectory data using maximum likelihood estimation (MLE). This procedure asks which combinations of the model parameters make the observed data most probable.
The crucial part of this procedure is to compute the likelihood function of the above model, which is a function of the model parameters. Given the model parameters, the function outputs the joint probability for the observed data. The fit is then obtained by maximizing the likelihood function. \akshit{maybe here we should say a few sentences about the gist of the likelihood approach; what we will seek to compute and how we will ultimately use it to infer the parameters. this is nice before you jump into the gory details of the procedure, at which point a naive reader will definitely miss the forest for the trees.}\kobe{added}

We shall compute the likelihood function using filtering algorithms. For an observed trajectory $u$ with reads $A_{u,t_i}$ at times $t_i$, the log-likelihood function conditioned on $s_\mathrm{inv}$, denoted by $l_u(\theta,s_\mathrm{inv})$ (where $\theta=\{\eta,N_\mathrm{eff},\sigma\}$ is the set of model parameters) can be decomposed into the log-likelihoods for each time point under one-step-ahead predictions
\begin{equation}
    l_u(\theta,s_\mathrm{inv})=\sum_i \log p\left(A_{u,t_i}|A_{u,t_1},\cdots, A_{u,t_{i-1}}\right)\,.
\end{equation}
Each term in the above can be computed as follows. To propagate from $A_{u,t_{i-1}}$ to $A_{u,t_i}$, we first need to convert them into the latent frequency variables $f_{u,t_{i-1}}$ and $f_{u,t_i}$. For this purpose, in the algorithm we maintain a posterior distribution of $f_{u,t_i}$
\begin{equation}
    p\left(f_{u,t_i}|A_{u,t_1},\cdots A_{u,t_i}\right)\,.
\end{equation}
Using this posterior, we propagate from $f_{u,t_{i-1}}$ to $f_{u,t_i}$ using the dynamics by
\begin{equation}
    p\left(f_{u,t_i}|A_{u,t_1},\cdots A_{u,t_{i-1}}\right)=\int_0^1 df_{u,t_{i-1}}p_f(f_{u,t_i}|f_{u,t_{i-1}})p\left(f_{u,t_{i-1}}|A_{u,t_1},\cdots A_{u,t_{i-1}}\right)\,,
\end{equation}
where $p_f$ is the probability distribution of $f$ from the dynamics in Eq.~\eqref{eq:frequencyDynamicsSI} with absorbing boundaries at $f=0,1$, obtained from integrating forward the Fokker-Planck equation in Eq.~\eqref{eq:FokkerPlanck} (see below for detail). Finally, we sample $A_{u,t_i}$ using the measurement noise model by
\begin{equation}
    p\left(A_{u,t_i}|A_{u,t_1},\cdots, A_{u,t_{i-1}}\right)=\int_0^1 df_{u,t_i}\,p_A(A_{u,t_i}|f_{u,t_i})p\left(f_{u,t_i}|A_{u,t_1},\cdots A_{u,t_{i-1}}\right)\,,
\end{equation}
where
\begin{equation}
    p_A(A_{u,t_i}|f_{u,t_i})=\int_{-\infty}^\infty dx\,\binom{D_{u,t_i}}{A_{u,t_i}}\frac{e^{A_{u,t_i}x}}{(1+e^x)^{D_{u,t_i}}}\cdot\frac{1}{\sigma}\varphi\left(\frac{x-\mathrm{logit}(f_{u,t_i})}{\sigma}\right)\,,
\end{equation}
is the probability distribution of $A$ from the sampling in Eq.~\eqref{eq:measurementNoise}, 
\begin{equation}
    \binom{D_{u,t_i}}{A_{u,t_i}}=\frac{D_{u,t_i}!}{A_{u,t_i}!(D_{u,t_i}-A_{u,t_i})!}\,,
\end{equation}
is the binomial coefficient,
and
\begin{equation}
    \varphi(x)=\frac{1}{\sqrt{2\pi}}e^{-x^2/2}\,,
\end{equation}
is the standard normal PDF.\kobe{defined more math functions} We first numerically find the unique peak of the integrand, then compute the integral around the peak till the integrand decays by a factor of $e^{-40}$. Note that the above is undefined at the two absorbing boundaries; we additionally define
\begin{equation}
    p_A(A_{u,t_i}|f_{u,t_i}=0)=\mathbf 1_{A_{u,t_i}=0}\,,\quad p_A(A_{u,t_i}|f_{u,t_i}=1)=\mathbf 1_{A_{u,t_i}=D_{u,t_i}}\,,
\end{equation}
where
\begin{equation}
    \mathbf 1_A=\begin{cases}
        1 & A\,\mathrm{is}\,\mathrm{true} \\
        0 & A\,\mathrm{is}\,\mathrm{false}
    \end{cases}\,,
\end{equation}
is the indicator function.
\akshit{define notation of 1 in the equation above here.}\kobe{added} To apply this algorithm, we first need to choose an initialization for the posterior $p(f_{u,t_1}|A_{u,t_1})$. For simplicity, we assume a uniform prior on $f$ before any observation, hence
\begin{equation}
    p(f_{u,t_1}|A_{u,t_1})\propto p_A(A_{u,t_0}|f_{u,t_0})\,.
\end{equation}
As the algorithm goes, we recursively update the posterior by
\begin{equation}
    p\left(f_{u,t_i}|A_{u,t_1},\cdots A_{u,t_i}\right)=\frac{p_A(A_{u,t_i}|f_{u,t_i})p\left(f_{u,t_i}|A_{u,t_1},\cdots A_{u,t_{i-1}}\right)}{p\left(A_{u,t_i}|A_{u,t_1},\cdots, A_{u,t_{i-1}}\right)}\,,
\end{equation}
which can be computed according to the above using the posterior at the previous time point. To perform all the above integrations in frequencies, we split the space of $f$ into $N=400$ cells.
Taking account of the absorbing boundaries, the fit would be more accurate with higher resolution of points near $f=0,1$. For this, we choose the points to be uniform in the variance-stabilizing coordinate $y=2\arcsin\sqrt f\in[0,\pi]$, i.e., the cells are the intervals $f\in [\sin^2((i-1)\pi/2N),\sin^2 (i\pi/2N)]$ for $1\leq i\leq N$.

After computing $l_u(\theta,s_\mathrm{inv})$ for each cluster trajectory $u$, we define the log-likelihood for a species as
\begin{equation}
    l_i(\theta,s_\mathrm{inv})=\sum_{u\in i}l_u(\theta,s_\mathrm{inv})\,.
\end{equation}
Note that there is no additional weight for each cluster, since their log-likelihoods are already scaled with the read counts through the measurement noise model. Now, we can integrate out $s_\mathrm{inv}$ to find the total log-likelihood
\begin{equation}
    l(\theta,\mu,\tau)=\sum_i\log \int ds_\mathrm{inv}\, e^{l_i(\theta,s_\mathrm{inv})}\cdot\frac{1}{\tau}\varphi\left(\frac{s_\mathrm{inv}-\mu}{\tau}\right)\,.
\end{equation}
The Gaussian integral is computed in the same manner as in $p_A$.

In the filtering algorithm, we obtain $p_f$ by numerically solving the Fokker-Planck equation in Eq.~\eqref{eq:FokkerPlanck} on the grid of $f$ with size $N$. Let $p_i(t)$ be the probability that $f$ is within cell $i$ at time $t$. Our goal is to convert Eq.~\eqref{eq:FokkerPlanck} into the form of master equation
\begin{equation}
    \frac{dp_i}{dt}=\lambda_{i-1}^+ p_{i-1}+\lambda_{i+1}^- p_{i+1}-(\lambda_i^+ + \lambda_i^-)p_i\,,
\end{equation}
where $\lambda_i^\pm$ are transition rates to be determined. For the interior cells with $1<i<N$, the transition rates can be found by discretizing the derivatives in Eq.~\eqref{eq:FokkerPlanck}. Let $f_i=\sin^2((i-1/2)\pi/2N)$ be the midpoint of cell $i$ and define the cell distances $h_i^-=h_{i-1}^+=f_i-f_{i-1}$ and $h_i^+=h_{i+1}^-=f_{i+1}-f_i$. For second derivatives, we use the central approximation
\begin{equation}
    \frac{\partial^2 g}{\partial f^2}\simeq \frac{2}{h_i^-(h_i^-+h_i^+)}g(f_{i-1})-\frac{2}{h_i^-h_i^+}g(f_i)+\frac{2}{h_i^+(h_i^-+h_i^+)}g(f_{i+1})\,,
\end{equation}
for some function $g$. For first derivatives, to ensure that the resulting transition rates are positive, we use the upwind approximation (instead of, e.g., the central approximation) \cite{kushner1990numerical}
\begin{equation}
    \frac{\partial g}{\partial f}=\begin{cases}
        \frac{g(f_{i+1})-g(f_i)}{h_i^+}\,, & s_\mathrm{inv}\leq \eta f_i\\
        \frac{g(f_i)-g(f_{i-1})}{h_i^-}\,, & s_\mathrm{inv}>\eta f_i\\
    \end{cases}\,.
\end{equation}
Substituting these derivatives into Eq.~\eqref{eq:FokkerPlanck} and comparing the coefficients, we then obtain the transition rates
\begin{align}
    \lambda_i^+&\simeq \frac{f_i(1-f_i)}{h_i^+}\left(\max(s_\mathrm{inv}-\eta f_i,0)+\frac{1}{N_\mathrm{eff}(h_i^-+h_i^+)}\right)\,,\\
    \lambda_i^-&\simeq\frac{f_i(1-f_i)}{h_i^-}\left(\max(\eta f_i-s_\mathrm{inv},0)+\frac{1}{N_\mathrm{eff}(h_i^-+h_i^+)}\right)\,.
\end{align}
However, note again that the above derivation is only valid for the interior cells. For the two end cells, we impose absorbing boundary conditions and set
\begin{equation}
    \lambda_1^\pm=\lambda_N^\pm=0\,.
\end{equation}
The above defines the full transition matrix on the grid of $f$. By exponentiating the transition matrix, we finally obtain $p_f$ used in the likelihood function.

\subsection{Maximum likelihood estimation}

We now maximize the log-likelihood $l(\theta,\mu,\tau)$ for each period. We first perform grid search on the parameters $\eta,N_\mathrm{eff},\sigma,\mu,\tau$, then refine the neighbor of grid points with maximum likelihood using optimization algorithms. We use the following grid:
\begin{itemize}
    \item $\eta\in[-0.25,0.75]$ in steps of $0.03125$ for the disease and recovery periods; $\eta\in[-0.25,0.75]$ in steps of $0.0625$, then $[0.75,3]$ in steps of $0.125$ for the baseline period;
    \item $N_\mathrm{eff}^{-1}\in[5.6\times 10^{-6},0.3]$ in $26$ logarithmic steps;
    \item $\sigma\in\{0,0.025,0.05,0.1,
0.15,0.2,0.25,0.3,0.4,0.5,0.6,0.7,0.8,0.9,1.05,1.25,1.5\}$;
    \item $\mu\in[-0.4,1.2]$ in $80$ uniform steps;
    \item $\tau\in[0.003,1]$ in $40$ logarithmic steps.
\end{itemize}
For integrating out $s_\mathrm{inv}$, we also compute $l_u(\theta,s_\mathrm{inv})$ with $s_\mathrm{inv}\in[-0.6,1.6]$ in steps of $0.01$. We use natural cubic splines to interpolate between grid points of $s_\mathrm{inv}$.

We refine the maximum likelihood from grid search as follows. First, for each given $\eta,N_\mathrm{eff},\sigma$, we find the optimal $\mu,\tau$ using Nelder-Mead method on the space $(\mu,\log \tau)$. This is computationally efficient since $l_u(\theta,s_\mathrm{inv})$ remains unchanged at every iteration of the method. Next, since $\eta$ is the main focus of this analysis, 
we compute its marginal likelihood (see Fig.~\ref{fig:dataAnalysis}c in main text)
\begin{equation}
    l(\eta)=\max_{N_\mathrm{eff},\sigma,\mu,\tau}l(\eta,N_\mathrm{eff},\sigma,\mu,\tau)\,,
\end{equation}
using the grid points of $N_\mathrm{eff},\sigma$ and optimized $\mu,\tau$. We then approximate the actual peak of $l(\eta)$ by fitting a smoothing spline around the maximum points. The remaining parameters are then fitted at the optimal $\eta$ with the above grid search, followed by Nelder-Mead method on the space $(\log N_\mathrm{eff},\sigma)$.

\subsection{Inference results and confidence}

As shown in Fig.~\ref{fig:dataAnalysis}c in main text, $l(\eta)$ has clear positive peaks significantly away from $\eta=0$ for the disease and recovery periods. On the other hand, the peak in $l(\eta)$ for the baseline period is not significantly different from $\eta=0$. Moreover, $l(\eta)$ for the baseline period reaches plateau instead of decaying for large $\eta$. Therefore, we cannot confidently identify the optimal $\eta$ for the baseline period. Below we focus on the disease and recovery periods.

\begin{table}[]
    \centering
    \begin{tabular}{|c|cc|}
        \hline
        & Disease & Recovery\\
        \hline
        $\eta$ & $0.174$ & $0.078$ \\
        bootstrap 50\% CI & $[0.127, 0.196]$ & $[0.058, 0.097]$ \\
        bootstrap 95\% CI & $[0.048, 0.268]$ & $[0.026, 0.137]$ \\
        \hline
        $N_\mathrm{eff}^{-1}$ & $6.39\times 10^{-3}$ & $3.93\times 10^{-3}$\\
        bootstrap 95\% CI & $[2.73, 14.1]\times 10^{-3}$ & $[1.64, 6.60]\times 10^{-3}$\\
        \hline
        $\alpha$ & $5.21$ & $4.44$\\
        bootstrap 95\% CI & $[2.6, 7.9]$ & $[2.4, 8.2]$\\
        \hline
        $\sigma$ & $0.473$ & $0.760$ \\
        bootstrap 95\% CI & $[0, 0.69]$ & $[0.51, 0.97]$\\
        \hline
        $\mu$ & $0.097$ & $0.040$ \\
        $\tau$ & $0.104$ & $0.022$ \\
        $P(0<s_\mathrm{inv}<\eta)$ & $0.59$ & $0.92$\\
        expected coexistence & $10.2/15$ & $15.9/17$\\
        \hline
    \end{tabular}
    \vspace{10px}
    \caption{\textbf{Fitted model parameters for the human gut microbiome data.} We fit the model of frequency-dependent selection to the human gut microbiome data \cite{roodgar2021longitudinal} and infer the optimal parameters $\eta,N_\mathrm{eff},\sigma,\mu,\tau$. Confidence intervals are obtained using bootstrapping on species. The inferred values indicate strong ecological feedback with parent-mutant coexistence in the community.}
    \label{tab:inferenceResult}
\end{table}

The optimal parameters are presented in Table~\ref{tab:inferenceResult}. The confidence intervals are obtained from bootstrapping on species. That is, we produce $B=4000$ communities of resampled species with replacement, then refit the optimal parameters. Compared with the original MLE procedure, the optimal $N_\mathrm{eff},\sigma$ are now also inferred using smoothing splines, so that the computational cost is controlled. This bootstrap approach is valid for the parameters $\eta,N_\mathrm{eff},\sigma$, but not for the distribution parameters $\mu,\tau$, which are expected to depend on the species composition.

With the inferred distribution of $s_\mathrm{inv}$, we can estimate the expected fraction of species that satisfies $0<s_\mathrm{inv}<\eta$, hence supports parent-mutant coexistence. A simple estimate for the fraction of coexistence is the probability $P(0<s_\mathrm{inv}<\eta)$ under the inferred Gaussian distribution, see Table~\ref{tab:inferenceResult}. However, we are more interested in the coexistence of the selected species, and the probabilities can be refined for each species using the trajectory data. Namely, we can compute the posterior distribution of $s_\mathrm{inv}$
\begin{equation}
\label{eq:posteriorSinv}
    p(s_\mathrm{inv}^i)=p(s_\mathrm{inv}|\mathrm{trajectories}_i)\propto e^{l_i(\theta,s_\mathrm{inv})}\cdot \frac{1}{\tau}\varphi\left(\frac{s_\mathrm{inv}-\mu}{\tau}\right)\,,
\end{equation}
where $l_i(\theta,s_\mathrm{inv})$ has been computed in the MLE. Then, the expected number of species with coexistence is
\begin{equation}
    \left<S_\mathrm{coexist}\right>=\sum_i P(0<s_\mathrm{inv}^i<\eta)\,.
\end{equation}
This number is presented in Table~\ref{tab:inferenceResult} and Fig.~\ref{fig:dataAnalysis}e in the main text.

\begin{table}[]
    \centering
    \begin{tabular}{|c|cc|}
        \hline
        & Disease & Recovery\\
        \hline
        $\eta$ & $0.174$ & $0.078$ \\
        jackknife range & $[0.148, 0.196]$ & $[0.064, 0.095]$ \\
        synthetic mean & $0.200$ & $0.097$ \\
        synthetic 95\% CI & $[0.113, 0.318]$ & $[0.058, 0.136]$ \\
        \hline
    \end{tabular}
    \vspace{10px}
    \caption{\textbf{Variations of frequency-dependent selection under refitting on jackknife and on synthetic data.} We check the robustness of the inferred $\eta$ with two tests. First, we refit $\eta$ on data with one species removed, i.e., the jackknife resampling. Second, we refit $\eta$ on $B=200$ set of synthetic data. The variations are well within the confidence intervals of $\eta$.}
    \label{tab:etaRobustness}
\end{table}

\begin{table}[]
    \centering
    \begin{tabular}{|lcccccc|}
    \hline
    \multicolumn{7}{|c|}{Disease} \\
    \hline
    $s_\mathrm{inv}$ distribution & $\eta$ & $10^3N_\mathrm{eff}^{-1}$ & $\sigma$ & $s_\mathrm{inv}$ mean & $s_\mathrm{inv}$ s.d. & coexistence \\
    \hline
    Gaussian & 0.174 & 6.39 & 0.47 & 0.097 & 0.104 & 10.2/15 \\
    Logistic & 0.171 & 6.19 & 0.48 & 0.081 & 0.102 & 10.4/15 \\
    Laplace & 0.169 & 5.96 & 0.49 & 0.066 & 0.104 & 10.5/15 \\
    Uniform & 0.181 & 6.66 & 0.46 & 0.144 & 0.111 & 9.9/15 \\
    Gumbel, right-skewed & 0.170 & 5.98 & 0.49 & 0.093 & 0.095 & 10.4/15 \\
    Gumbel, left-skewed & 0.173 & 6.74 & 0.46 & 0.085 & 0.143 & 10.0/15 \\
    \hline
    \multicolumn{7}{|c|}{Recovery} \\
    \hline
    $s_\mathrm{inv}$ distribution & $\eta$ & $10^3N_\mathrm{eff}^{-1}$ & $\sigma$ & $s_\mathrm{inv}$ mean & $s_\mathrm{inv}$ s.d. & coexistence \\
    \hline
    Gaussian & 0.078 & 3.93 & 0.76 & 0.040 & 0.022 & 15.9/17 \\
    Logistic & 0.074 & 3.79 & 0.76 & 0.039 & 0.023 & 15.5/17 \\
    Laplace & 0.070 & 3.65 & 0.76 & 0.040 & 0.024 & 15.1/17 \\
    Uniform & 0.083 & 4.13 & 0.76 & 0.040 & 0.021 & 17.0/17 \\
    Gumbel, right-skewed & 0.085 & 4.27 & 0.76 & 0.044 & 0.025 & 16.4/17 \\
    Gumbel, left-skewed & 0.069 & 3.60 & 0.76 & 0.036 & 0.022 & 15.5/17 \\
    \hline
    \end{tabular}
    \vspace{10px}
    \caption{\textbf{Variations of fit under different assumed distributions of $s_\mathrm{inv}$ across species.} To check whether our results are sensitive to the Gaussian assumption on the distribution of invasion fitness $s_\mathrm{inv}$, we perform the same fit using several other two-parameter distributions. In both periods, the model parameters and the statistics of $s_\mathrm{inv}$ vary little across different choices of the distribution.}
    \label{tab:sinvRobustness}
\end{table}

We also perform a number of independent checks for the robustness of our results, as summarized in Tables~\ref{tab:etaRobustness} and~\ref{tab:sinvRobustness}. First, we apply jackknife resampling, i.e., refit $\eta$ on data with one of the species removed. We go through all possible one-species removals and see little variation in $\eta$ (second row in Table~\ref{tab:etaRobustness}), \akshit{might be worth specifying some numbers here.}\kobe{it is in the table, clarified} indicating that the positive $\eta$ is not driven by just one of the species. Second, we refit $\eta$ by generating $B=200$ set of synthetic data using the optimal parameters (see the detail of parametric bootstrap below). The refitted mean slightly deviates from the optimal $\eta$, but is still well within the confidence intervals (third and fourth rows in Table~\ref{tab:etaRobustness}). The bias does not bring the intervals across $\eta=0$ either. Therefore, our inference method has reliable level of accuracy. Finally, in Table~\ref{tab:sinvRobustness}, we show that our results are robust under different choices of the distribution of $s_\mathrm{inv}$.

\subsection{Goodness of fit}

Apart from the above confidence intervals, we can test the goodness of fit of our model by asking how well the distribution of $A_{u,t_i}$ from our fitted model matches with the data distribution. Specifically, we focus on the goodness of one-step-ahead predictions for each time point. In the data, there may be more correlations across multiple days, such as memory effects beyond the current frequencies. Our model does not take these effects into account, and we do not measure these effects in the statistical test.

To assess the goodness of fit, there have been well-established statistical tests that compare data with continuous and common distributions, such as a uniform distribution. However, our data are discrete and follow distributions of arbitrary shape.
To better utilize those statistical tests, we thus choose to use the
randomized probability integral transform (PIT) to convert the data into new coordinates $x_{u,t_i}$, which we expect to follow a standard uniform distribution if our model is the true model underlying the data. We can therefore test our model by comparing the observed data in $x$-space with a standard uniform distribution using standard statistical tests.

The randomized PIT is performed as follows. Consider the cumulative distribution
\begin{equation}
    F_{u,t_i}(A)=\sum_{A'=0}^A p\left(A_{u,t_i}=A'|A_{u,t_1},\cdots, A_{u,t_{i-1}}\right)\,,
\end{equation}
where the probabilities were already computed in the likelihood function. Since $A_{u,t_i}$ are integer data, the possible values of $F_{u,t_i}$ are also discrete. To obtain a continuous coordinate, we further smooth these spikes in $F_{u,t_i}$ by interpolation with uniform random noise. Define the variables
\begin{equation}
    x_{u,t_i}=F_{u,t_i}(A_{u,t_i}-1)+U_{u,t_i}(F_{u,t_i}(A_{u,t_i})-F_{u,t_i}(A_{u,t_i}-1))\,,
\end{equation}
where $U_{u,t_i}$ are independent standard uniform random variables. Intuitively, since observations with higher probabilities now occupy larger intervals in $x$, $x_{u,t_i}$ should follow a uniform distribution. \akshit{we should clarify and explain why a bit more here. it is not clear what exactly you are trying to do. maybe it is worth saaying that we have good statistical tests for continuous uniform distributions, but the data are discrete distributions of arbitrary shape, which we do not have good tests for. so we choose to instead convert the data to transformed data coordinates where we would instead expect a continuous uniform distribution. this can be done by choosing the cumulative distribution F and focusing on the variable x. and maybe say a bit that read counts with larger probabilities of observation will instead get converted to bigger intervals in the transformed space of x, resulting in a uniform distribution over x. thus by comparing how likely the observed data in x-space are consistent with a uniform distribution, we can test the goodness of fit of our model.}\kobe{added motivation and clarified various stuff}

We collect a set of $x_{u,t_i}$ using $20$ realizations of $U_{u,t_i}$. To focus on one-step-ahead predictions, we split the set of $x_{u,t_i}$ by the time points $t_i$ and compute the test statistics for each $t_i$. We compare $x_{u,t_i}$ to standard uniform distribution using the Cram\'er-von Mises (CvM) statistics $\omega^2$ and the Anderson-Darling (AD) statistics $A^2$.

We further calibrate the $p$-values for the tests using parametric bootstrap. Namely, we generate $B=200$ sets of synthetic data using the fitted model, then refit the parameters and compute the test statistics, so that we obtain the null distributions for the tests. For the synthetic data, we use the fitted optimal $\eta,N_\mathrm{eff},\sigma$, and the posterior mean for each $s_\mathrm{inv}^i$ (see Eq.~\eqref{eq:posteriorSinv}), so that the tests focus on the selected species. For consistency, the synthetic trajectories are drawn using the same distributions as in the likelihood computations. To keep the computational cost under control, we keep the same $N_\mathrm{eff},\sigma$ from the original model and only refit $\eta,\mu,\tau$ for each set of synthetic data.

\begin{table}[]
    \centering
    \begin{tabular}{|ccc||ccc|}
        \hline
        Time & CvM & AD & Time & CvM & AD \\
        \hline
        Disease & 0.597 & 0.318 & Recovery & 0.149 & 0.149 \\
        46 & 0.149 & 0.080 & 80 & 0.741 & 0.701 \\
        49 & 0.363 & 0.512 & 92 & 0.443 & 0.274 \\
        67 & 0.726 & 0.617 & 110 & 0.219 & 0.184 \\
        69 & 0.428 & 0.368 & 118 & 0.025 & 0.025 \\
        & & & 129 & 0.378 & 0.308 \\
        & & & 153 & 0.443 & 0.398 \\
        \hline
    \end{tabular}
    \vspace{10px}
    \caption{\textbf{Goodness of fit on one-step-ahead predictions.} We perform Cram\'er-von Mises (CvM) and Anderson-Darling (AD) tests for our model of frequency-dependent selection at each predicted time point in the disease and recovery periods. The $p$-values shown are calibrated using parametric bootstrap. The first row shows the Bonferroni-corrected overall $p$-value for each period.}
    \label{tab:pValues}
\end{table}

The resulting $p$-values for each time point in the disease and recovery periods (except the first time points at which there are no predictions) are listed in Table~\ref{tab:pValues}. We also obtain an overall $p$-value per period using the Bonferroni correction:
\begin{equation}
    p_\mathrm{period}=\min\left(1,(n_\mathrm{days}-1)\cdot\min_{t_i}p_{t_i}\right)\,,
\end{equation}
where $n_\mathrm{days}=5,7$ are the number of days in each period. We see that there is no $p$-value rejecting our model except on day 118. The overall $p$-value cannot rule out that this rejection possibly occurs only by chance. Therefore, we conclude that our model is calibrated for one-step-ahead predictions, showing that frequency-dependent selection can indeed describe the data.

\end{document}